%% file: main.tex
\documentclass[12pt]{iopart}

\usepackage{iopams}  
\expandafter\let\csname equation*\endcsname\relax
\expandafter\let\csname endequation*\endcsname\relax
\usepackage{amsmath}

\input{texcode/packages.tex}

\input{texcode/packages_settings.tex}

\input{graphs/graphs_memeft.tex}

\input{graphs/graphs_masters.tex}

\input{graphs/graphs_topologies.tex}

\hypersetup{
colorlinks,
linkcolor={red!50!black},
citecolor={blue!50!black},
urlcolor={blue!80!black}
}

\begin{document}

\begin{fmffile}{fmfmain}
\input{texcode/graphs_settings.tex}
\input{texcode/shortcuts.tex}
\input{plots/plots.tex}

\title{Field-theory approach to flat polymerized membranes}

\author{
Simon Metayer\textsuperscript{1,2$\star$} and
Sofian Teber\textsuperscript{1$\dagger$}
}


\address{
$^1$Sorbonne Universit\'e, CNRS, Laboratoire de Physique Th\'eorique et Hautes Energies (LPTHE), F-75005 Paris, France \\
$^2$Institute of Nuclear and Particle Physics (INPAC), Shanghai Jiao Tong University (SJTU), 200240 Shanghai, China
}
\ead{$^\star$smetayer@lpthe.jussieu.fr, $^\dagger$teber@lphte.jussieu.fr}
\vspace{10pt}
\begin{indented}
\item[]February 2025
\end{indented}

\begin{abstract}
We review the field-theoretic renormalization-group approach to critical properties of flat polymerized membranes. We start with a presentation of the flexural effective model that is entirely expressed in terms of a transverse (flexural) field with non-local interactions. We then provide a detailed account of the full three-loop computations of the renormalization-group functions of the model within the dimensional regularization scheme. The latter allows us to consider the general case of a $d$-dimensional membrane embedded in $D$-dimensional space. Focusing on
the critical flat phase of two-dimensional membranes ($d=2$) in three-dimensional space ($D=3$), we analyse the corresponding flow diagram and present the derivation of the anomalous stiffness. The latter controls all the other critical exponents of the theory such as the roughness exponent and the scaling of the elastic constants.
State-of-the-art four-loop results as well as discussions on the structure of the perturbative series and comparison with other approaches are also provided. 
\end{abstract}

\section{Introduction and motivations}

The statistical mechanics of membranes has attracted continuous interest for decades, see the review \cite{Bowick:2001}. These random $d$-dimensional surfaces embedded in a $D$-dimensional space are complex classical systems whose physical properties are dominated by thermal fluctuations. In the following, the case of interest will correspond to two-dimensional surfaces ($d=2$) in three-dimensional space ($D=3$). Membranes are also characterized by different types of microscopic orders such as crystalline, fluid (with vanishing shear modulus) and hexatic orders. We will focus on crystalline (also referred to as tethered or polymerized) membranes with a broken translational invariance in the plane and a fixed connectivity due to rigidly bound monomers, see \cite{Wiese:2020} for a review. There are many concrete realizations of them in nature which greatly enhances the significance of their study. A natural occurrence of a polymerized membrane is the cytoskeleton of cell surfaces and in particular red blood cells. Other, more artificial, realizations are inorganic crystalline membranes, the most popular example nowadays being graphene \cite{katsnelson2012graphene} and graphene-like materials, see the review \cite{Amorim:2016}.

Polymerized membranes have a rich set of universality classes and have been extensively studied since the 80s especially with the theoretical discovery of the (low-temperature) flat phase \cite{Nelson:1987,Aronovitz:1988,Aronovitz:1989, David:1988,Guitter:1988,Guitter:1989}. Such a flat phase is characterized by a long-range orientational order that is stabilized by an anomalously large bending rigidity together with softened elastic constants due to long-range interactions that evade the Mermin-Wagner theorem in two dimensions. Interestingly, such a polymerized membrane has also been predicted to have a negative Poisson ratio \cite{LeDoussal:1992} making the corresponding materials auxetic. It is for the
cytoskeleton of red blood cells that early light scattering experiments (measuring the roughness exponent) gave evidence for a critical flat phase \cite{Schmidt:1993}. The interest has been considerably boosted since the discovery of graphene whose out-of-plane deformations, the so-called ripples, may be responsible for the stability of a flat phase in the case of room temperature free-standing graphene subject to a weak tension. Although, the case of graphene seems to be more complicated (with ripples not fully thermal in origin and subtle effects arising from disorder, boundaries together with the presence of electrons) there has been some experimental evidence for anomalous elastic effects, see \eg, the review \cite{Amorim:2016}.

From a theoretical perspective, the anomalous elasticity at the origin of the flat phase is a remarkable example of renormalization. Statistical field theory, see, \eg, the textbook \cite{Vasilev:2004yr}, has therefore been the framework of choice since the early studies devoted to polymerized membranes. The basic model involves two massless fields: a $d$-dimensional (in-plane) phonon field $\vec u$ 
and a $(D-d)$-dimensional (out-of-plane) flexural field $\vec h$ together with coupling constants that depend on the Lam\'e coefficients, $\lambda$ and $\mu$, that encode the microscopic elastic properties of the system; it has an upper critical dimension of $4$ and a lower critical dimension below $2$. The anomalous elasticity that stabilises the two-dimensional flat phase arises from a non-trivial renormalization of the bending rigidity and elastic constants due to the long-range (elastic) interactions among the $\vec u$ and $\vec h$ fields. This in turn means that the flat phase is an attractive fixed point of the renormalization-group (RG) flow. It is characterized by power-law behaviors for the phonon-phonon and flexural-flexural correlation functions \cite{Nelson:1987,Aronovitz:1988,Aronovitz:1989, David:1988,Guitter:1988,Guitter:1989}
\begin{equation}
\bra u(p) u(-p) \ket\sim p^{-(2+\eta_u)} \hspace{0.7cm} {\hbox{and}} \hspace{0.7cm} \bra h(p) h(-p) \ket\sim {p^{-(4-\eta)}},
\label{mem:correlation}
\end{equation}
where the elasticity softening exponent $\eta_u$ and the anomalous stiffness exponent $\eta$ are two nontrivial anomalous dimensions related by the Ward identity 
\be
\eta_u=4-d-2\eta\, .
\ee
Note that such an identity (together with hyperscaling relations) implies that we can focus on the determination of only one of the exponents, in our case $\eta$, from which all other exponents of interest may be deduced. As an example, the roughness exponent that measures the fluctuations transverse to the flat directions, can be deduced from $\eta$ with the help of 
\be
\zeta=(4-d-\eta)/2 \, .
\label{eq:mem:roughness}
\ee
In short, the stiffness exponent $\eta$ governs all infra-red properties of membranes in the flat phase, where parameters are length-scale-dependent, like the softened Lam\'e elastic moduli $\mu(p)\sim \lambda(p) \sim p^{4-d-2\eta}$ and the enhanced bending energy $\kappa(p)\sim p^{-\eta}$.

In this context, a major challenge lies in the accurate theoretical determination of the exponent $\eta$ at the stable fixed point. In the Gaussian case (no interactions), $\eta=0$ obviously. Upon turning on interactions, the seminal study of Nelson and Peliti \cite{Nelson:1987} revealed that, in a self-consistent 
approximation, the anomalous stiffness becomes positive, $\eta \approx 1$, stabilizing the flat phase in the low temperature limit. They also argued that, beyond their approximation,
the value of $\eta$ should decrease. This was confirmed by the one-loop computation of Aronovitz and Lubensky \cite{Aronovitz:1988,Aronovitz:1989} who found a slight decrease preserving the positivity of the exponent, $\eta_{\text{1-loop}}=0.96$, and hence the stability of the flat phase. Because of the complexity of calculations, higher orders remained inaccessible for decades. It was also thought that the perturbative results would badly converge (which, as we will see, turns out to be incorrect) due to the distance between the upper critical dimension, $d_{\text{uc}}=4$, and the physical dimension, $d=2$, as the results take the form of series in $\eps=2-d/2$ with the case of interest corresponding to $\eps=1$. In order to circumvent the perturbative analysis, early works have focused on alternate expansion schemes as well as various non-perturbative approaches that are able to tackle the physics directly in dimension $d=2$. Early attempts included an expansion in the large embedding space $(D=3)$ dimension, \cite{Guitter:1988,Guitter:1989}, yielding $\eta_{\text{large-D}}=2/D\approx0.67$. This was then followed by a series of works on a $1/d_c$ expansion \cite{David:1988,Guitter:1988,Aronovitz:1989,Guitter:1989,Gornyi:2015,Saykin:2020},
where $d_c$ is the co-dimension of the membrane ($d_c=D-d$). In the limit $d_c=1$, this method does not seem to be reliable as the results obtained for $\eta$ are larger that $1$. More recently, the critical properties of membranes have been studied by means of the self-consistent screening approximation (SCSA), see, \eg, \cite{Gazit:2009,Zakharchenko:2010,Roldan:2011,LeDoussal:1992,LeDoussal:2018} as well as the so-called non-perturbative renormalization group (NPRG), see, \eg, \cite{Kownacki:2009,Braghin:2010,Hasselmann:2011,Essafi:2014,Coquand:2016a}. These approaches have produced roughly compatible results: $\eta^{\text{LO}}_{\text{SCSA}} \simeq 0.821$ \cite{LeDoussal:1992,LeDoussal:2018} at leading order ($\eta^{\text{NLO}}_{\text{SCSA}} \simeq 0.789$ at controversial next-to-leading order \cite{Gazit:2009}) and $\eta_{\text{NPRG}}=0.849$ \cite{Kownacki:2009}. As for Monte Carlo simulations of membranes, some of the notable results include, \eg, $\eta=0.81(3)$ \cite{Zhang:1993}, $\eta= 0.750(5)$ \cite{Bowick:1996}, and $\eta=0.795(10)$ \cite{Troster:2013} and Monte Carlo simulations of graphene, $\eta\simeq 0.85$ \cite{Los:2009}. 

On the experimental side, the early study \cite{Schmidt:1993} on the spectrin skeleton of red blood cells measured a roughness exponent of $\zeta=0.65\pm0.10$, which in terms of anomalous stiffness, using \eqref{eq:mem:roughness}, yields $\eta_{\text{blood-cells}}=0.7\pm0.2$. A similar result has been obtained using X-ray scattering on amphiphilic films (arachidic acid) \cite{Gourier:1997} with the result $\eta_{\text{amphiphilic-films}}=0.7\pm0.2$. Still on the biophysics side, very recent measurement of nuclear wrinkling during egg development in the fruit fly (Drosophila melanogaster) \cite{Jackson:2022} obtained measurements and numerical simulations compatible with $\eta_{\text{vesicles}}\approx0.8$. On the condensed matter physics side, precision measurements of the elastic critical properties of flat materials seem difficult to carry out. In graphene and graphene-like materials, the measurement of the anomalous stiffness exponent usually leads to rough estimations in between $\eta_{\text{graphene}}\approx0.8$ and $\eta_{\text{graphene}}\approx1$, see \eg, the review \cite{Amorim:2016} and also the experimental study \cite{LopezPolin:2015} where $\eta_{\text{graphene}}\approx0.82$.

Going back to theory, there has recently been a renewed interest in both pure \cite{Mauri:2020,Coquand:2020a} and disordered membranes \cite{Metayer:2022rdp} going beyond leading order. In the pure case that will be of interest in this manuscript, the two-loop order approach performed in \cite{Coquand:2020a} revealed an intriguing agreement between the perturbative and non-perturbative approaches in the vicinity of the upper critical dimension. Moreover, the value of the two-loop order anomalous dimension in $d=2$, $\eta_{\text{2-loop}}= 0.9139$ \cite{Coquand:2020a}, when compared to the one-loop order one, $\eta_{\text{1-loop}}=0.96$ \cite{Aronovitz:1988,Guitter:1988,Aronovitz:1989,Guitter:1989}, has been found to move in the right direction when referring to the generally accepted values that lie in the range $[0.7,0.9]$, see table \ref{tab:mem:literatureeta}, page \pageref{tab:mem:literatureeta}.

In this review, we provide the general multi-loop framework to systematically (order by order in perturbation theory) compute $\eta$. We also provide a detailed account of our full three-loop calculations \cite{Metayer:2021kxm} and a brief description of our much lengthier state-of-the art four-loop ones \cite{Metayer:2024}. All our calculations will be performed within the flexural effective model of a polymerized membrane obtained by integrating out exactly the phonon field. As we will seen in the following, the 
agreement with numerical simulations and non-perturbative approaches starts to manifest at three loops, $\eta_{\text{3-loop}}=0.8872$, and is further confirmed by the recent four-loop result, $\eta_{\text{4-loop}}=0.8670$ \cite{Metayer:2024}. The latter is also in perfect agreement with the one first obtained by Pikelner \cite{Pikelner:2021} within the two-field model. These results are all exact without any resummation involved. At four loops, there are enough terms in the perturbative series to reliably resum it and this leads to the all order estimate $\eta=0.8347$ which is well within the range of accepted values. 

The outline of the review is the following. Starting from a short presentation of the two-field model, we will introduce the flexural effective model of a polymerized membrane in Sec.~\ref{sec:model}. We will then present the perturbative setup in Sec.~\ref{sec:setup} which is central to this review and provides the framework in which to systematically extract the critical exponents of the model. A brief discussion of the fate of infra-red divergences will also be included. In Sec.~\ref{sec:PT}, we present a detailed diagrammatic analysis of the model up to three-loops and add a brief description of the four-loop computations. These results are then used in Sec.~\ref{sec:RGflows+FP}, to analyze the flow diagram and compute $\eta$ at various fixed points including the flat phase one. In Sec.~\ref{sec:benchmarks} we briefly compare the perturbative results to those obtained with other methods. The conclusion is given in Sec.~\ref{sec:conclusion}. Additional details related to Feynman diagrams, their integrals and related topologies can be found in \ref{chap:appendix}.


\section{The flexural effective model}
\label{sec:model}

We consider a $d$-dimensional homogeneous and isotropic membrane embedded in an Euclidean $D$-dimensional space. 
Each mass point of the membrane is indexed by $\vec x \in \R^d$. In $\R^D$, the unperturbed state of the membrane is the flat phase 
where each of these mass points is indexed by $\vec R^{(0)}(\vec x) = (\vec x, \vec 0_{d_c})$ where $\vec 0_{d_c}$ is the null vector
of dimension $d_c = D-d$ (the co-dimension of the manifold). In the following, latin indices run from $1$ to $d$, \eg, $\{a,b\}=1,\cdots,d$. Similarly,
greek indices run from $1$ to $d_c$, \eg, $\{\al, \beta \}=1,\cdots,d_c$.

The displacements inside the membrane are parameterized by a phonon field $\vec u(\vec x) \in \R^d$ and a flexuron field $\vec h(\vec x) \in \R^{d_c}$ such that the 
perturbed mass points are located at
\be
\vec R(\vec x) = \big(\vec x + \vec u(\vec x), \vec h(\vec x) \big)\, ,
\ee
which is the Monge parametrization. The induced metric is then defined as 
\be
g_{ab} = \partial_a \vec R(\vec x) \cdot \partial_b \vec R(\vec x)\, ,
\ee
where $g_{ab}^{(0)}=\delta_{ab}$ in the unperturbed phase. Though the case of interest corresponds to a two-dimensional membrane ($d=2$) in a three-dimensional embedding space ($D=3$), we shall keep these dimensions arbitrary in the following. This is because we will work in dimensional regularization with $d=4-2\eps$. Moreover, it is also useful to keep $d_c$ arbitrary to compare results with other methods such as large-$d_c$ techniques, see section \ref{sec:mem:largedc}.


The strain tensor $T$ that encodes the local deformations with respect to the flat configuration $\vec R^{(0)}(\vec x)=(\vec x,\vec 0_{d_c})$ is defined as
\ba
T_{ab} &= \frac{1}{2}\left(g_{ab}-g_{ab}^{(0)}\right) 
= \frac{1}{2} \left(\partial_a \vec R(\vec x) \cdot \partial_b \vec R(\vec x) - \delta_{ab} \right) 
= \frac{1}{2} \big(\partial_a u_b + \partial_b u_a + \partial_a h_\alpha \partial_b h_\alpha + \partial_a u_c \partial_b u_c \big)\, .
\ea
Neglecting the non-linearities in the phonon field (since they are irrelevant in the RG sense by simple canonical power counting) yields 
\be
T_{ab}\approx\frac{1}{2} \left(\partial_a u_b + \partial_b u_a + \partial_a h_\alpha \partial_b h_\alpha \right)\, .
\ee
From there, the Euclidean low-energy action of the membrane, see, \eg, \cite{Nelson:1987,Aronovitz:1988,Aronovitz:1989, David:1988,Guitter:1988,Guitter:1989}, reads 
\ba
S[\vec u, \vec h ] = \int \D^d x \left[ \frac{\kappa}{2} (\partial^2 h_\alpha)^2 + \frac{\lambda}{2} T_{aa}^2 + \mu T_{ab}^2 \right]\, ,
\label{mem:S}
\ea
where self-avoidance has been neglected (this is the case of a so-called phantom membrane relevant to the flat phase), $\kappa$ is the bending rigidity and $\lambda$ and $\mu$ are the Lam\'e parameters ($\mu$ is sometimes referred to as the shear modulus). As we will see in the following, these two parameters will act as coupling constants for the field theory and are simply related to the other elastic moduli of the membrane via the relations (in arbitrary dimension)
\ba
& \text{Bulk modulus:} \quad B=\lambda +2 \mu/d\, , 
&& \text{Poisson ratio:} \quad \nu
=\cfrac{\lambda}{(d-1)\lambda + 2\mu}\, , \nonumber \\
& \text{Young modulus:} \quad Y=\cfrac{2\mu(d \lambda+2\mu)}{(d-1)\lambda+2\mu}\, , 
&& \text{p-wave modulus:} \quad W=\lambda+2\mu\, . 
\label{mem:moremoduli}
\ea

Using the action \eqref{mem:S} in which quadratic, irrelevant in the RG sense, terms in the phonon field $\vec u$ are neglected, yields the expanded form
\ba
S[\vec u, \vec h ] &= \frac{1}{2} \int \D^d x \left[ \kappa (\partial^2 h_\alpha)^2 
+ \lambda \left((\partial_a u_a)^2 + \partial_a u_a (\partial_b h_\alpha)^2 + \frac{1}{4} (\partial_a h_\alpha)^4 \right) \right .
\nonum \\
&\hspace{1.8cm}+ \left. \mu \left((\partial_a u_b)^2 + \partial_a u_b \partial_b u_a + 2 \partial_a u_b (\partial_a h_\alpha \partial_b h_\alpha) + 
\frac{1}{2} (\partial_a h_\alpha \partial_b h_\beta)^2 \right) \right ]\, , 
\label{mem:Sexp}
\ea
which contains all relevant operators as can be checked with the help of a dimensional analysis in $d=4-2\eps$. Incidentally, the canonical dimensions
\be
[u]=d-3=1-2\eps, \qquad [h]=\frac{d}{2}-2=-\eps, \qquad [\mu]=[\lambda]=4-d=2\eps \, ,
\ee
also reveal the renormalizability of the model in 4 dimensions.
In addition, a term $(\Delta u_a)^2$ has been dropped as it is negligible in comparison
with $\lambda (\partial_a u_a)^2$ and $\mu (\partial_a u_b)^2$ at small momenta. The action \eqref{mem:Sexp} is therefore a massless and highly derivative scalar two-field and two-coupling theory.

Based on the fact that the action \eqref{mem:Sexp} is quadratic in the phonon field, we may integrate over it exactly. This leads to the so-called effective flexural theory (EFT) approach. The resulting effective action depends only on the flexuron field. In Fourier space, it reads \cite{Nelson:1987,LeDoussal:1992}
\be
S_{\text{EFT}}[\vec h ] = \frac{\kappa}{2} \int_p p^4 |h_\alpha(\vec p)|^2 +\frac{1}{4} \int_{p_i} h_\alpha(\vec p_1) h_\alpha(\vec p_2) R_{abcd}^{(0)}(\vec p) p_1^a p_2^b p_3^c p_4^d h_\beta(\vec p_3) h_\beta(\vec p_4) \, ,
\label{mem:SEFT}
\ee
where the Euclidean momenta are $p_i=\{p_1,p_2,p_3,p_4\}$ with $\vec p = \vec p_1 + \vec p_2 = - \vec p_3 - \vec p_4$. We also use the shorthand notation $\int_p=\int[\D^d p]$ and $\int_{p_i}=\int[\D^d p_1][\D^d p_2][\D^d p_3][\D^d p_4]$ where brackets stands for $[\D^d p]=\D^d p/(4\pi)^d$. 

The action \eqref{mem:SEFT} is the model we will consider in the rest of this review. It is equivalent to the two-field model defined in \eqref{mem:S} but with a more intricate tensor structure and a non-local interaction. The rank-four tensor entering the four-flexuron term is given by
\be
R_{abcd}^{(0)}(\vec p) = \mu M_{abcd}(\vec p) + b(d) N_{abcd}(\vec p)\, , 
\label{mem:Rabcd} 
\ee
and is decomposed onto two $M$ and $N$ tensors that are defined as
\bs
\label{mem:MNTensors}
\ba
&M_{abcd}(\vec p) = \frac{1}{2} \left[P_{ac}^{(\bot)}(\vec p) P_{bd}^{(\bot)}(\vec p) + P_{ad}^{(\bot)}(\vec p) P_{bc}^{(\bot)}(\vec p) \right] - N_{abcd}(\vec p),
\label{mem:Nabcd} \\
&N_{abcd}(\vec p) = \frac{1}{d-1} P_{ab}^{(\bot)}(\vec p) P_{cd}^{(\bot)}(\vec p), \qquad P_{ab}^{(\bot)}(\vec p)= \delta_{ab}-\frac{p_a p_b}{p^2}\, .
\label{mem:Mabcd}
\ea
\es
Using the definitions \eqref{mem:MNTensors}, it is straightforward to derive the full contractions 
\be
M_{abcd}(\vec p)M^{abcd}(\vec p) = \frac{(d+1)(d-2)}{2}, \quad
N_{abcd}(\vec p)N^{abcd}(\vec p) = 1, \quad 
M_{abcd}(\vec p)N^{abcd}(\vec p) = 0\, , 
\ee
from which we can define the normalized projectors
\be
P^{M}_{abcd}(\vec p)= \frac{2}{(d+1)(d-2)}M_{abcd}(\vec p), \qquad
P^{N}_{abcd}(\vec p)=N_{abcd}(\vec p)\, ,
\label{mem:eq:projectors:normalized}
\ee
that are particularly handy to project out tensorial quantities onto their $M$ and $N$ components.
Therefore, in the effective flexural theory, while $\mu$ is still our first coupling, the second one is not $\lambda$ anymore, but is replaced by $b(d)$ that we introduced in \eqref{mem:Rabcd} and reads 
\be
b(d) = \frac{\mu (d\lambda + 2\mu)}{\lambda + 2\mu}.
\label{eq:mem:bdef}
\ee
This new coupling is proportional to the $d$-dimensional bulk modulus $B$, or equivalently to the Young modulus $Y$, see \eqref{mem:moremoduli}, \ie, 
\be
b(d)=\frac{\mu d}{W}B = \frac{\lambda}{2 W \nu}Y.
\ee
Because the tensors $M$ and $N$ are mutually orthogonal under tensor multiplication, we expect that $\mu$ and $b(d)$ will renormalize independently of each other and should therefore be considered as independent couplings. Indeed, as noticed in \cite{Guitter:1989}, setting $\mu=0$ in \eqref{mem:Rabcd} when expressed in terms of $\mu$ and $\lambda$ yields a zero vertex, \eg, a free flexuron field; but it is know, already from the one-loop analysis of the two-field model, that the $\mu=0$ limit leads to non-zero renormalization constants. Following, \cite{Coquand:2020a,Metayer:2021kxm}, we will not only consider $b$ as independent
of $\lambda$ and $\mu$ but also independent of the dimension $d$. Recovering the results of the two-field model from the flexural effective approach is therefore a non-trivial check of their validity.

For completeness, we provide the elastic moduli \eqref{mem:moremoduli} of the membrane in terms of the new $(\mu,b)$ variables in arbitrary dimension $d$, see \eqref{eq:mem:bdef}
\ba
\label{mem:moremodulimub}
& \text{Bulk modulus:} \quad B=\frac{2 b (d-1) \mu }{d (d \mu -b)}, 
&& \text{Poisson ratio:} \quad \nu
=\frac{b-\mu }{b (d-2)+\mu }, \nonumber \\
& \text{Young modulus:} \quad Y=\frac{2 b (d-1) \mu }{b (d-2)+\mu }, 
&& \text{p-wave modulus:} \quad W=\frac{2 (d-1) \mu ^2}{d \mu -b}, 
\ea
and the mechanical stability of the model is given by 
\be
\mu>0, \qquad b>0.
\label{eq:constrmemeft}
\ee
Note also that, in the following, we will remove the trivial bending rigidity, $\kappa$, by rescaling the field and couplings as
\be 
h_\alpha\rightarrow h_\alpha \kappa^{-1/2}, \qquad \mu\rightarrow\mu \kappa^2, \qquad b\rightarrow b \kappa^2 \, .
\ee
This is equivalent to working in natural units and setting $\kappa=1$ in the action (\ref{mem:SEFT}). Retrospectively, we can also set $u_a\rightarrow u_a \kappa^{-1}$ and $\lambda\rightarrow\lambda\kappa^2$ in the original action \eqref{mem:Sexp} to get rid of $\kappa$ in the same way.

\section{Perturbative setup and conventions}
\label{sec:setup}

In this section we present the general perturbative setup which provides the framework in which to systematically extract the critical exponents of the effective flexural model \eqref{mem:SEFT} at fixed co-dimension $d_c$ and near the upper critical dimension $d_{\text{uc}} = 4$ where the model is renormalizable.

\subsection{Feynman rules}
\label{mem:subsec:model:feynrules}

From the action \eqref{mem:SEFT}, the free massless flexuron propagator reads
\be
S_{\al \beta}^{(0)}(\vec p) =\bra h_\alpha(\vec p) h_\beta(-\vec p) \ket_0= \flexuronprop= \frac{\delta_{\al\beta}}{p^4}\, .
\label{mem:FR:S}
\ee
%
Similarly, the four-point flexuron vertex reads
\ba
V_{\al \beta \gamma \delta}^{(0)}(\vec p_i) = \bra h_\alpha(\vec p_1) h_\beta(\vec p_2) h_\gamma(\vec p_3) h_\delta(\vec p_4) \ket_0 = \hspace{-0.5cm} \fourpointvertex \hspace{-0.5cm} =
-2 R_{abcd}^{(0)}(\vec p) \delta_{\al\beta} \delta_{\gamma\delta} p_1^a p_2^b p_3^c p_4^d\, , 
\label{mem:eq:Vdef} \nonum \\[-0.5cm]
\ea
%
In \eqref{mem:eq:Vdef}, the front factor, $-2$, is made of three contributions. First, the usual minus sign associated with quartic interactions. Second, the $1/4$ factor in the action (\ref{mem:SEFT}). And third, the vertex factor, which is $8$, so that $-8/4=-2$. Let us remark that the vertex factor is indeed 8 and not $4!=24$, as one might expect. This is due to the apparent asymmetry of the four-point coupling $V_{\al \beta \gamma \delta}^{(0)}(\vec p_i)$ that couples only bi-flexuron pairs. We represent graphically this particularity with a dashed line carrying momentum $\vec p=\vec p_1+\vec p_2=-\vec p_3-\vec p_4$ (all momenta are incoming). 

In order to emphasize this asymmetry of the four-point coupling, one can decompose it and define an alternative but equivalent set of Feynman rules based on a three-point vertex. While the free flexuron propagator is kept identical, the vertex interaction is decomposed in two parts, the effective free $R$-propagator 
\be 
R^{(0)}_{abcd}(\vec p) = \effprop = \mu M_{abcd}(\vec p) + b N_{abcd}(\vec p), 
\label{mem:effprop}
\ee
and a three-point interaction reading
\be
\Gamma_{\alpha\beta}^{ab (0)}(\vec p_1,\vec p_2)= \hspace{-0.3cm} \threepointvertex \hspace{0.1cm} = \I \sqrt{2} \delta_{\alpha\beta} p_1^a p_2^b.
\label{mem:threepointint}
\ee
The factor $\I \sqrt{2}$ is designed in such a way that $(\I \sqrt{2})^2=-8/4=-2$ from (\ref{mem:eq:Vdef}). Therefore, the four-point interaction (vertex factor $8$) is equivalent to a multiplication of two three-point interactions (vertex factor $2$) and an effective $R$-propagator, \ie
\be
V^{(0)}_{\alpha\beta\gamma\delta}(\vec p_i)=
\Gamma_{\alpha\beta}^{ab (0)}(\vec p_1,\vec p_2) 
R^{(0)}_{abcd}(\vec p) 
\Gamma_{\gamma\delta}^{cd (0)}(\vec p_3,\vec p_4),
\label{mem:eq:VandGammaequivalency}
\ee
where $\vec p=\vec p_1+\vec p_2=-\vec p_3-\vec p_4$. With these definitions, the two sets of Feynman rules, (\ref{mem:FR:S}) and (\ref{mem:eq:Vdef}) on the one hand and (\ref{mem:FR:S}), \eqref{mem:effprop} and \eqref{mem:threepointint} on the other hand, are equivalent. The second set of Feynman rules is very convenient because of its graphical similarity to quantum electrodynamics (QED) provided one identifies flexurons with fermions and the $R$-propagators with photons. Indeed, it allows us to automate diagram generation using codes similar to those for QED based on the Fortran tool \textsc{Qgraf} \cite{Nogueira:1993,Nogueira:2021wfp}. The drawback of the use of the three-point vertex is that it results in more contractions over the Euclidean space (latin) indices, see \eqref{mem:eq:VandGammaequivalency}. However, it is not an issue since we carry all our computations in a completely automated way using \textsc{Mathematica}, with homemade codes to perform efficiently the contractions.

\subsection{Fate of infra-red divergences}

The form of the flexuron propagator ($\sim 1/p^4$), see (\ref{mem:FR:S}), suggests that the theory may be plagued by severe infra-red (IR) singularities,
thus invalidating the renormalization prescription. It turns out that this is fortunately not the case and that the renormalization
constants are determined by ultraviolet (UV) poles only.

In order to prove this statement in a non-perturbative way, let us consider the following correlation function
\be
G_{\al \beta}(\vec p) = \langle \partial_j h_\al (\vec p) \,\partial_j h_\beta (-\vec p) \rangle \,.
\ee
Obviously, $G_{\al \beta}(\vec p)$ is an IR-safe function with respect to loop integrals. It turns out that it is simply related to the flexuron propagator,
$S_{\al \beta}(\vec p) = \langle h_\al (\vec p) h_\beta (-\vec p) \rangle$ and the relation reads
\be
G_{\al \beta}(\vec p) = p^2 S_{\al \beta}(\vec p) \,.
\ee
This simple identity shows that the flexuron propagator is directly related to an IR-safe quantity. Hence, it is itself IR-safe. 

At a more practical level, IR singularities do show up in the course of the perturbative computations. Their appearance is due to the ambiguous nature of the
massless tadpole, which is zero in dimensional regularization as a consequence of a subtle cancellation between IR and UV singularities. However, because the flexuron propagator
is IR-safe these IR poles are harmless and do not require any special treatment as they simply cancel each-other order by order in perturbation theory. In this case,
we may proceed with dimensional regularization in the conventional way as if all poles were of UV type, see, \eg, \cite{Chetyrkin:1983} for a proof of this statement.

\subsection{Dyson equations}
\label{mem:subsec:model:dyson-eq}

We shall consider the Dyson equation for the dressed flexuron propagator reading
\be
S_{\al \beta}(\vec p) = S_{\al \beta}^{(0)}(\vec p) + S_{\al \gamma}^{(0)}(\vec p) \Sigma_{\gamma \delta}(\vec p) S_{\delta \beta}(\vec p),
\ee
where $\Sigma_{\al \beta}(\vec p)$ is the 1-particle irreducible flexuron self-energy, $S_{\alpha\beta}$ is the fully dressed flexuron propagator and $S^{(0)}_{\alpha\beta}$ the free one, \eqref{mem:FR:S}. It is convenient to project out the tensor structure of the various quantities of interest in order to deal with simpler scalar functions. The task is straightforward for the flexuron propagator because
\be
S_{\alpha\beta}(\vec p)=\delta_{\alpha\beta}S(\vec p), \qquad 
\Sigma_{\al \beta}(\vec p)= \delta_{\al \beta} \Sigma(\vec p)\, ,
\ee
which slightly simplifies the Dyson equation as
\ba
S(\vec p) & =S^{(0)}(\vec p)+S^{(0)}(\vec p) \Sigma(\vec p) S(\vec p) = \frac{S^{(0)}(\vec p)}{1-\Sigma(\vec p) S^{(0)}(\vec p)}\, .
\ea
Moreover, since $S^{(0)}(\vec p)=p^{-4}$, a convenient form is given by
\be
S(\vec p) = \frac{1}{p^4} \frac{1}{1-\tilde{\Sigma}(p^2)}, \qquad \tilde{\Sigma}(p^2)= p^{-4} \Sigma(\vec p) \, .
\label{mem:S+Sigma}
\ee

Similarly, the dressed four-point vertex may be expressed with the help of the dressed $R$-propagator $R_{abcd}(\vec p)$ assuming that, to all orders, the following holds
\ba
V_{\al \beta \gamma \delta}(\vec p_i) 
& = \Gamma_{\alpha\beta}^{ab (0)}(\vec p_1,\vec p_2) 
R_{abcd}(\vec p) \Gamma_{\gamma\delta}^{cd (0)}(\vec p_3,\vec p_4) 
= - 2 R_{abcd}(\vec p) \delta_{\al\beta} \delta_{\gamma\delta} p_1^a p_2^b p_3^c p_4^d\, .
\ea
This implies that all corrections are in $R_{abcd}(\vec p)$ which satisfies the following Dyson equation
\ba
R_{abcd}(\vec p) = R_{abcd}^{(0)}(\vec p) + R_{abef}^{(0)}(\vec p) \Pi_{efgh}(\vec p) R_{ghcd}(\vec p)\, ,
\label{mem:DysonR}
\ea
where $\Pi_{abcd}(\vec p)$ is a 1-particle irreducible self-energy of the effective $R$-propagator, \ie, corresponding to a vacuum polarization. As will be shown in the following, 
the fact that the Dyson equation for the four-point vertex is entirely encapsulated in (\ref{mem:DysonR}) is not an approximation and allows us to reproduce exactly the non-trivial two-field results. This also reinforces the use of the second set of Feynman rules that consists of a free flexuron propagator, a free $R$-propagator and the triple vertex made of a $R$-propagator and two flexuron propagators. At this point, it is convenient to decompose the polarization operator and the vertex function on the basis of the tensors $M$ and $N$ \eqref{mem:MNTensors} in order to solve the Dyson equation for the $R$-propagator \eqref{mem:DysonR}. This yields
\bs
\label{mem:R:RM+RN}
\ba
&R_{abcd}(\vec p) = R_M(\vec p) M_{abcd}(\vec p) + R_N(\vec p) N_{abcd}(\vec p),
\label{mem:R:gen} \\
&\Pi_{abcd}(\vec p) = \Pi_M(\vec p) M_{abcd}(\vec p) + \Pi_N(\vec p) N_{abcd}(\vec p) \, ,
\label{mem:Pi:gen}
\ea
\es
where the scalar functions in factor of the tensors take the form:
\bs
\ba
&& &R_M(\vec p) = \frac{\mu}{1 - \tilde{\Pi}_M(\vec p)}, && \tilde{\Pi}_M(\vec p) = \mu \Pi_M(\vec p), & 
\label{mem:RN} \\
&& &R_N(\vec p) = \frac{b}{1 - \tilde{\Pi}_N(\vec p)}, && \tilde{\Pi}_N(\vec p) = b \Pi_N(\vec p). & 
\label{mem:RM} 
\ea
\es

\subsection{Renormalization conventions}
\label{mem:subsec:model:renormalization-convention}

We are now in a position to introduce the renormalization constants associated with the model (\ref{mem:SEFT}):
\be
\label{mem:ren-constants-EFT}
\vec h = Z^{1/2} \vec h_r, \qquad \mu = Z_\mu \mu_r M^{2\eps}, \qquad b = Z_b b_r M^{2\eps}\, ,
\ee
where the subscript $r$ denotes renormalized quantities and the renormalization scale, $M$, has been introduced in such a way that $\mu_r$ and $b_r$ 
are dimensionless in $d=4-2\eps$ dimensions. The latter is related to the corresponding parameter $\overline{M}$ in the modified minimal subtraction ($\overline{\text{MS}}$) scheme with the help of
\be
\overline{M}^{ 2} = 4\pi e^{-\gamma_E} M^2\, ,
\label{mem:MSbar}
\ee
where $\gamma_E$ is Euler's constant. In such a scheme, the renormalization constants take the form of Laurent series in $\eps$
\be
Z_x(\mu_r,b_r) = 1 + \delta Z_x(\mu_r,b_r) = 1 + \sum_{l=1}^\infty \sum_{j=1}^l Z_x^{(l,j)}(\mu_r,b_r) \frac{1}{\eps^j}\, ,
\label{mem:Zx}
\ee
where $x \in \{\mu,b \}$ and they do not depend on momentum (or mass which is absent in the present model). Furthermore, the dependence on $M$ is only through $\mu_r$ and $b_r$. So the renormalization constants $Z_x$ depend only on $\mu_r(M)$, $b_r(M)$ and $\eps$.
They also relate renormalized and bare propagators as follows
\bs
\label{mem:renormalized-propagators}
\ba
S_{\al \beta}(\vec p ;\mu,b) & = Z(\mu_r,b_r) S_{\al \beta,r}(\vec p ;\mu_r,b_r,M)\, , \label{mem:renormalized-propagatorsa} \\
R_{abcd}(\vec p ;\mu,b) & = M^{2 \eps} Z^{-2}(\mu_r,b_r) R_{abcd,r} (\vec p ;\mu_r,b_r,M)\, ,
\label{mem:renormalized-propagatorsb}
\ea
\es
where the bare propagators do not depend on $M$. Then, projecting out the tensor structure of each one of these functions (and dropping the arguments for simplicity) yields
\be
S =Z S_r \qquad 
R_M = M^{2\eps} Z^{-2} R_{M,r}, \qquad
R_N = M^{2\eps} Z^{-2} R_{N,r}\, .
\label{eq:memeft:Zdefs}
\ee

We may now explain our renormalization technique, \ie, our method to determine the renormalization constants $\{Z,Z_\mu,Z_b\}$. It has to be underlined here that we are not using counter-terms or any other advanced renormalization method. Instead, for the whole manuscript, we will use a more pragmatic approach that is very efficient for high-order calculations. It consists in directly expressing the renormalization constants $\{Z,Z_\mu,Z_b\}$ in terms of the self-energy and polarizations $\{\Sigma,\Pi_M,\Pi_N\}$. We now detail this procedure for the flexuron. From \eqref{eq:memeft:Zdefs} together with the Dyson equation \eqref{mem:S+Sigma}, we can first write 
\be
S=Z S_r \quad \implies \quad \frac{1}{p^4} \frac{1}{1-\tilde\Sigma} = Z \frac{1}{p^4}\frac{1}{1-\tilde\Sigma_r}\, ,
\ee
from which the renormalized self-energy can be expressed as
\be
\tilde\Sigma_r=1-(1-\tilde\Sigma)Z\, .
\label{mem:sigma-ren}
\ee
Using the fact that the renormalized self-energy is finite, this equation then allows one to
completely determine $Z$ once $\tilde\Sigma$ has been computed. Mathematically, this is achieved using the fact that $\KK{\tilde{\Sigma}_r(\vec p)}=0$, where $\mathcal{K}$ is the operator used to take the divergent part of the series and is defined as
\be
\mathcal{K}\left[ \sum_{n=-\infty}^{+\infty} \frac{c_n}{\eps^n} \right] = \sum_{n=1}^{+\infty} \frac{c_n}{\eps^n}\, .
\label{mem:K:def}
\ee
Using the parametrization
\bs
\ba
& \tilde\Sigma=\tilde\Sigma_1+\tilde\Sigma_2+\tilde\Sigma_3+..., \\
& Z=1+\delta Z_1+\delta Z_2+\delta Z_3+...,
\ea
\es
the relation
\be
0=\mathcal{K}\big[(1-\tilde\Sigma)Z\big]\, , 
\label{eq:mem:renflex}
\ee
can be solved order by order in the loop expansion and completely defines $Z$ recursively as a function of $\tilde\Sigma$. Explicitly and up to three loops this leads to
\bs
\label{mem:deltaZ:gen}
\ba
& \text{1-loop:}\quad \delta Z_1= \mathcal{K}(\tilde \Sigma_1), \\
& \text{2-loop:}\quad \delta Z_2= \mathcal{K}(\delta Z_1 \tilde\Sigma_1) +\mathcal{K}(\tilde\Sigma_2), \\
& \text{3-loop:}\quad \delta Z_3=\mathcal{K}(\delta Z_2 \tilde\Sigma_1) + \mathcal{K}(\delta Z_1 \tilde\Sigma_2) +\mathcal{K}(\tilde\Sigma_3).
\ea
\es
This way of proceeding is particularly useful in the present case due to the intricate tensor structure of the effective model that is very demanding computationally. Once $Z$ is computed, it is also possible to compute $\tilde\Sigma_r$ by going back to (\ref{mem:sigma-ren}); the poles of $Z$ and $\tilde\Sigma$ should cancel out leaving only a finite result in the limit $\eps \rightarrow 0$ order by order in perturbation theory.

Similarly, the computation of the renormalization constants $Z_\mu$ and $Z_b$ are derived from the renormalization of the vertex parts. From (\ref{mem:R:RM+RN}) and (\ref{mem:ren-constants-EFT}), we have
\bs
\label{mem:Pi-ren}
\ba
&\tilde{\Pi}_{M,r}(\vec p) = 1+\Big(1-\tilde{\Pi}_M(\vec p)\Big)Z_{\Gamma_\mu}^{-1}\, ,
\label{mem:Pi-ren:N}\\
&\tilde{\Pi}_{N,r}(\vec p) = 1+\Big(1-\tilde{\Pi}_N(\vec p)\Big)Z_{\Gamma_b}^{-1}\, ,
\label{mem:Pi-ren:M}
\ea
\es
where we have introduced two intermediate renormalization functions
\be
Z_{\Gamma_\mu} = Z_\mu Z^2, \qquad \quad Z_{\Gamma_b} = Z_b Z^2\, .
\label{mem:ZG}
\ee
Then, using the fact that $\KK{\tilde{\Pi}_{M,r}(\vec p) } = 0$ and $\KK{\tilde{\Pi}_{N,r}(\vec p)} = 0$, it is possible to extract the general expressions of $Z_{\Gamma_\mu}$ and $Z_{\Gamma_b}$ from
\bs
\label{eq:memeft:renidetntitescouplings}
\ba
& 0 = \KK{\Big(1-\tilde{\Pi}_M(\vec p)\Big)Z_{\Gamma_\mu}^{-1}}\, , \\
& 0 = \KK{\Big(1-\tilde{\Pi}_N(\vec p)\Big)Z_{\Gamma_b}^{-1}}\, .
\ea
\es
By combining \eqref{eq:memeft:renidetntitescouplings} with the expression of $Z$ (\ref{mem:deltaZ:gen}), we are able to deduce $Z_\mu$ and $Z_b$ from (\ref{mem:ZG}).

Once the renormalization constants are determined, we are in a position to compute the renormalization-group functions that are defined as
\be
\beta_\mu = \frac{\D \mu_r}{\D \log M}\bigg|_B, \qquad
\beta_b = \frac{\D b_r}{\D \log M}\bigg|_B, \qquad
\eta = \frac{\D \log Z}{\D \log M}\bigg|_B \, ,
\label{mem:ren-functions:def}
\ee
where the subscript $B$ indicates that bare parameters, which do not depend on the renormalization scale $M$, are fixed. The system of beta functions to be solved perturbatively is then
\bs
\label{mem:ren-functions:def+:betas}
\ba
&\beta_\mu = -2 \eps \mu_r-\mu_r \mathcal{D} Z_\mu\, ,
\label{mem:ren-functions:def+:betamu}\\
&\beta_b = -2 \eps b_r-b_r \mathcal{D} Z_b\, ,
\label{mem:ren-functions:def+:betalambda}
\ea
\es
where we introduced the differential operator
\be
\mathcal{D} X = \beta_\mu \partial_{\mu_r} \log X +
\beta_b \partial_{b_r} \log X\, .
\ee
The solution to the linear system of equation is then given, in matrix form, by
\be
\begin{pmatrix}
\beta_\mu \\
\beta_b \\
\end{pmatrix}
=-2\eps
\begin{pmatrix}
\mu_r \partial_{\mu_r}\log\mu_r Z_{\mu}
& \mu_r \partial_{b_r}\log \mu_r Z_{\mu} \\
b_r \partial_{\mu_r}\log b_r Z_{b}
& b_r \partial_{b_r}\log b_r Z_{b} \\
\end{pmatrix}
^{-1}
\begin{pmatrix}
\mu_r \\
b_r \\
\end{pmatrix}\, .
\ee
Finally, the field anomalous dimension associated with the flexuron field reads
\be
\eta = \mathcal{D} Z.
\label{mem:ren-functions:def+:gamma}
\ee
Let us recall that, physically, it corresponds to the anomalous stiffness induced by long-range correlations, such that the dressed flexuron propagator scales in the IR as \eqref{mem:correlation}.
%
%

Upon solving perturbatively the RG-functions (\ref{mem:ren-functions:def+:betas}) and (\ref{mem:ren-functions:def+:gamma}), it is well known that all of them are determined only by the coefficients of the simple poles of the renormalization constants. 
However, in the following, we shall use once again a more pragmatic approach, consisting in computing the RG-functions ($\beta_\mu$, $\beta_b$, $\eta$) directly with the complete (containing all kinds of poles) expressions of the renormalization constants ($Z$, $Z_\mu$, $Z_b$). Then, if the RG-functions are finite (pole free), it implies that the full set of constraints is satisfied. The finiteness of the RG-functions together with the locality of the renormalization constants (no momentum dependence), constitute a strong check on the validity of our results.

Finally, once the RG functions ($\beta_\mu$, $\beta_b$, $\eta$) have been determined as a function of the renormalized couplings $\mu_r$ and $b_r$, we will search for the fixed points of the theory. Indeed, since the beta functions characterize the scaling of the coupling with respect to the renormalization scale $M$, one can search for the specific points where the theory is scale invariant, \ie, by solving the system
\be 
\beta_\mu (\mu^*,b^*)=0, \qquad \beta_{b}(\mu^*,b^*)=0,
\label{eq:mem:betasys}
\ee
and obtain various fixed point coordinates $(\mu^*,b^*)$ where the theory exhibits universal scaling behaviors and such that $\eta(\mu^*,b^*)$ is a physical universal number characterizing the corresponding phase, \ie, a critical exponent. Since canceling one of the couplings leads to trivial solutions, we expect to find 4 types of fixed points. First, a trivial Gaussian fixed point (P$_1$) at which $\mu_1^*=b_1^*=0$ such that the theory is non-interacting. Second, a shearless fixed point (P$'_2$)\footnote{\label{footnote:P'2} We refer to it as P$'_2$ in order to distinguish it from P$_2$ that is proper to the two-field model as the anomalous dimension $\eta(\text{P}'_2)$ in the flexural model differs from $\eta(\text{P}_2)$ in the two-field model.} at which $\mu_2^*=0$ and $b_2^*\neq0$. Third, a fixed point P$_3$ with $\mu_3^* \neq 0$ and $b_3^*=0$, \ie, a vanishing bulk modulus (infinitely compressible) fixed point. And finally, a fourth, non-trivial fixed point (P$_4$) at which both couplings are non-zero, $\mu_4^*\neq0$ and $b_4^*\neq0$; this is the most interesting fixed point since it corresponds to the fully interacting theory. Moreover, the mechanical stability of the membrane, see \eqref{eq:constrmemeft}, requires a positive shear modulus $\mu^*>0$ as well as a positive bulk modulus $\lambda^*+2\mu^*/d>0$, \ie, $b^*>0$. The RG stability of these fixed points can be studied by searching the eigenvalues of the stability matrix 
\be
\mathcal{S}=
\begin{pmatrix}
\D\beta_{\mu}/\D\mu_r & \D\beta_{\mu}/\D b_r \\
\D\beta_{b}/\D\mu_r & \D\beta_{b}/\D b_r
\end{pmatrix},
\label{eq:mem:stabilitymatrix}
\ee
such that positive (respectively negative) eigenvalues indicates an IR stable (respectively unstable) fixed point. We recall that an IR stable fixed point is attractive as the theory renormalizes down to the IR and therefore controls the long-distance behavior of the model.

\section{Perturbative calculations}
\label{sec:PT}

In this section, we present the full three-loop computations of the renormalization constants $Z$, $Z_\mu$ and $Z_b$, 
\eqref{mem:ren-constants-EFT}, of the effective model \eqref{mem:SEFT} derived from our general setup. In order to clarify the presentation, we shall proceed
order by order and, at each order, consider all the Feynman diagrams entering the self-energy of the flexuron propagator \eqref{mem:S+Sigma} and the polarization of the $R$-propagator \eqref{mem:DysonR}.

\subsection{One-loop analysis}
\label{mem:sec:one-loop}

\begin{figure}[h]
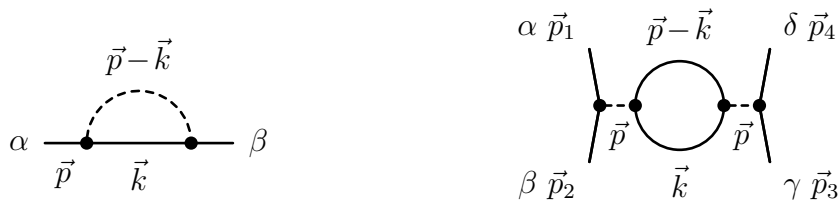

\begin{subfigure}{.5\textwidth}
\centering
\GraphSigmaOne
\vspace{-0.5cm}
\caption{Flexuron self-energy $\Sigma^{(1)}_{\alpha\beta}(\vec p)$, ${\text{S}=1}$.}
\label{mem:fig:one-loop:a}
\end{subfigure}
\hspace{-1cm}
\begin{subfigure}{.5\textwidth}
\centering
\GraphVOne
\caption{Vertex self-energy $V_{\al \beta \gamma \delta}^{(1)}(\vec p_i)$, ${\text{S}=1/2}$.}
\label{mem:fig:one-loop:b}
\end{subfigure}
\caption{One-loop diagrams and their associated symmetry factors (S).}
\label{mem:fig:one-loop}
\end{figure}

The one-loop, two-point and four-point self-energy diagrams are displayed on figure \ref{mem:fig:one-loop}, with their corresponding symmetry factors (S). 
\vspace{-0.5cm}

\subsubsection{One-loop flexuron self-energy}
\label{mem:subsec:one-loop:flexuron}

The one-loop flexuron self-energy, figure \ref{mem:fig:one-loop:a}, has a symmetry factor of $1$ and is defined as
\be
\Sigma_{\al \beta}^{(1)}(\vec p)
= \int [\D^d k] 
V_{\al \al_1 \beta_1 \beta}^{(0)}(\vec p, -\vec k, \vec k, -\vec p) S_{\al_1 \beta_1}^{(0)}(\vec k),
\ee
%
or, equivalently, with the second set of Feynman rules
\be
\Sigma_{\al \beta}^{(1)}(\vec p)= \int [\D^d k] \Gamma_{\alpha\alpha_1}^{ab (0)}(\vec p, -\vec k) R^{(0)}_{abcd}(\vec p-\vec k) \Gamma_{\beta_1\beta}^{cd (0)}(\vec k,-\vec p) S_{\al_1 \beta_1}^{(0)}(\vec k).
\ee
After using one of the two equivalent sets of Feynman rules and contracting part of the internal indices, the one-loop flexuron self-energy reads
\be
\Sigma_{\al \beta}^{(1)}(\vec p)=-2 \delta_{\alpha\beta} p^a p^b\int \frac{[\D^d k]}{k^4} k^c k^d R^{(0)}_{abcd}(\vec p-\vec k).
\ee
Using techniques of massless Feynman diagram calculations, see, \eg, \cite{Kotikov:2018wxe} for a review, the integral is straightforward to compute and the result reads
\be
\tilde{\Sigma}^{(1)}(p^2)= -\frac{(b+(d-2) \mu)}{8(d-1)} \frac{(p^2)^{d/2-2}}{(4\pi)^{d/2}} (d+1)(d-2) G(d,1,1),
\label{mem:EFT:Sigma1-exact2}
\ee
where $G(d,\alpha,\beta)$ is the dimensionless master integral at one loop defined in \ref{chap:appendix}. 

Upon expanding (\ref{mem:EFT:Sigma1-exact2}) in $\eps$-series, the one-loop flexuron self-energy in $\overline{\text{MS}}$-scheme then reads
\be
\tilde{\Sigma}^{(1)}(p^2)= \frac{5 (b+2 \mu)}{12 (4\pi)^2 M^{2\eps}} \left(-\frac{1}{\eps} + L_p - \frac{4}{15} - \frac{b}{b+2 \mu} + \Ord(\eps) \right),
\label{mem:Sigma1-expanded}
\ee
where $L_p = \log(p^2/\overline{M}^2)$ and only the first terms of the series expansion are displayed for simplicity. Note that due to the presence of a $d$-dependent combination of coupling constants in (\ref{mem:EFT:Sigma1-exact2}), the intuitive combination $b+2\mu$ does not factor in the expanded expression, and the finite term gets a non-trivial combination of coupling constants.

Combining (\ref{mem:Sigma1-expanded}) and (\ref{mem:deltaZ:gen}) yields the one-loop field renormalization constant
\be
\delta Z^{(1)} = \KK{\tilde{\Sigma}^{(1)}(p^2)} = 
-\frac{5 (b_r +2 \mu_r)}{12 (4\pi)^2 \eps},
\label{mem:Z:one-loop}
\ee
where we performed the trivial replacements $\mu\rightarrow M^{2\eps} \mu_r$ and $b\rightarrow M^{2\eps} b_r$ which is enough at the present order.
Also, using (\ref{mem:Z:one-loop}) as well as the first terms of the expansion of (\ref{mem:sigma-ren}), we have a straightforward access to the renormalized flexuron self-energy that we add for completeness
\be
\tilde{\Sigma}^{(1)}_r=\tilde{\Sigma}^{(1)}-\delta Z^{(1)}=\frac{5 (b_r+2 \mu_r)}{12 (4\pi)^2} \left(L_p - \frac{4}{15} - \frac{b_r}{b_r+2 \mu_r}+\Ord(\eps) \right).
\ee

\subsubsection{One-loop vertex self-energy}
\label{mem:subsec:one-loop:vertex}

The one-loop vertex self-energy, figure \ref{mem:fig:one-loop:b}, has a symmetry factor of $1/2$ and is defined as
\ba
V_{\al \beta \gamma \delta}^{(1)}(\vec p_i) 
& = \frac{1}{2} \int [\D^d k] 
V_{\al \beta \gamma_1 \delta_1}^{(0)}(\vec p_1, \vec p_2, -\vec k, -\vec p + \vec k) 
S_{\delta_1 \al_1}^{(0)}(\vec p-\vec k) 
S_{\gamma_1 \beta_1}^{(0)}(\vec k) V_{\al_1 \beta_1 \gamma \delta}^{(0)}(\vec p - \vec k, \vec k, \vec p_3, \vec p_4),
\ea
where all $\vec{p}_i$ are defined as in-going, \ie, ${\vec p = \vec p_1+\vec p_2=-\vec p_3-\vec p_4}$. This diagram can equivalently be expressed with the second set of Feynman rules
\ba
& V_{\al \beta \gamma \delta}^{(1)}(\vec p_i) = \frac{1}{2} \int [\D^d k] \Gamma^{ab (0)}_{\alpha\beta}(\vec p_1,\vec p_2) R^{(0)}_{abcd}(\vec p) \Gamma^{cd (0)}_{\delta_1\gamma_1}(-\vec p+\vec k,-\vec k) S^{(0)}_{\delta_1\alpha_1}(\vec p-\vec k) 
\nonum \\
& \hspace{5cm} \times S^{(0)}_{\gamma_1\beta_1}(\vec k) \Gamma^{ef (0)}_{\alpha_1\beta_1}(\vec p-\vec k,\vec k) R^{(0)}_{efgh}(\vec p) \Gamma^{gh (0)}_{\gamma\delta}(\vec p_3,\vec p_4),
\ea
which makes obvious the factorization of the external legs out of the integration, \ie
\ba
& V_{\al \beta \gamma \delta}^{(1)}(\vec p_i) = \Gamma^{ab (0)}_{\alpha\beta}(\vec p_1,\vec p_2) R^{(0)}_{abcd}(\vec p) \bigg[ 
\frac{1}{2} \int [\D^d k] \Gamma^{cd (0)}_{\delta_1\gamma_1}(-\vec p+\vec k,-\vec k) S^{(0)}_{\delta_1\alpha_1}(\vec p-\vec k)
\nonum \\
& \hspace{6cm} \times S^{(0)}_{\gamma_1\beta_1}(\vec k) \Gamma^{ef (0)}_{\alpha_1\beta_1}(\vec p-\vec k,\vec k) \bigg] R^{(0)}_{efgh}(\vec p) \Gamma^{gh (0)}_{\gamma\delta}(\vec p_3,\vec p_4).
\ea
Here, the central term in brackets, is by definition the polarization operator $\Pi^{(1)}_{cdef}(\vec p)$.
This procedure holds at all loop orders since only the polarization operator is taking loop corrections
\ba
V_{\al \beta \gamma \delta}^{(L)}(\vec p_i) & = \Gamma^{ab (0)}_{\alpha\beta}(\vec p_1,\vec p_2) R^{(0)}_{abcd}(\vec p) \Pi^{(L)}_{cdef}(\vec p) R^{(0)}_{efgh}(\vec p) \Gamma^{gh (0)}_{\gamma\delta}(\vec p_3,\vec p_4),
\ea
where $L$ stands for the $L$-loop order. Using the Feynman rules and performing the contractions yields the one-loop integral
\be
\Pi_1^{cdef}(\vec p) = -d_c \int \frac{[\D^d k]}{k^4 (\vec p - \vec k)^4} (\vec p - \vec k)^c k^d (\vec p - \vec k)^e k^f.
\ee
At this point, we project out the tensor structure
with the help of (\ref{mem:eq:projectors:normalized}) defining
\bs
\ba
\Pi_M^{(1)}(p^2) & = P^M_{abcd} \Pi^{(1)}_{abcd},\\
\Pi_N^{(1)}(p^2) & = P^N_{abcd} \Pi^{(1)}_{abcd}.
\ea
\es
Performing the contractions, reduction and integration, yields
\bs
\label{mem:tPi1-EFT-exact2}
\ba
&\tilde{\Pi}_{M}^{(1)}(p^2) = -\frac{d_c \mu}{8} \frac{(p^2)^{d/2-2}}{(4\pi)^{d/2}} \frac{d-2}{d-1} G(d,1,1),
\\ 
&\tilde{\Pi}_{N}^{(1)}(p^2) = -\frac{d_c b}{16} \frac{(p^2)^{d/2-2}}{(4\pi)^{d/2}} \frac{(d-2) (d+1)}{d-1} G(d,1,1),
\ea
\es
where $\tilde{\Pi}_{M}(p^2)$ and $\tilde{\Pi}_{N}(p^2)$ were defined in (\ref{mem:R:RM+RN}).
In expanded form, the one-loop polarization operator in the $\overline{\text{MS}}$-scheme then reads
\bs
\label{mem:tPi1-EFT-expanded}
\ba
&\tilde{\Pi}^{(1)}_{M}(p^2) = \frac{d_c \mu}{12 (4\pi M^\eps)^2} \left(-\frac{1}{\eps} + L_p - \frac{5}{3} + \Ord(\eps) \right),
\\
&\tilde{\Pi}^{(1)}_{N}(p^2) = \frac{5 d_c b}{24 (4\pi M^\eps)^2} \left(-\frac{1}{\eps} + L_p - \frac{19}{15} + \Ord(\eps) \right).
\ea
\es
Combining (\ref{mem:tPi1-EFT-expanded}) and (\ref{mem:Pi-ren}) yields the one-loop intermediate renormalization constant
\bs
\label{mem:ZGmu+ZGb-one-loop}
\ba
&\delta Z_{\Gamma_\mu}^{(1)} = - \KK{\tilde{\Pi}^{(1)}_{M}(p^2)} = \frac{d_c \mu_r}{12 (4\pi)^2 \eps},
\\
&\delta Z_{\Gamma_b}^{(1)} = -\KK{\tilde{\Pi}^{(1)}_{N}(p^2)}= \frac{5 d_c b_r}{24 (4\pi)^2 \eps},
\ea
\es
and hence the field renormalization constants
\bs
\label{mem:Zmu+Zb-one-loop}
\ba
&\delta Z_{\mu}^{(1)} = \delta Z_{\Gamma_\mu}^{(1)} - 2 \delta Z^{(1)} = \frac{10 b_r+(d_c+20) \mu_r}{12 (4\pi)^2 \eps},
\\
&\delta Z_{b}^{(1)} = \delta Z_{\Gamma_b}^{(1)} - 2 \delta Z^{(1)} = \frac{5 ((d_c+4) b_r+8 \mu_r)}{24 (4\pi)^2 \eps}.
\ea
\es
For completeness, we add the one-loop renormalized polarization operator in projected form:
\bs
\ba
&\tilde{\Pi}^{(1)}_{M,r}(p^2) = \tilde{\Pi}_M^{(1)}+\delta Z_{\Gamma_\mu}^{(1)}=- \frac{d_c \mu_r (5-3 L_p)}{36 (4\pi)^2}+\Ord(\eps), \\
&\tilde{\Pi}^{(1)}_{N,r}(p^2) = \tilde{\Pi}_N^{(1)}+\delta Z_{\Gamma_b}^{(1)} = -\frac{d_c b_r (19-15 L_p)}{72 (4\pi)^2}+\Ord(\eps).
\ea
\es

\subsection{Two-loop analysis}

\subsubsection{Two-loop flexuron self-energy}
\label{mem:subsec:two-loop:flexuron}

\begin{figure}[h]
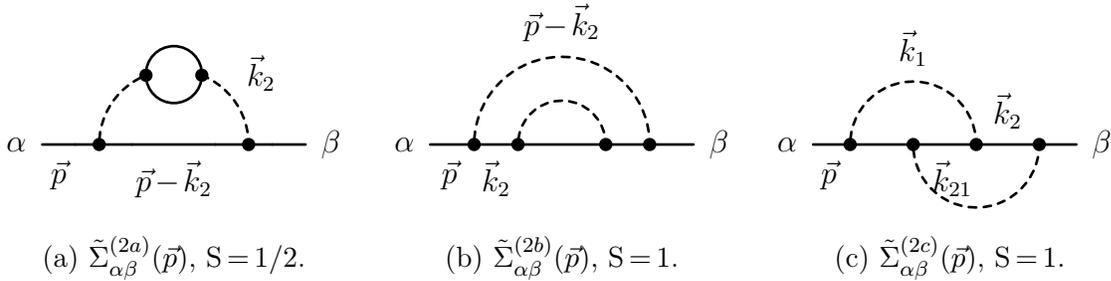

\begin{subfigure}{.32\textwidth}
\centering
\GraphSigmaTwoA
\caption{$\tilde{\Sigma}^{(2a)}_{\alpha\beta}(\vec p)$, $\text{S}=1/2$.}
\label{mem:fig:two-loop:Sigma:a}
\end{subfigure}
\begin{subfigure}{.32\textwidth}
\centering
\GraphSigmaTwoB
\caption{$\tilde{\Sigma}^{(2b)}_{\alpha\beta}(\vec p)$, $\text{S}=1$.}
\label{mem:fig:two-loop:Sigma:b}
\end{subfigure}
\begin{subfigure}{0.32\textwidth}
\centering
\GraphSigmaTwoC
\vspace{-0.5cm}
\caption{$\tilde{\Sigma}^{(2c)}_{\alpha\beta}(\vec p)$, $\text{S}=1$.}
\label{mem:fig:two-loop:Sigma:c}
\end{subfigure}
\caption{Two-loop flexuron self-energy diagrams and symmetry factors (S). }
\label{mem:fig:two-loop:Sigma}
\end{figure}

At two-loop, the flexuron self-energy has three corrections represented by the diagrams in figure \ref{mem:fig:two-loop:Sigma}, labeled $(a)$, $(b)$ and $(c)$.
%

We first consider the diagram of figure \ref{mem:fig:two-loop:Sigma:a}, defined as
%
\ba
\Sigma^{(2a)}_{\al \beta}(\vec p)
& = \frac{1}{2} \int [\D^d k_1] [\D^d k_2] 
V_{\al \al_1 \gamma_1 \gamma_2}^{(0)}(\vec p, -\vec p + \vec k_2, \vec k_1 -\vec k_2, - \vec k_1) 
S_{\al_1 \beta_1}^{(0)}(\vec p - \vec k_2) 
\nonum \\
&\hspace{1cm}
\times 
V_{\delta_1 \delta_2 \beta_1 \beta}^{(0)}(- \vec k_1+\vec k_2,\vec k_1, \vec p -\vec k_2, -\vec p) 
S_{\gamma_1 \delta_1}^{(0)}(- \vec k_1 + \vec k_2) S_{\gamma_2 \delta_2}^{(0)}(\vec k_1).
\ea
Performing the contractions, reduction and integration, yields the exact result
\ba
\tilde{\Sigma}^{(2a)}(p^2) &= - \frac{d_c}{24} \frac{(p^2)^{d-4}}{(4\pi)^d} 
\frac{(d-3) (d-2) (d+1)}{(d-1) (d-6) (d-4)} \big[(d+1) b^2+2(d-2) \mu^2 \big] G(d,1,1) G(d,1,2-d/2).
\ea
The corresponding $\eps$-expansion in the $\overline{\text{MS}}$-scheme reads:
\ba
\tilde{\Sigma}^{(2a)}(p^2) =\frac{d_c (5 b^2+4 \mu^2)}{288 (4\pi M^{\eps})^4} \biggr[ \frac{1}{2 \eps^2} 
+ \frac{1}{\eps} \biggr(\frac{53}{60}- L_p+\frac{3 b^2}{10 b^2+8 \mu^2}\biggl)+\Ord(\eps^0)\biggl].
\label{mem:Sigma_2a-exp2}
\ea

We then consider the diagram of figure \ref{mem:fig:two-loop:Sigma:b}, with symmetry factor $1$ and is defined as
\ba
\Sigma^{(2b)}_{\al \beta}(\vec p)
& = \int [\D^d k_1] [\D^d k_2] 
V_{\al \al_1 \beta_2 \beta}^{(0)}(\vec p, -\vec k_2, \vec k_2, -\vec p) 
S_{\al_1 \al_2}^{(0)}(\vec k_2) 
\nonum \\
& \hspace{1cm}\times V_{\al_2 \gamma_1 \gamma_2 \beta_1}^{(0)}(\vec k_2, - \vec k_1, \vec k_1, -\vec k_2) 
S_{\beta_1 \beta_2}^{(0)}(\vec k_2) S_{\gamma_1 \gamma_2}^{(0)}(\vec k_1).
\ea
Performing the contractions,reduction and integration, yields the exact result
\ba
\label{mem:Sigma_2b-EFT2}
\tilde{\Sigma}^{(2b)}(p^2) &= - \frac{\big(b+(d-2) \mu\big)^2}{12} \frac{(p^2)^{d-4}}{(4\pi)^d} \frac{(d-3) (d-2) (d+1)^2}{(d-6) (d-4) (d-1)} G(d,1,1) G(d,1,2-d/2),
\ea
and the corresponding $\eps$-expansion in the $\overline{\text{MS}}$-scheme reads
\ba
&\tilde{\Sigma}^{(2b)}(p^2) = \frac{25 (b+2 \mu)^2}{144 (4\pi M^{\eps})^4} \biggr[\frac{1}{2 \eps^2}+\frac{1}{\eps} \biggr(\frac{11}{60}-L_p+\frac{b}{b+2 \mu}\biggl)+\Ord(\eps^0) \biggl].
\label{mem:Sigma_2b-exp2}
\ea

Finally, we consider the diagram of figure \ref{mem:fig:two-loop:Sigma:c} where $\vec k_{21} = \vec k_2 - \vec k_1$, which has a symmetry factor of $1$ and is defined as
\ba
\Sigma^{(2c)}_{\al \beta}(\vec p) 
& = \int [\D^d k_1] [\D^d k_2] 
V_{\al \al_1 \beta_2 \beta_3}^{(0)}(\vec p, -\vec p + \vec k_1,\vec k_2 - \vec k_1, - \vec k_2) 
S_{\al_1 \al_2}^{(0)}(\vec p - \vec k_1) \
\nonum \\
&\hspace{1cm}\times V_{\al_2 \al_3 \beta_1 \beta}^{(0)}(\vec p - \vec k_1, -\vec k_2 + \vec k_1, \vec k_2, -\vec p) 
S_{\al_3 \beta_2}^{(0)}(\vec k_2 - \vec k_1) S_{\beta_3 \beta_1}^{(0)}(\vec k_2).
\ea
Performing the contractions, reduction and integration, yields
\ba
& \tilde{\Sigma}^{(2c)}(p^2) =\frac{1}{256}\frac{(p^2)^{d-4}}{(4\pi)^d} \frac{1}{(d-6)(d-4)(d-1)^3} G^2(d,1,1) \nonum \\ 
& \times \bigg[ (d^7-28 d^6+313 d^5-1686 d^4+4388 d^3-4864 d^2+976 d+960) b^2 \nonum \\ 
& + 2(d^8-26 d^7+277 d^6-1556 d^5+4956 d^4-8832 d^3+7680 d^2-1408 d-1152) b \mu \nonum \\
& + (d^9-24 d^8+245 d^7-1394 d^6+4936 d^5-11464 d^4+17008 d^3-14048 d^2+4032 d+768) \mu^2 \bigg] \nonum \\
& -\frac{1}{12}\frac{(p^2)^{d-4}}{(4\pi)^d} \frac{d-3}{(d-6)^2(d-4)^2(d-1)^2} G(d,1,1) G(d,1,2-d/2) \nonum \\ 
& \times \bigg[ (2 d^6-26 d^5+93 d^4+35 d^3-604 d^2+524 d+336) b^2 \nonum \\ 
& + 2(d^7-14 d^6+64 d^5-57 d^4-366 d^3+1012 d^2-616 d-384) b \mu \nonum \\ 
& + (d^8-15 d^7+82 d^6-188 d^5+44 d^4+788 d^3-1776 d^2+1232 d+192) \mu^2 \bigg].
\ea
The corresponding $\eps$-expansion in the $\overline{\text{MS}}$-scheme reads
\ba
&\tilde{\Sigma}^{(2c)}(p^2) = \frac{5 \big(121 b^2-56 b \mu+52 \mu^2\big)}{5184 (4\pi M^{\eps})^4\eps} + \Ord(\eps^0),
\label{mem:Sigma_2c-exp2}
\ea
which contains only a simple pole and no non-local (proportional to $L_p$) term at this order. This comes from the fact that this diagram does not have any divergent subdiagram.



Summing all the previous contributions, the total two-loop self-energy reads
\ba
\tilde{\Sigma}^{(2)} 
& =\tilde{\Sigma}^{(2a)}+\tilde{\Sigma}^{(2b)}+\tilde{\Sigma}^{(2c)} \nonum \\
& =\frac{5 (5 b^2 (d_c+2)+40 b \mu+4 \mu^2 (d_c+10))}{288 (4\pi M^\eps)^4} \bigg[
\frac{1}{2 \eps^2} \nonum \\
& + \frac{1}{\eps}\bigg(-L_p+\frac{5 b^2 (213 d_c+668)+4360 b \mu +4 \mu^2 (159 d_c+460)}{180 (5 b^2 (d_c+2)+40 b \mu
+4 \mu^2 (d_c+10))}\bigg) +\Ord(\eps^0)\bigg].
\label{mem:eq:Sigma_2_tot}
\ea
Together with the one-loop results of the previous section, this allows us to access the two-loop renormalization constant, (\ref{mem:deltaZ:gen}), that can be written in the form
\be
\delta Z^{(2)} =\KK{\tilde{\Sigma}^{(2)}} + \KK{\tilde{\Sigma}^{(1)} \KK{\tilde{\Sigma}^{(1)}}} \, .
\ee
Straightforward substitutions lead to
\ba
\delta Z^{(2)} & = \frac{-5}{576 (4\pi)^4} \bigg[\frac{1}{\eps^2} \Big(
5 (d_c + 2) b_r^2 + 40 b_r \mu_r + 4 (d_c + 10) \mu_r^2 \Big)
\nonum \\
&+\frac{1}{90 \eps} \Big(
5 (15 d_c - 212) b_r^2 +1160 b_r \mu_r - 4 (111 d_c-20) \mu_r^2 \Big) \bigg],
\label{mem:Z_2:exp2}
\ea
where all the non-local terms cancelled out as expected in the $\overline{\text{MS}}$-scheme. 

We may also access the renormalized self-energy, (\ref{mem:sigma-ren}), that we add here at two-loop order
for completeness:
\ba
\tilde{\Sigma}_r^{(2)} & = \tilde{\Sigma}^{(2)}-\delta Z^{(2)}+ \delta Z^{(1)} \tilde{\Sigma}^{(1)} \nonum \\
& = \frac{1}{20736 (4\pi)^4}\bigg[(3239 d_c-6912 \zeta_3+19024) b_r^2+8 (432 \zeta_3+1651) b_r \mu_r \nonum \\
& + 4 (805 d_c-6912 \zeta_3+12708) \mu_r^2+ 180 \Big(5 (d_c+2) b_r^2+40 b_r \mu_r+4 (d_c+10) \mu_r^2\Big) L_p^2 \nonum \\
& - 20 \Big(11 (9 d_c+40) b_r^2 +320 b_r \mu_r+4 (27 d_c+44) \mu_r^2\Big) L_p\bigg] \, .
\label{mem:eq:deltaZ2}
\ea
%


\subsubsection{Two-loop vertex self-energy}
\label{mem:subsec:two-loop:vertex}

At two-loop, the self-energy of the flexuron four-point interaction (or equivalently the $R$-propagator polarization) has two corrections represented by the diagrams in figure \ref{mem:fig:two-loop:vertex}, labeled $(a)$ and $(b)$.

\begin{figure}[h]
\begin{subfigure}{.5\textwidth}
\centering
\GraphVTwoA
\caption{$V_{\al \beta \gamma \delta}^{(2a)}(\vec p_1, \vec p_2, \vec p_3, \vec p_4)$, $\text{S}=1$.}
\label{mem:fig:two-loop:vertex:a}
\end{subfigure}
\begin{subfigure}{.5\textwidth}
\centering
\GraphVTwoB
\caption{$V_{\al \beta \gamma \delta}^{(2b)}(\vec p_1, \vec p_2, \vec p_3, \vec p_4)$, $\text{S}=1/2$.}
\label{mem:fig:two-loop:vertex:b}
\end{subfigure}
\caption{Two-loop vertex self-energy diagrams with their symmetry factors (S). \\ Note that $\vec k_{12} = \vec k_1 - \vec k_2$ and $\vec p = \vec p_1 + \vec p_2=-\vec p_3-\vec p_4$.}
\label{mem:fig:two-loop:vertex}
\end{figure}

We first consider the diagram of figure \ref{mem:fig:two-loop:vertex:a} with symmetry factor of $1$:
\ba
V_{\al \beta \gamma \delta}^{(2a)}(\vec p_i)=
& \int [\D^d k_1] [\D^d k_2] 
V_{\al \beta \gamma_1 \delta_1}^{(0)}(\vec p_1, \vec p_2, -\vec p + \vec k_2, - \vec k_2) 
S_{\delta_1 \al_2}^{(0)}(\vec k_2) V_{\al_2 \beta_2 \gamma_2 \delta_2}^{(0)}(\vec k_2, -\vec k_1, \vec k_1, - \vec k_2)
\nonum \\
&\hspace{1cm}\times 
S_{\beta_2 \gamma_2}^{(0)}(\vec k_1) S_{\delta_2 \al_1}^{(0)}(\vec k_2)
V_{\al_1 \beta_1 \gamma \delta}^{(0)}(\vec k_2, \vec p - \vec k_2, \vec p_3,\vec p_4) S_{\gamma_1 \beta_1}^{(0)}(\vec p - \vec k_2)\,.
\ea
It can be re-expressed in the following form
\bs
\label{mem:V2a-expr}
\ba
&V_{\al \beta \gamma \delta}^{(2a)}(\vec p_i) = \Gamma^{ab (0)}_{\alpha\beta}(\vec p_1,\vec p_2)
R_{abcd}^{(2a)}(\vec p) \Gamma^{cd(0)}_{\gamma\delta}(\vec p_3,\vec p_4),
\\
&R_{abcd}^{(2a)}(\vec p) = R_{abef}^{(0)}(\vec p) \Pi^{(2a)}_{efgh}(\vec p) R_{ghcd}^{(0)}(\vec p),
\ea
\es
where $\Pi^{(2a)}_{efgh}(\vec p)$ is the two-loop polarization associated with the diagram. Using the Feynman rules and performing the contractions leads to
\ba
\Pi^{(2a)}_{efgh}(\vec p) & = \int \frac{4 d_c[\D^d k_1] [\D^d k_2]}{k_1^8 (\vec p - \vec k_1)^4 (\vec k_1 - \vec k_2)^4} 
(\vec k_1 - \vec p)_e (k_1)_f (\vec k_1 - \vec p)_g (k_1)_h R_{ijlm}^{(0)}(-\vec k_2) k_1^i (\vec k_1 - \vec k_2)^j k_1^l (\vec k_1 - \vec k_2)^m.
\ea
Decomposing this self-energy on the basis of the irreducible tensors and evaluating all the integrals, one arrives at the exact results
\bs
\label{mem:tPi2a-EFT-exact2}
\ba
&\tilde{\Pi}^{(2a)}_{M}(p^2) = -\frac{d_c \mu (b+(d-2) \mu)}{6} \frac{(p^2)^{d-4}}{(4\pi)^{d}} \frac{(d-3) (d-2) (d+1)}{(d-6) (d-4) (d-1)} G(d,1,1) G(d,1,2-d/2),
\\
&\tilde{\Pi}^{(2a)}_{N}(p^2) = -\frac{d_c b (b+(d-2) \mu)}{12} \frac{(p^2)^{d-4}}{(4\pi)^{d}} \frac{(d-3) (d-2) (d+1)^2}{(d-6) (d-4) (d-1)} G(d,1,1) G(d,1,2-d/2),
\ea
\es
where $\tilde{\Pi}_{M}(p^2)$ and $\tilde{\Pi}_{N}(p^2)$ were defined in (\ref{mem:R:RM+RN}). In expanded form 
\bs
\label{mem:tPi2a-EFT-expanded2}
\ba
&\tilde{\Pi}^{(2a)}_{M}(p^2) = \frac{5 d_c \mu (b+2\mu)}{72 (4\pi M^{\eps})^4} \bigg[\frac{1}{2 \eps^2}+\frac{1}{\eps} \left(\frac{53}{60}-L_p+\frac{b}{2 (b+2 \mu)}\right) + \Ord(\eps^0) \bigg],
\\
&\tilde{\Pi}^{(2a)}_{N}(p^2) = \frac{25 d_c b (b+2\mu)}{144 (4\pi M^{\eps})^4} \bigg[\frac{1}{2 \eps^2}+\frac{1}{\eps} \left(\frac{41}{60}-L_p+\frac{b}{2 (b+2 \mu)}\right) + \Ord(\eps^0) \bigg].
\ea
\es

We then consider the diagram figure \ref{mem:fig:two-loop:vertex:b}, with symmetry factor $1/2$ and defined as
\ba
V_{\al \beta \gamma \delta}^{(2b)}(\vec p_i)
&= \frac{1}{2} \int [\D^d k_1] [\D^d k_2] 
V_{\al \beta \gamma_1 \delta_1}^{(0)}(\vec p_1, \vec p_2, -\vec k_1, - \vec p + \vec k_1) 
S_{\delta_1 \delta_2}^{(0)}(\vec p - \vec k_1) \\
&\hspace{-1cm}\times V_{\gamma_2 \alpha_2 \delta_2 \beta_2}^{(0)}(\vec k_1, - \vec k_2, \vec p - \vec k_1, - \vec p + \vec k_2)
S_{\al_2 \al_1}^{(0)}(\vec k_2) S_{\gamma_1 \gamma_2}^{(0)}(\vec k_1) V_{\al_1 \beta_1 \gamma \delta}^{(0)}(\vec k_2, \vec p - \vec k_2, \vec p_3,\vec p_4) S_{\beta_2 \beta_1}^{(0)}(\vec p - \vec k_2). \nonum
\ea
Decomposing this self-energy on the basis of the irreducible tensors, performing the reduction and the integration, yields the two projections of the polarization
\bs
\label{mem:Pi2b-EFT-expanded1}
\ba
\tilde{\Pi}^{(2b)}_{M}(p^2) & = \frac{d_c \mu}{128}\frac{(p^2)^{d-4}}{(4\pi)^d}\frac{1}{(d-6)(d-4)(d-1)^3(d+1)} G^2(d,1,1) \nonum \\ 
& \times \big[(d^5-13 d^4+54 d^3+60 d^2-744 d+672) b \nonum \\
& + (d^6-15 d^5+80 d^4+16 d^3-608 d^2+1456 d-960) \mu \big] \nonum \\
& - \frac{d_c \mu}{24}\frac{(p^2)^{d-4}}{(4\pi)^d} \frac{d-3}{(d-6)^2(d-4)^2(d-1)^2(d+1)}G(d,1,1) G(d,1,2-d/2) \nonum \\
& \times \big[(d^6-10 d^5+27 d^4+10 d^3-164 d^2+184 d+672) b \nonum \\
& + (d^7-8 d^6+39 d^5-160 d^4+208 d^3+16 d^2-240d-576) \mu\big],
\\
\tilde{\Pi}^{(2b)}_{N}(p^2) & = \frac{d_c b}{512}\frac{(p^2)^{d-4}}{(4\pi)^d}\frac{1}{(d-6)(d-4)(d-1)^3} G^2(d,1,1) \nonum \\ 
& \times \big[(d^7-28 d^6+313 d^5-1686 d^4+4388 d^3-4864 d^2+976 d+960) b \nonum \\ 
& + (9 d^8-174 d^7+1377 d^6-5768d^5+13784 d^4-18936 d^3+13584 d^2-2784 d-1152) \mu\big] \nonum \\
& - \frac{d_c b}{24}\frac{(p^2)^{d-4}}{(4\pi)^d}\frac{d-3}{(d-6)^2(d-4)^2(d-1)^2} G(d,1,1) G(d,1,2-d/2) \nonum \\
& \times \big[(2 d^6-26 d^5+93 d^4+35 d^3-604 d^2+524 d+336) b \nonum \\ 
& +(2 d^7-22 d^6+57 d^5+157 d^4-982 d^3+1700d^2-984 d-288) \mu\big].
\ea
\es
In expanded form and in $\overline{\text{MS}}$-scheme, these results read
\bs
\ba
&\tilde{\Pi}^{(2b)}_{M}(p^2) = -\frac{5 d_c \mu (b+2\mu)}{5184 (4\pi M^{\eps})^4} \biggl[ \frac{1}{\eps} + \Ord(\eps^0) \biggr],
\\
&\tilde{\Pi}^{(2b)}_{N}(p^2) = \frac{5 d_c b (b+2\mu)}{10368 (4\pi M^{\eps})^4} \biggl[ \frac{121}{\eps} + \Ord(\eps^0) \biggr].
\ea
\label{mem:Pi2b-EFT-expanded2}
\es
Similarly to the diagram 2b in the flexuron self-energy, the results \eqref{mem:Pi2b-EFT-expanded2} shows only simple poles.


Summing the two contributions, (\ref{mem:tPi2a-EFT-expanded2}) and (\ref{mem:Pi2b-EFT-expanded2}), yields the total two-loop polarization projections
\bs
\ba
\tilde{\Pi}^{(2)}_{M}(p^2) & =\tilde{\Pi}^{(2a)}_{M}(p^2)+\tilde{\Pi}^{(2b)}_{M}(p^2) \nonum \\
& = \frac{5 d_c \mu (b+2\mu)}{72 (4\pi M^{\eps})^4} \biggl[\frac{1}{2 \eps^2}+\frac{1}{\eps} \left(\frac{313}{360}-L_p+\frac{b}{2 (b+2 \mu)}\right) + \Ord(\eps^0) \biggr],
\\
\tilde{\Pi}^{(2)}_{N}(p^2) & = \tilde{\Pi}^{(2a)}_{N}(p^2)+\tilde{\Pi}^{(2b)}_{N}(p^2) \nonum \\ 
& = \frac{25 d_c b (b+2\mu)}{144 (4\pi M^{\eps})^4} \biggl[\frac{1}{2 \eps^2}+\frac{1}{\eps} \left(\frac{367}{360}-L_p+\frac{b}{2 (b+2 \mu)}\right) + \Ord(\eps^0) \biggr].
\ea
\es
From the above results, we compute the two-loop contribution to the intermediary renormalization constant
\bs
\label{mem:ZGmu+ZGb-two-loop}
\ba
\delta Z_{\Gamma_\mu}^{(2)} & = -\KK{\tilde{\Pi}_M^{(2)}} + \KK{{\delta Z_{\Gamma_\mu}^{(1)}}^2} + \KK{\delta Z_{\Gamma_\mu}^{(1)} \tilde{\Pi}_M^{(1)}} \nonum \\
& = \frac{d_c \mu_r}{5184 (4\pi)^4}\bigg[\frac{36}{\eps^2}\Big(5 b_r+(10+d_c) \mu_r\Big)+\frac{1}{\eps}(107 b_r+574 \mu_r)\bigg],
\\
\delta Z_{\Gamma_b}^{(2)} & = -\KK{\tilde{\Pi}_N^{(2)}} + \KK{{\delta Z_{\Gamma_b}^{(1)}}^2} + \KK{\delta Z_{\Gamma_b}^{(1)} \tilde{\Pi}_N^{(1)}} \nonum \\
& = \frac{d_c b_r}{576 (4\pi)^4}\bigg[\frac{25}{\eps^2}\Big((2+d_c) b_r+4 \mu_r\Big)-\frac{5}{18 \eps}\Big(91 b_r-178 \mu_r\Big)\bigg].
\ea
\es
Hence, we can compute the coupling renormalization constants
\bs
\label{mem:Zmu+Zb-two-loop}
\ba
\delta Z_{\mu}^{(2)} & = \delta Z_{\Gamma_\mu}^{(2)}-2 \delta Z^{(2)}-2 \delta Z^{(1)} \delta Z_{\Gamma_\mu}^{(1)}+3 {\delta Z^{(1)}}^2 \nonum \\
& = \frac{1}{5184 (4\pi)^4}\bigg[ \frac{18}{\eps^2}\Big(10(3 d_c+80) b_r \mu_r+25 (d_c+8) b_r^2+2(d_c+20)^2 \mu_r^2\Big) \nonum \\ 
& \hspace{1cm} + \frac{1}{\eps}\Big((107 d_c+1160) b_r \mu_r+5(15 d_c-212) b_r^2+10 (13 d_c+8) \mu_r^2\Big)\bigg],
\\
\delta Z_{b}^{(2)} & = \delta Z_{\Gamma_b}^{(2)}-2 \delta Z^{(2)}-2 \delta Z^{(1)} \delta Z_{\Gamma_b}^{(1)}+3 {\delta Z^{(1)}}^2 \nonum \\
& = \frac{1}{576 (4\pi)^4} \bigg[ \frac{5}{\eps^2} \Big(20(3 d_c+16) b_r \mu_r+5(d_c+4)^2 b_r^2 +8(d_c+40) \mu_r^2\Big) \nonum \\
& \hspace{0.75cm} + \frac{1}{18 \eps} \Big(10 (89 d_c+232) b_r \mu_r-5(61 d_c+424) b_r^2-8(111 d_c-20) \mu_r^2 \Big)\bigg]\, ,
\ea
\es
where, once again, all the $L_p$ contributions cancel out.

These results also allow us to compute the two-loop contribution to the renormalized polarization projections that we add here for completeness
\bs
\ba
\tilde{\Pi}^{(2)}_{M,r}(p^2) & = \tilde{\Pi}_M^{(2)}+\delta Z_{\Gamma_\mu}^{(2)} - \delta Z_{\Gamma_\mu}^{(1)} \tilde{\Pi}_M^{(1)} - {\delta Z_{\Gamma_\mu}^{(1)}}^2 \\
& = -\frac{d_c \mu_r}{51840 (4\pi)^4}\Big[2(13187 + 260 L_p + 1800 L_p^2 - 12096 \zeta_3) \mu_r \nonum \\[-0.25cm] 
& \hspace{2.75cm} - (6863 - 3860 L_p + 1800 L_p^2 - 864 \zeta_3) b_r\Big], \nonum \\
\tilde{\Pi}^{(2)}_{N,r}(p^2) & = \tilde{\Pi}_N^{(2)}+\delta Z_{\Gamma_b}^{(2)} - \delta Z_{\Gamma_b}^{(1)} \tilde{\Pi}_N^{(1)} - {\delta Z_{\Gamma_b}^{(1)}}^2 \\
& = \frac{d_c b_r}{20736 (4\pi)^4}\Big[2(7065 - 2780 L_p + 1800 L_p^2 - 6048 \zeta_3) \mu_r \nonum \\[-0.25cm] 
& \hspace{2.55cm} + (12751 - 6380 L_p + 1800 L_p^2 - 3456 \zeta_3) b_r \Big]. \nonum
\ea
\es

\subsection{Three-loop analysis}
\label{mem:sec:three-loop}

\subsubsection{Three-loop flexuron self-energy}
\label{mem:subsec:three-loop:flexuron}

\begin{figure}[htp]
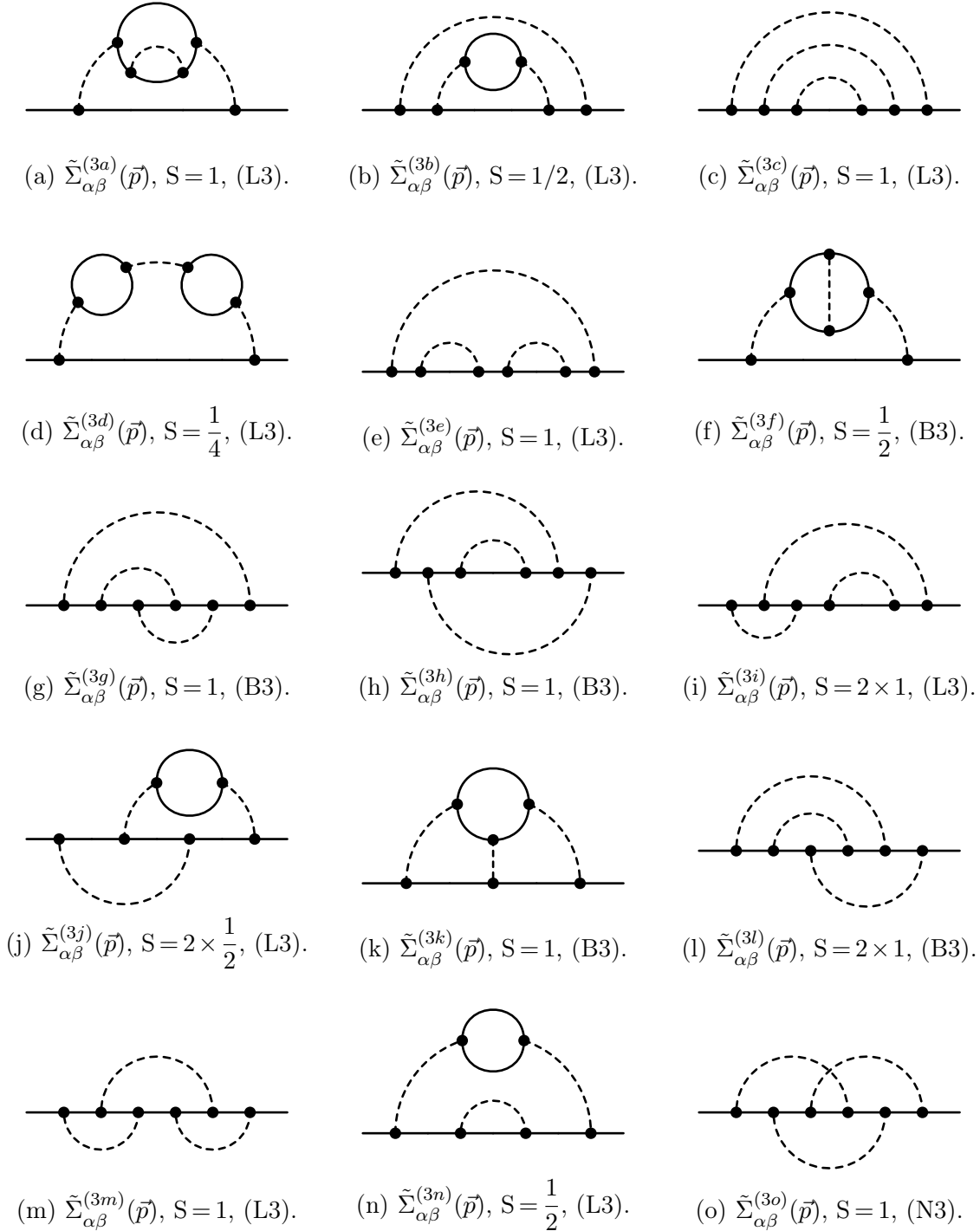

\centering
\begin{subfigure}{.32\textwidth}
\centering
\GraphSigmaThreeA
\caption{$\tilde{\Sigma}^{(3a)}_{\alpha\beta}(\vec p)$,~$\text{S}=1$,~(L3).}
\end{subfigure}
\begin{subfigure}{.32\textwidth}
\centering
\GraphSigmaThreeB 
\caption{$\tilde{\Sigma}^{(3b)}_{\alpha\beta}(\vec p)$,~$\text{S}=1/2$,~(L3).}
\end{subfigure}
\begin{subfigure}{.32\textwidth}
\centering
\GraphSigmaThreeC 
\caption{$\tilde{\Sigma}^{(3c)}_{\alpha\beta}(\vec p)$,~$\text{S}=1$,~(L3).}
\end{subfigure}
\\
\begin{subfigure}{.32\textwidth}
\centering
\GraphSigmaThreeD
\caption{$\tilde{\Sigma}^{(3d)}_{\alpha\beta}(\vec p)$,~$\text{S}=\dfrac{1}{4}$,~(L3).}
\end{subfigure}
\begin{subfigure}{.32\textwidth}
\centering
\GraphSigmaThreeE 
\caption{$\tilde{\Sigma}^{(3e)}_{\alpha\beta}(\vec p)$,~$\text{S}=1$,~(L3).}
\end{subfigure}
\begin{subfigure}{.32\textwidth}
\centering
\GraphSigmaThreeF
\caption{$\tilde{\Sigma}^{(3f)}_{\alpha\beta}(\vec p)$,~$\text{S}=\dfrac{1}{2}$,~(B3).}
\end{subfigure}
\\
\begin{subfigure}{.32\textwidth}
\centering
\GraphSigmaThreeG
\caption{$\tilde{\Sigma}^{(3g)}_{\alpha\beta}(\vec p)$,~$\text{S}=1$,~(B3).}
\end{subfigure}
\begin{subfigure}{.32\textwidth}
\centering
\GraphSigmaThreeH
\caption{$\tilde{\Sigma}^{(3h)}_{\alpha\beta}(\vec p)$,~$\text{S}=1$,~(B3).}
\end{subfigure}
\begin{subfigure}{.32\textwidth}
\centering
\GraphSigmaThreeI
\caption{$\tilde{\Sigma}^{(3i)}_{\alpha\beta}(\vec p)$,~$\text{S}=2\times1$,~(L3).}
\end{subfigure}
\\
\begin{subfigure}{.32\textwidth}
\centering
\GraphSigmaThreeJ
\caption{$\tilde{\Sigma}^{(3j)}_{\alpha\beta}(\vec p)$,~$\text{S}=2\times\dfrac{1}{2}$,~(L3).}
\end{subfigure}
\begin{subfigure}{.32\textwidth}
\centering
\GraphSigmaThreeK
\caption{$\tilde{\Sigma}^{(3k)}_{\alpha\beta}(\vec p)$,~$\text{S}=1$,~(B3).}
\end{subfigure}
\begin{subfigure}{.32\textwidth}
\centering
\GraphSigmaThreeL
\caption{$\tilde{\Sigma}^{(3l)}_{\alpha\beta}(\vec p)$,~$\text{S}=2\times1$,~(B3).}
\end{subfigure}
\\
\begin{subfigure}{.32\textwidth}
\centering
\GraphSigmaThreeM
\caption{$\tilde{\Sigma}^{(3m)}_{\alpha\beta}(\vec p)$,~$\text{S}=1$,~(L3).}
\end{subfigure}
\begin{subfigure}{.32\textwidth}
\centering
\GraphSigmaThreeN
\caption{$\tilde{\Sigma}^{(3n)}_{\alpha\beta}(\vec p)$,~$\text{S}=\dfrac{1}{2}$,~(L3).}
\end{subfigure}
\begin{subfigure}{.32\textwidth}
\centering
\GraphSigmaThreeO
\caption{$\tilde{\Sigma}^{(3o)}_{\alpha\beta}(\vec p)$,~$\text{S}=1$,~(N3).}
\end{subfigure}
\caption{Three-loop flexuron self-energy diagrams and their associated symmetry factors (S). Momenta arrows have been dropped to keep it light.}
\label{mem:fig:three-loop:sigma}
\end{figure}

We now consider the three-loop flexuron self-energy that consists of $15$ independent diagrams, all displayed in figure \ref{mem:fig:three-loop:sigma} and labeled in alphabetical order from $a$ to $o$. There are 9 diagrams ($a,b,c,d,e,i,j,m,n$) of the Ladder (L3) topology, 5 diagrams ($f,g,h,k,l$) of the Benz (B3) topology and one diagram ($o$) of the Non-planar (N3) topology, see \ref{chap:appendix} for more details on the three-loop topologies. All of them are displayed with their corresponding symmetry factor (S) that are either $1$, $1/2$ or $1/4$. Moreover, by symmetry, 3 diagrams ($i,j,l$) should be taken into account twice. We therefore add an explicit factor $2$ to their symmetry factor (S).

Proceeding along the lines of the one and two-loop cases, \ie, performing carefully the contractions, reduction, integration, and $\eps$ expansion of the 15 diagrams as well as summing all of them together yields the complete three-loop flexuron self-energy
\ba
& \tilde{\Sigma}^{(3)}(p^2) = \tilde{\Sigma}^{(3a)}(p^2)+\tilde{\Sigma}^{(3b)}(p^2)+\dots+\tilde{\Sigma}^{(3o)}(p^2) \nonum \\
& =- \frac{e^{-3 \eps L_p}}{20736 (4\pi M^{\eps})^6}\bigg[\frac{5}{\eps^3}\Big(
50(5 d_c+18) b^2\mu +25(d_c+2)(d_c+3) b^3 +100(d_c+18) b \mu^2 \nonum \\
& + 8(d_c+10)(d_c+15) \mu^3 \Big) + \frac{4}{18 \eps^2}\Big(25(565 d_c+2604) b^2\mu+50((39 d_c+265)+442) d_c b^3 \nonum \\ 
& + 20(296 d_c+2595) b \mu^2+4((147 d_c+2105)+5050) d_c \mu^3\Big) \nonum \\
& - \frac{1}{162 \eps}\Big(12 \left(10125 (d_c+18) \zeta_2+648 (44 d_c+597) \zeta_3-144560 d_c-1933147 \right)b \mu ^2 \nonum \\
& + \left(30375 (d_c+2) (d_c+3) \zeta_2+2592 (171 d_c+524) \zeta_3-d_c (346320 d_c+2837779)-5182746 \right)b^3\nonum \\
& +8\left(1215 (d_c+10) (d_c+15) \zeta_2+648 (48 d_c+2669) \zeta_3-2 d_c (6516 d_c+152741)-3328149 \right) \mu ^3 \nonum \\
& +6 \left(10125 (5 d_c+18) \zeta_2+5184 (5 d_c+71) \zeta_3-516735 d_c-2336048 \right)b^2 \mu\Big)+\Ord(\eps^0)\bigg],
\ea
where the $L_p$ dependence has been factorized out for the sake of brevity. However, all computations were carried out with explicit $L_p$ terms such that their cancellation in the next steps provides a non-trivial check of the computations. 

Combining our three-loop result with the one- and two-loop results of the previous sections, yields the following three-loop contribution to the field renormalization constant
\ba
\delta Z^{(3)} & = \KK{\tilde{\Sigma}^{(3)}} + \KK{\delta Z^{(2)} \tilde{\Sigma}^{(1)}} + \KK{\delta Z^{(1)} \tilde{\Sigma}^{(2)}} \nonum \\
& = - \frac{1}{20736 (4\pi)^6}\bigg[\frac{5}{\eps^3}\Big(50(5 d_c+18) b_r^2 \mu_r+100(d_c+18) b_r \mu_r^2+25(d_c+2)(d_c+3) b_r^3 \nonum \\ 
& + 8(d_c+10)(d_c+15) \mu_r^3 \Big) + \frac{1}{18 \eps^2}\Big(50(311 d_c-480) b_r^2 \mu_r-20(283 d_c-3000) b_r \mu_r^2 \nonum \\ 
& + 25(15 d_c^2-319 d_c-1060) b_r^3-8(3 d_c+5)(37 d_c-100) \mu_r^3 \Big) \nonum \\
& - \frac{1}{324 \eps}\Big(6\big(18144(5 d_c+2) \zeta_3-56445 d_c+221204\big) b_r^2 \mu_r+12\big(1296(50 d_c+57) \zeta_3\nonum \\
& - 82681 d_c-108974\big) b_r \mu_r^2-\big(5184(9 d_c+16) \zeta_3-41625 d_c^2+180563 d_c+516252\big) b_r^3 \nonum \\ 
& + 8((1395 d_c-124416 \zeta_3+188605) d_c+659664 \zeta_3-652398) \mu_r^3\Big)+\Ord(\eps^0)\bigg].
\ea
For completeness, we add the three-loop contribution to the renormalized flexuron self-energy
\ba
& \tilde{\Sigma}_r^{(3)}(p^2) = \tilde{\Sigma}^{(3)}-\delta Z^{(3)}+\delta Z^{(2)}\tilde{\Sigma}^{(1)}+\delta Z^{(1)}\tilde{\Sigma}^{(2)} \nonum \\
& = \frac{1}{13436928 (4\pi)^6}\bigg[2((41014512 \zeta_3+816480 \zeta_4-55987200 \zeta_5+7390987) d_c \nonum \\ 
& + 32 (3750624 \zeta_3+10206 \zeta_4-6006960 \zeta_5+1500773)) b_r^2 \mu_r \nonum \\ 
& + 4((216345168 \zeta_3+583200 \zeta_4-324725760 \zeta_5+75106109) d_c+2(82306368 \zeta_3 \nonum \\ 
& + 332424 \zeta_4-122472000 \zeta_5+27546817)) b_r \mu_r^2-((3 (54000 \zeta_3+52087) d_c-4774032 \zeta_3 \nonum \\ 
& + 139968 \zeta_4+11876687) d_c-12(2260872 \zeta_3-20736 \zeta_4-1658880 \zeta_5-2141599)) b_r^3 \nonum \\ 
& - 8((3(2160 \zeta_3+7681) d_c-243603504 \zeta_3+373248 \zeta_4+369515520 \zeta_5-90673907) d_c \nonum \\
& - 2 (103991040 \zeta_3+989496 \zeta_4-164384640 \zeta_5+44356427)) \mu_r^3 \nonum \\ 
& - 12\Big(6(2592(15 d_c+52) \zeta_3-108355 d_c-452468) b_r^2 \mu_r+12(1944(16 d_c+79) \zeta_3 \nonum \\
& - 67768 d_c-483607) b_r \mu_r^2+(2592(51 d_c+164) \zeta_3-52065 d_c^2-780274 d_c-1663086) b_r^3 \nonum \\ 
& - 8((2925 d_c+31104 \zeta_3+12997) d_c-796392 \zeta_3+1183509) \mu_r^3\Big) L_p \nonum \\
& - 540\Big(30(91 d_c+632) b_r^2 \mu_r+12(163 d_c+820) b_r \mu_r^2+5(99 d_c^2+813 d_c+1532) b_r^3 \nonum \\ 
& + 8(3(9 d_c+91) d_c+640) \mu_r^3\Big) L_p^2 +3240\Big(50(5 d_c+18) b_r^2 \mu_r+100(d_c+18) b_r \mu_r^2 \nonum \\ 
& + 25(d_c+2)(d_c+3) b_r^3+8(d_c+10)(d_c+15) \mu_r^3 \Big) L_p^3 \bigg].
\ea

\vspace{-0.5cm}

\subsubsection{Three-loop vertex polarization}

\begin{figure}[htp]
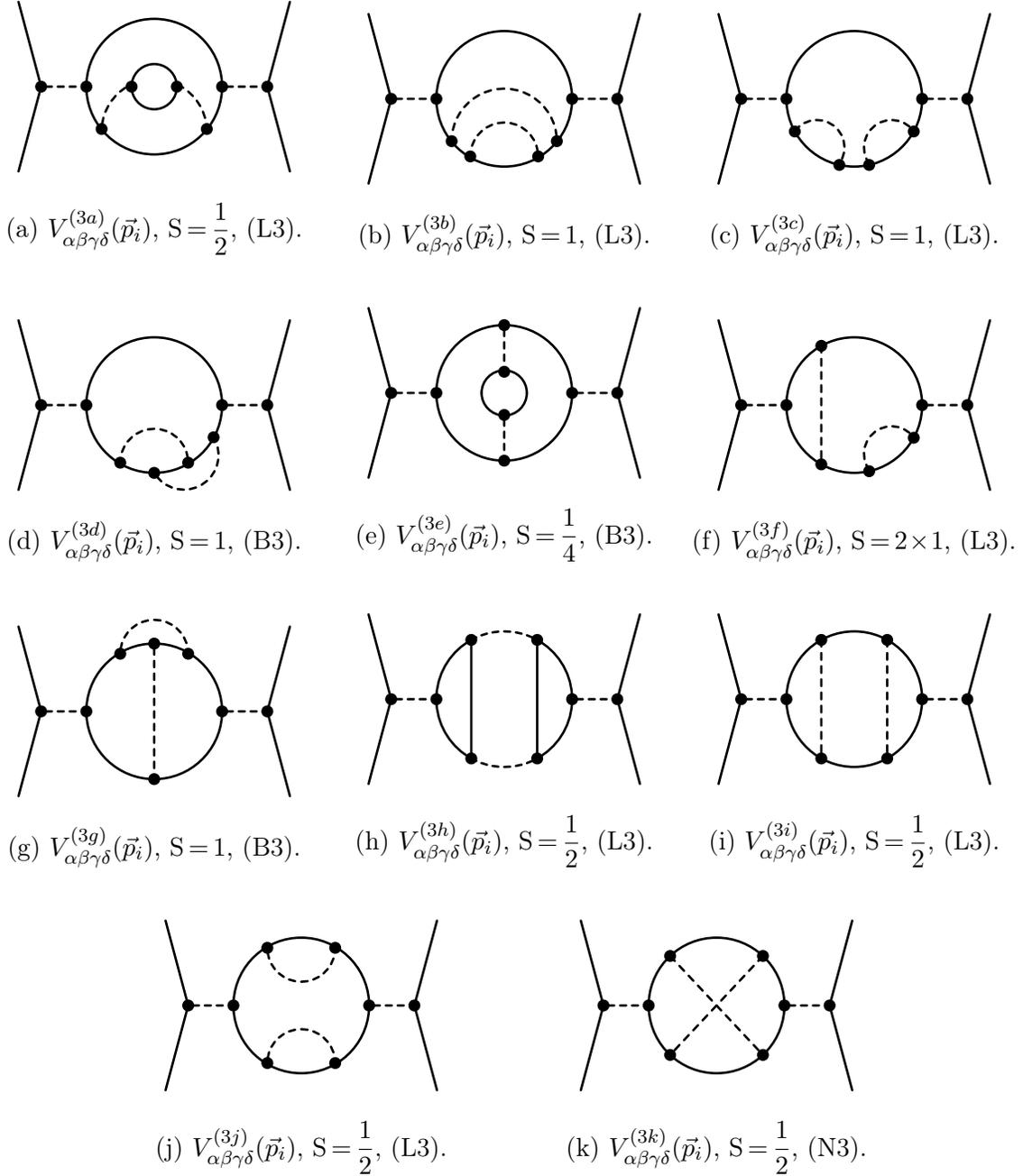

\centering
\begin{subfigure}{.32\textwidth}
\centering
\GraphVThreeA
\caption{$V_{\alpha\beta\gamma\delta}^{(3a)}(\vec p_i)$,~$\text{S}=\dfrac{1}{2}$,~(L3).}
\end{subfigure}
\begin{subfigure}{.32\textwidth}
\centering
\GraphVThreeB 
\caption{$V_{\alpha\beta\gamma\delta}^{(3b)}(\vec p_i)$,~$\text{S}=1$,~(L3).}
\end{subfigure}
\begin{subfigure}{.32\textwidth}
\centering
\GraphVThreeC 
\caption{$V_{\alpha\beta\gamma\delta}^{(3c)}(\vec p_i)$,~$\text{S}=1$,~(L3).}
\end{subfigure}
\vspace*{-0.5cm}\\
\begin{subfigure}{.32\textwidth}
\centering
\GraphVThreeD
\caption{$V_{\alpha\beta\gamma\delta}^{(3d)}(\vec p_i)$,~$\text{S}=1$,~(B3).}
\end{subfigure}
\begin{subfigure}{.32\textwidth}
\centering
\GraphVThreeE 
\caption{$V_{\alpha\beta\gamma\delta}^{(3e)}(\vec p_i)$,~$\text{S}=\dfrac{1}{4}$,~(B3).}
\end{subfigure}
\begin{subfigure}{.32\textwidth}
\centering
\GraphVThreeF
\caption{$V_{\alpha\beta\gamma\delta}^{(3f)}(\vec p_i)$, $\text{S}=2\!\times\!1$, (L3).}
\end{subfigure}
\vspace*{-0.5cm}\\
\begin{subfigure}{.32\textwidth}
\centering
\GraphVThreeG
\caption{$V_{\alpha\beta\gamma\delta}^{(3g)}(\vec p_i)$,~$\text{S}=1$,~(B3).}
\end{subfigure}
\begin{subfigure}{.32\textwidth}
\centering
\GraphVThreeH
\caption{$V_{\alpha\beta\gamma\delta}^{(3h)}(\vec p_i)$,~$\text{S}=\dfrac{1}{2}$,~(L3).}
\end{subfigure}
\begin{subfigure}{.32\textwidth}
\centering
\GraphVThreeI
\caption{$V_{\alpha\beta\gamma\delta}^{(3i)}(\vec p_i)$,~$\text{S}=\dfrac{1}{2}$,~(L3).}
\end{subfigure}
\vspace*{-0.5cm}\\ 
\centering
\begin{subfigure}{.5\textwidth}
\centering
\GraphVThreeJ
\caption{$V_{\alpha\beta\gamma\delta}^{(3j)}(\vec p_i)$,~$\text{S}=\dfrac{1}{2}$,~(L3).}
\end{subfigure}
\hspace{-2cm}
\begin{subfigure}{.5\textwidth}
\centering
\GraphVThreeK
\caption{$V_{\alpha\beta\gamma\delta}^{(3k)}(\vec p_i)$,~$\text{S}=\dfrac{1}{2}$,~(N3).}
\end{subfigure}
\caption{Three-loop vertex self-energy diagrams and their associated symmetry factors (S). All momentum parametrizations and indices have been dropped to keep it light. All external momenta $\vec p_i=\{\vec p_1, \vec p_2, \vec p_3, \vec p_4\}$ are defined incoming.} 
\label{mem:fig:three-loop:V}
\end{figure}

We now consider the three-loop vertex self-energy that consists of $11$ independent diagrams, all displayed in figure \ref{mem:fig:three-loop:V} and labeled in alphabetical order from $a$ to $k$. There are 7 diagrams ($a,b,c,f,h,i,j$) of the Ladder (L3) topology, 3 diagrams ($d,e,g$) of the Benz (B3) topology and one diagram ($k$) of the Non-planar (N3) topology. All of them are displayed with their corresponding symmetry factor (S) that are either $1$, $1/2$ or $1/4$. Moreover, by symmetry, one diagram ($f$) should be taken into account twice. We therefore add an explicit factor $2$ to its symmetry factor (S).

Proceeding along the lines of the one- and two-loop cases, \ie, performing carefully the projections, contractions, reductions, integrations and expansions of the 11 diagrams as well as summing all of them together yields%
\bs
\ba
\tilde{\Pi}^{(3)}_{M}(p^2) & = \tilde{\Pi}^{(3a)}_{M}(p^2)+\tilde{\Pi}^{(3b)}_{M}(p^2)+\dots+\tilde{\Pi}^{(3k)}_{M}(p^2) \nonum \\
& =- \frac{5 \mu d_c e^{-3 \eps L_p}}{5184 (4\pi M^{\eps})^6}\bigg[
\frac{1}{2 \eps^3}\Big(5(d_c+8) b^2+160 b \mu +4(d_c+40) \mu^2\Big) \nonum \\
& +\frac{1}{45 \eps^2}\Big(5(82 d_c+739) b^2+9830 b \mu+2(137 d_c+3250) \mu^2\Big) \nonum \\ 
& -\frac{1}{8100 \eps}\Big(\left(5 d_c \left(6075 \zeta_2+5832 \zeta_3-89497 \right)+30 \left(8100 \zeta_2+15120 \zeta_3-124853
\right)\right)b^2 \nonum \\
& + \left(4 d_c \left(6075 \zeta_2-81648 \zeta_3+19591\right)+360 \left(2700 \zeta_2-37404 \zeta_3+21511\right)\right)\mu ^2 \nonum\\
& + \left(240 \left(4050 \zeta_2-25731 \zeta_3-10517\right)-39200 d_c\right)b \mu \Big)+\Ord(\eps^0)\bigg], \\
\tilde{\Pi}^{(3)}_{N}(p^2) & = \tilde{\Pi}^{(3a)}_{N}(p^2)+\tilde{\Pi}^{(3b)}_{N}(p^2)+\dots+\tilde{\Pi}^{(3k)}_{N}(p^2) \nonum \\ 
& =-\frac{5 b d_c e^{-3 \eps L_p}}{5184 (4\pi M^{\eps})^6}\bigg[
\frac{5}{4 \eps^3}\Big(5(d_c+8) b^2+160 b \mu+4(d_c+40) \mu^2 \Big) \nonum \\
& +\frac{1}{36 \eps^2}\Big(5(209 d_c+1838) b^2+26860 b \mu+8(91 d_c+2525) \mu^2\Big) \nonum \\ 
& - \frac{1}{3240 \eps}\Big( \left(5 d_c \left(6075 \zeta_2+23328 \zeta_3-131779
\right)+6 \left(40500 \zeta_2+165024 \zeta_3-827623\right)\right)b^2\nonum\\
& + \left(4 d_c \left(6075 \zeta_2+40824 \zeta_3-118343\right)+288 \left(3375 \zeta_2+21204 \zeta_3-42083
\right)\right)\mu ^2 \nonum\\
& + \left(120 \left(8100 \zeta_2+44280 \zeta_3-126979\right)-3920 d_c\right)b \mu \Big)+\Ord(\eps^0)\bigg],
\ea
\es
where, once again, the $L_p$ dependency has been factorized out for the sake of simplicity. 

From all the previous results, we are now in a position to first compute the three-loop contribution to the intermediate renormalization constants
\bs
\ba
\delta Z_{\Gamma\mu}^{(3)} & = -\KK{\tilde{\Pi}_M^{(3)}} +\KK{\delta Z_{\Gamma_\mu}^{(1)} \tilde{\Pi}_M^{(2)}} +\KK{\delta Z_{\Gamma_\mu}^{(2)} \tilde{\Pi}_M^{(1)}} - \KK{{\delta Z_{\Gamma_\mu}^{(1)}}^2 \tilde{\Pi}_M^{(1)}} +2 \KK{\delta Z_{\Gamma_\mu}^{(1)} \delta Z_{\Gamma_\mu}^{(2)}} - \KK{{\delta Z_{\Gamma_\mu}^{(1)}}^3} \nonum \\
& = \frac{d_c \mu}{10368 (4\pi)^6}\bigg[\frac{1}{\eps^3}\Big(
20(3 d_c+40) b_r \mu_r+25(d_c+8) b_r^2+2(d_c+10)(3 d_c+40) \mu_r^2 \Big) \nonum \\
& +\frac{1}{36 \eps^2}\Big(4(321 d_c+9040) b_r \mu_r+5(167 d_c+8) b_r^2+20 (313 d_c+2312) \mu_r^2\Big) \nonum \\ 
& +\frac{1}{648 \eps}\Big(32(980 d_c+2754 \zeta_3+3801) b_r \mu_r+(34987 d_c-384(459 \zeta_3-296)) b_r^2 \nonum \\
& +4(5317 d_c+80352 \zeta_3+93600)\Big) \mu_r^2+\Ord(\eps^0)\bigg] , \\
\delta Z_{\Gamma_b}^{(3)} & = -\KK{\tilde{\Pi}_N^{(3)}} +\KK{\delta Z_{\Gamma_b}^{(1)} \tilde{\Pi}_N^{(2)}} +\KK{\delta Z_{\Gamma_b}^{(2)} \tilde{\Pi}_N^{(1)}} - \KK{{\delta Z_{\Gamma_b}^{(1)}}^2 \tilde{\Pi}_N^{(1)}} +2 \KK{\delta Z_{\Gamma_b}^{(1)} \delta Z_{\Gamma_b}^{(2)}} - \KK{{\delta Z_{\Gamma_b}^{(1)}}^3} \nonum \\
& = \frac{d_c b_r}{41472 (4\pi)^6}\bigg[\frac{25}{\eps^3}\Big(
40(3 d_c+8) b_r \mu_r+5(d_c+2)(3 d_c+8) b_r^2+8(d_c+40) \mu_r^2 \Big) \nonum \\
& +\frac{25}{18 \eps^2}\Big(4(267 d_c+224) b_r \mu_r-(577 d_c+1576) b_r^2-4(71 d_c-728) \mu_r^2\Big) \nonum \\ 
& +\frac{1}{324 \eps}\Big(2240(7 d_c-6 (54 \zeta_3+5)) b_r \mu_r+(371495 d_c-228096 \zeta_3+614832) b_r^2 \nonum \\ 
& + 4(87893 d_c+425088 \zeta_3-248616) \mu_r^2\Big)+\Ord(\eps^0)\bigg]\, ,
\ea
\es
and then extract the three-loop contribution to the renormalization constant of the couplings that read
\bs
\ba
\delta Z_\mu^{(3)} & = \delta Z_{\Gamma_\mu}^{(3)}-2 \delta Z^{(3)} -2 \delta Z^{(1)} \delta Z_{\Gamma_\mu}^{(2)}-2 \delta Z^{(2)} \delta Z_{\Gamma_\mu}^{(1)} + 3 {\delta Z^{(1)}}^2 \delta Z_{\Gamma_\mu}^{(1)}+6 \delta Z^{(1)} \delta Z^{(2)} - 4 {\delta Z^{(1)}}^3 \nonum \\
& = \frac{1}{10368 (4\pi)^6}\bigg[\frac{1}{\eps^3}\Big(
100(d_c+10)(d_c+36) b_r^2 \mu_r+40((3 d_c+145) d_c+1800) b_r \mu_r^2 \nonum \\ & + 125(d_c+6)(d_c+8) b_r^3+6(d_c+20)^3 \mu_r^3 \Big) \nonum \\
& +\frac{1}{36 \eps^2}\Big(5((257 d_c+8940) d_c-26880) b_r^2 \mu_r+12((107 d_c+3260) d_c+28000) b_r \mu_r^2 \nonum \\ 
& + 50((15 d_c-184) d_c-2968) b_r^3+140(d_c+20)(13 d_c+8) \mu_r^3 \Big) \nonum \\ 
& -\frac{1}{648 \eps}\Big((20736(61 d_c+21) \zeta_3-34987 d_c^2-791004 d_c+2654448) b_r^2 \mu_r \nonum \\ 
& + 8(648(283 d_c+342) \zeta_3-3920 d_c^2-263247 d_c-326922) b_r \mu_r^2 \nonum \\ 
& - 2(5184(9 d_c+16) \zeta_3-41625 d_c^2+180563 d_c+516252) b_r^3 \nonum \\ 
& + 4((263 d_c-578016 \zeta_3+660820) d_c+2638656 \zeta_3-2609592) \mu_r^3\Big)+\Ord(\eps^0)\bigg], \\
\delta Z_b^{(3)} & = \delta Z_{\Gamma_b}^{(3)}-2 \delta Z^{(3)} -2 \delta Z^{(1)} \delta Z_{\Gamma_b}^{(2)}-2 \delta Z^{(2)} \delta Z_{\Gamma_b}^{(1)} + 3 {\delta Z^{(1)}}^2 \delta Z_{\Gamma_b}^{(1)}+6 \delta Z^{(1)} \delta Z^{(2)} - 4 {\delta Z^{(1)}}^3 \nonum \\
& = \frac{1}{41472 (4\pi)^6}\bigg[\frac{5}{\eps^3}\Big(160(d_c (d_c+62)+360) b_r \mu_r^2+400((3 d_c+29) d_c+72) b_r^2 \mu_r \nonum \\ 
& + 75(d_c+4)^3 b_r^3+32(d_c+30)(d_c+40) \mu_r^3\Big) \nonum \\
& +\frac{1}{18 \eps^2}\Big(300((89 d_c+484) d_c-896) b_r^2 \mu_r+20((3972-1021 d_c) d_c+33600) b_r \mu_r^2 \nonum \\ 
& - 175(d_c+4)(61 d_c+424) b_r^3-32((111 d_c+4880) d_c-1400) \mu_r^3\Big) \nonum \\ 
& -\frac{1}{324 \eps}\Big(8(36288(10 d_c+3) \zeta_3-1960 d_c^2-160935 d_c+663612) b_r^2 \mu_r \nonum \\ 
& + 4(5184(68 d_c+171) \zeta_3-87893 d_c^2-743556 d_c-1307688) b_r \mu_r^2 \nonum \\
& + (41472(d_c-8) \zeta_3-204995 d_c^2-1337084 d_c-2065008) b_r^3 \nonum \\ 
& + 32 ((1395 d_c-124416 \zeta_3+188605) d_c+659664 \zeta_3-652398) \mu_r^3\Big) +\Ord(\eps^0)\bigg].
\ea
\es
All the above results also allow us to compute the renormalized three-loop contributions to the $R$-propagator polarization projections that we add here for completeness: 
\bs
\ba
\tilde{\Pi}^{(3)}_{M,r}(p^2) & = \tilde{\Pi}_M^{(3)} + \delta Z_{\Gamma_\mu}^{(3)} - \delta Z_{\Gamma_\mu}^{(1)} \tilde{\Pi}_M^{(2)}+\left({\delta Z_{\Gamma_\mu}^{(1)}}^2-\delta Z_{\Gamma_\mu}^{(2)}\right) \tilde{\Pi}_M^{(1)} -2 \delta Z_{\Gamma_\mu}^{(1)} \delta Z_{\Gamma_\mu}^{(2)}+{\delta Z_{\Gamma_\mu}^{(1)}}^3 \nonum \\
& = \frac{\mu_r d_c}{67184640 (4\pi)^6}\bigg[8(78783408 \zeta_3 d_c-77760(1545 d_c+812) \zeta_5+29826539 d_c\nonum \\ 
& + 28920240 \zeta_3-165240 \zeta_4+30851634) b_r \mu_r+(28429488 \zeta_3 d_c\nonum \\
& - 622080(63 d_c+146) \zeta_5+ 4895207 d_c+77296032 \zeta_3+2643840 \zeta_4-11407688) b_r^2\nonum \\ 
& + 4(9 (43511376 \zeta_3-66614400 \zeta_5+16738993) d_c+8(167805432 \zeta_3-150660 \zeta_4\nonum \\
& - 275484240 \zeta_5+84287719)) \mu_r^2 + 12 \bigg(80(980 d_c+53298 \zeta_3-29007) b_r \mu_r \nonum \\
& - 5(1296(3 d_c+92) \zeta_3-43877 d_c-367500) b_r^2\nonum \\
& + 4(1296(42 d_c+1835) \zeta_3-12749 d_c-2145660) \mu_r^2 \bigg) L_p \nonum \\
& - 2700 \bigg((163 d_c+1968) b_r^2+2832 b_r \mu_r+4(47 d_c+96) \mu_r^2\bigg) L_p^2 \nonum \\
& + 32400 \bigg(5(d_c+8) b_r^2+160 b_r \mu_r+4 (d_c+40) \mu_r^2\bigg) L_p^3\bigg] , \\
\tilde{\Pi}^{(3)}_{N,r}(p^2) & = \tilde{\Pi}_N^{(3)} + \delta Z_{\Gamma_b}^{(3)} - \delta Z_{\Gamma_b}^{(1)} \tilde{\Pi}_N^{(2)}+\left({\delta Z_{\Gamma_b}^{(1)}}^2-\delta Z_{\Gamma_b}^{(2)}\right) \tilde{\Pi}_N^{(1)} -2 \delta Z_{\Gamma_b}^{(1)} \delta Z_{\Gamma_b}^{(2)}+{\delta Z_{\Gamma_b}^{(1)}}^3 \nonum \\
& = \frac{b_r d_c}{26873856 (4\pi)^6}\bigg[64(1235520 \zeta_3 d_c-77760(24 d_c+49)\zeta_5+450089 d_c+3847500 \zeta_3\nonum \\ 
& +34020 \zeta_4-933261) b_r \mu_r+(3(930672 \zeta_3+622080 \zeta_5-2835595) d_c+ 8(4245372 \zeta_3\nonum \\ 
& +85536 \zeta_4-2488320 \zeta_5-5320957))b_r^2+4((144862128 \zeta_3-219749760 \zeta_5\nonum \\
& + 53345485) d_c+8 (22129200 \zeta_3-159408 \zeta_4-34972560 \zeta_5+8943373)) \mu_r^2 \nonum \\ 
& + 12\bigg(160(49 d_c-23652 \zeta_3+42723) b_r \mu_r-(5184(15 d_c+142) \zeta_3 - 455545 d_c\nonum \\ 
& - 2844276) b_r^2-4(1296(21 d_c+676) \zeta_3-101047 d_c-1142832) \mu_r^2\bigg) L_p \nonum \\ 
& - 540 \bigg(5(289 d_c+2976) b_r^2+34320 b_r \mu_r+4(361 d_c+5520) \mu_r^2\bigg) L_p^2 \nonum \\ 
& + 32400 \bigg(5(d_c+8) b_r^2+160 b_r \mu_r+4(d_c+40) \mu_r^2\bigg) L_p^3 \bigg].
\ea
\es

\subsection{Four-loop calculations}
\label{mem:subsec:four-loop}

The four-loop flexuron self-energy $\tilde\Sigma^{(4)}$ consists in 155 diagrams and the vertex self-energy in 91 diagrams for both $\tilde\Pi_M^{(4)}$ and $\tilde\Pi_N^{(4)}$. This makes a total of $155+2\times91=389$ diagrams to compute, out of which only 276 are independent thanks to 113 topological relations. Each diagram belongs to one of the 11 topologies of four-loop massless 2-point Feynman integrals, including 6 planar and 5 non-planar topologies, see \ref{chap:appendix} for more details. The symmetry factors for all diagrams are from the set $S=\{1/8, 1/4, 1/2, 1\}$. 

Following the same procedure as for lower loop orders, one generates all 389 diagram expressions and perform all projections and contractions over space and co-space indices. At this stage, more than $57$ million multi-loop integrals need to be computed. These can be analytically evaluated using the same IBP technique as for lower orders but require the use of much faster automated algebra codes such as \textsc{Fire} \cite{Smirnov:2008iw,Smirnov:2013dia,Smirnov:2014hma,Smirnov:2019qkx} as well as a homemade optimized version of \textsc{LiteRed} \cite{Lee:2012cn,Lee:2013mka}. Most of the numerator contractions and IBP computations require the use of a supercomputer. 
All integrals are then reduced to linear combinations of $39$ known master integrals, see \ref{chap:appendix}, hence, after simplification of the expressions, finishing the computation of all the diagrams.

One can then use the 113 topological relations between diagrams to check the results, extract the renormalization constants $\delta Z^{(4)}$, $\delta Z_\mu^{(4)}$ and $\delta Z_b^{(4)}$, allowing the derivation of the renormalized self-energies $\tilde\Sigma_r^{(4)}$, $\tilde\Pi_{M,r}^{(4)}$ and $\tilde\Pi_{N,r}^{(4)}$, the beta functions $\beta_\mu$, $\beta_b$ and finally, the four-loop anomalous elasticity $\eta^{(4)}$. 
Since not a single four-loop expression would fit in one page, and that the computation essentially follows the three-loop method, we shall not give further details about the four-loop computations. In the rest of this review, we display most of the intermediary results at three-loop only and restore the four-loop contribution in the last stages of the analysis. 


\section[RG and fixed point]{renormalization-group functions and fixed points}
\label{sec:RGflows+FP}

In the previous section, we computed explicitly the three-loop renormalization constants of the flexuron field and the couplings within the flexural effective model. In this section, we use these results in order to derive the renormalization-group functions of the flexural model and analyse its fixed point structure. Once again, the three-loop contributions will be made explicit in intermediary formulas but, due to their length, the four-loop results will only be included in the final equations.



\subsection{Beta functions and anomalous dimension}


The beta functions are obtained by solving the system \eqref{mem:ren-functions:def+:betas}. Up to four-loop order, they yields:
\bs
\label{eq:mem:betafuncs}
\ba
\beta_\mu & =-2 \mu_r \eps +2 \mu_r \eta+\frac{d_c\mu_r^2}{6(4\pi)^2} + \frac{d_c\mu_r^2 (574\mu_r+107b_r)}{2^{4}3^{4}(4\pi)^4} +\frac{d_c\mu_r^2}{2^{9}3^{7}(4\pi)^6}\bigg[\mu_r^2 (52 (409d_c+7200) \nonum \\
& +321408 \zeta_3) +\mu_rb_r (224 (140 d_c+543)+88128 \zeta_3) + b_r^2 (34987 d_c-176256 \zeta_3+113664) \bigg] \nonum \\ &
+ \frac{d_c \mu_r^2}{2^{12}3^{10}(4\pi)^8} \bigg[\mu_r^3 \big(8 (29169 d_c^2 +4970049980 d_c+10779226092) -6912(135 d_c^2 \nonum \\ & 
-15314905 d_c-30857183) \zeta_3 -186624 (65 d_c-1439) \zeta_4 -622080 (258828 d_c+531313) \zeta_5\big) \nonum \\ &
+\mu_r^2 b_r \big(24 (224 d_c^2+685931199 d_c+657009061)+5184 (8443323 d_c+7107725) \zeta_3 \nonum \\ &
+46656 (317 d_c+3222) \zeta_4-933120 (71352 d_c+62111) \zeta_5\big) +\mu_r b_r^2 \big(12 (48160 d_c^2 \nonum \\ & 
+143406743 d_c+307944968)+20736 (222020 d_c+404643) \zeta_3 +23328 (505 d_c-2892) \zeta_4 \nonum \\ 
& -16796160 (416 d_c+795) \zeta_5\big) +b_r^3 \big(3059319 d_c^2+60235892 d_c+271405878-432 (6750 d_c^2 \nonum \\
& -273983 d_c-1506511) \zeta_3 -186624 (47 d_c+263) \zeta_4-311040 (582 d_c+3031)\zeta_5\big) 
\bigg] +\Ord\Big(\!(\mu_r+b_r)^6\!\Big) \,,
\\
\beta_b & = -2 b_r \eps +2 b_r \eta +\frac{5 d_c b_r^2}{12(4\pi)^2} + \frac{5 d_c b_r^2 (178\mu_r-91 b_r)}{2^{5}3^{4}(4\pi)^4} +\frac{d_cb_r^2}{2^{10}3^{7}(4\pi)^6} \bigg[ \mu_r^2 (4 (87893 d_c-248616) \nonum \\ & 
+1700352 \zeta_3) +\mu_rb_r (2240 (7 d_c-30)-725760 \zeta_3) +b_r^2 (371495 d_c-228096 \zeta_3+614832) \bigg] \nonum \\ &
+\frac{d_c b_r^2}{2^{13}3^{10}(4\pi)^8} \bigg[ 
\mu_r^3 \big(8 (34413 d_c^2 +3552053866 d_c+6505946424) -3456 (1350 d_c^2 \nonum \\ & 
-22155283 d_c-34097318) \zeta_3 -186624 (316 d_c-7465) \zeta_4-1244160 (93111 d_c+150844) \zeta_5\big) \nonum \\ &
+\mu_r^2 b_r \big(48 (1036 d_c^2+206496285 d_c+158637586)+5184 (5188449 d_c+3003346) \zeta_3 \nonum \\ & 
+466560 (136 d_c+315) \zeta_4-1866240 (21915 d_c+13712) \zeta_5\big) +\mu_r b_r^2 \big(12 (33880 d_c^2 \nonum \\ 
& +102216310 d_c+441948121)+2592 (1149311 d_c+3758781) \zeta_3 +933120 (35 d_c-321) \zeta_4 \nonum \\ & 
-16796160 (262 d_c+933) \zeta_5\big) +b_r^3 \big(23058315 d_c^2+31986562 d_c+123308526-432 (33750 d_c^2 \nonum \\ & 
-32533 d_c-1226795) \zeta_3 -466560 (31 d_c+148) \zeta_4-155520 (1446 d_c+5839) \zeta_5\big) 
\bigg] +\Ord\Big(\!(\mu_r+b_r)^6\!\Big)\,,
\ea
\es
where $\eta$ is the anomalous dimension of the flexuron field (\ref{mem:ren-functions:def+:gamma}). The latter reads, up to four loops:
\ba
\eta & =\frac{5 (2 \mu_r+b_r)}{6(4\pi)^2} + \frac{-4\mu_r^2(111 d_c-20) +1160 \mu_r b_r+5 b_r^2 (15 d_c-212)}{2^{5}3^{4}(4\pi)^4} \nonum \\
& +\frac{1}{2^{9}3^{7}(4\pi)^6} \bigg[ 
\mu_r^3 \big(-8 (1395 d_c^2+188605 d_c-652398) + 10368 (96 d_c-509) \zeta_3\big) \nonum \\
& + \mu_r^2 b_r (12 (82681 d_c+108974)-15552 (50 d_c+57) \zeta_3) + \mu_r b_r^2 (6 (56445 d_c-221204) \nonum \\
& -108864 (5 d_c+2) \zeta_3) + b_r^3 (-41625 d_c^2+180563 d_c+516252+5184 (9 d_c+16) \zeta_3)
\bigg] \nonum \\
& +\frac{1}{2^{13}3^{10}(4\pi)^8} 
\bigg[ 
\mu_r^4 \big(
-16 (20763 d_c^3-445985328 d_c^2-15370535368 d_c-11680556284) \nonum \\ & 
+6912 (135 d_c^3+2722314 d_c^2+90983931 d_c+66318202) \zeta_3 +186624 (288 d_c^2+2003 d_c \nonum \\ &
-20360) \zeta_4 -1244160 (23130 d_c^2+777029 d_c+569386) \zeta_5 \big) +\mu_r^3 b_r \big(16 (178219467 d_c^2 \nonum \\ & 
+5548743275 d_c+8561040707) +3456 (2196402 d_c^2+67864998 d_c+88562149) \zeta_3 \nonum \\ & 
-373248 (75 d_c^2+668 d_c+6800) \zeta_4 -622080 (18540 d_c^2+574178 d_c+778045) \zeta_5 \big) \nonum \\ & 
+ \mu_r^2 b_r^2\big( 24 (141866953 d_c^2+1890278050 d_c+1106116087) +5184 (1737678 d_c^2 \nonum \\ &
+20174394 d_c+15225185) \zeta_3 -139968 (320 d_c^2+5433 d_c+3400) \zeta_4  -933120 (14634 d_c^2 \nonum \\ & 
+176815 d_c+126437) \zeta_5 \big) +\mu_r b_r^3 \big(20 (29547339 d_c^2+94268974 d_c+437788822) \nonum \\ &
+1728 (634335 d_c^2+4251180 d_c+9336377) \zeta_3 -2332800 (21 d_c^2+64 d_c+8) \zeta_4 \nonum \\ &
-311040 (5760 d_c^2+30430 d_c+82109) \zeta_5 \big)+b_r^4 \big( -8083125 d_c^3+26884554 d_c^2\nonum \\ & 
-71433074 d_c+126576784+432 (16875 d_c^3-16896 d_c^2+627483 d_c+1500340) \zeta_3 \nonum \\ & 
+233280 (27 d_c^2+164 d_c+128) \zeta_4 +311040 (90 d_c^2-1991 d_c-4993)\zeta_5 \big) \bigg] +\Ord\Big(\!(\mu_r+b_r)^5\!\Big)\,.
\ea
These are the main results of this review. For convenience, they are also available in computer readable files as ancillary files to the arXiv version of the letter \cite{Metayer:2024}\footnote{Note that the $4\pi$ conventions are different in \cite{Metayer:2024} and the corresponding ancillary file, as all $4\pi$ factors are absorbed in the definitions of $\mu_r$ and $b_r$, however leading in the end to the same results for $\eta$.}. We now proceed on analyzing them. For simplicity, part of the the following analysis will still be displayed at three-loop order only.

\subsection{Fixed points}

From the $\beta$-functions, we now compute the fixed points by solving the system $\beta_x(\mu^*,b^*)=0$, $x=\mu,b$. This is done perturbatively, by taking an ansatz of the form $\mu=\mu^{(1)}\eps+\mu^{(2)}\eps^2+\mu^{(3)}\eps^3+\cdots$ and similarly for $b$, and determining the coefficients of the epsilon-series order by order. Up to (and including) four loops, we find 4 fixed points as advertised in the discussion below equation \eqref{eq:mem:betasys}. 

\subsubsection{Gaussian fixed point P\titlemath{_1}}

The Gaussian fixed point P$_1$ is characterized up to four loops by
\ba
\text{P}_1 : \quad 
&{\begin{rcases}
\mu_1^* = 0+ \Ord(\eps^5) ~~
\label{mem:FP:3l:P1} \\
b_1^*=0+ \Ord(\eps^5) ~~
\end{rcases}
\text{~~(Gaussian)}\, .}
\ea
With vanishing couplings, this fixed point is trivial. It corresponds to a free membrane without any elastic interactions and therefore a vanishing flexuron anomalous dimension
\ba
&\eta(\text{P}_1) = 0+ \Ord(\eps^5)\, .
\label{eq:mem:etap1}
\ea
The Gaussian fixed point is twice unstable in the RG flow sense, \ie, the eigenvalues of the stability matrix \eqref{eq:mem:stabilitymatrix} are both negative, implying that this fixed point is repulsive in all directions as the renormalization flow goes to lower energies.

\subsubsection{Shearless fixed point P\titlemath{'_2}}

The shearless fixed point P$'_2$,\footnote{See footnote \ref{footnote:P'2}} is characterized by 
%
\bs
\ba
\text{P}'_2 : \quad 
& \mu_2^* = 0 + \Ord(\eps^4) \text{~~(Shearless)},
\label{mem:FP:3l:P2} \\ 
&b_2^*= (4\pi)^2 \bigg[\frac{24 \eps}{5 (d_c+4)}+\bigg(\frac{96}{5 (d_c+4)^3}+\frac{488}{75 (d_c+4)^2}\bigg) \eps^2 + \bigg(\frac{768}{5 (d_c+4)^5}\\
& \hspace{2cm} 
-\frac{32 (10368 \zeta_3-18371)}{5625 (d_c+4)^4} 
+ \frac{8 (3456 \zeta_3+37643)}{5625 (d_c+4)^3}
-\frac{81998}{3375(d_c+4)^2} \bigg)\eps^3+ \Ord(\eps^4)\bigg] \nonum , 
\ea
\es
with vanishing shear modulus $\mu^*_2$ (including at four-loops) and a non-trivial value for $b^*_2$ (the four-loop contribution of which is too lengthy to be displayed). The absence of shear is a characteristic property of fluid membranes, although a dynamical connectivity would also be needed, which is not the case here \cite{Guitter:1988}. According to \cite{LeDoussal:2018}, such a shearless phase may also be realized in nematic elastomer membranes, see \cite{Xing:2003,Xing:2003b,Stenull:2003,Stenull:2004,Xing:2005,Xing:2008,Xing:2008b}. The two-loop order correction to this fixed point has first been computed by Mauri and Katsnelson \cite{Mauri:2020}. 
As for the eigenvalues of the stability matrix \eqref{eq:mem:stabilitymatrix}, it reveals that it is unstable in the $\mu_r$ direction and stable in the $b_r$ direction. The corresponding anomalous dimension of the flexuron field is also non-trivial and reads 
\ba
\label{eq:mem:etap2}
&\eta(\text{P}'_2) = \frac{4 \eps}{d_c + 4}+\bigg(
\frac{16}{(d_c+4)^3}
-\frac{20}{3 (d_c+4)^2}
+\frac{2}{3 (d_c+4)} \bigg) \eps^2
+ \bigg(
\frac{128}{(d_c+4)^5} \\
& \hspace{0.5cm}-\frac{16 (10368 \zeta_3+2029)}{3375(d_c+4)^4}
-\frac{4 (5184 \zeta_3+58177)}{3375 (d_c+4)^3}
+ \frac{4 (3888 \zeta_3+27239)}{3375 (d_c+4)^2}
-\frac{37}{9 (d_c+4)} \bigg)\eps^3+ \Ord(\eps^4)\,. 
\nonum 
\ea
This result displays an interesting structure in the perturbative series with denominators in powers of $1/(4+d_c)$. In the physical case $d_c=1$, including explicitly the four-loop contribution, it simplifies as
\ba
\eta(\text{P}'_2) & = \frac{4 \eps}{5} - \frac{2 \eps^2}{375} + \frac{(119232 \zeta_3-120079) \eps^3}{2109375} \\
& + \frac{(-51994931+7803552 \zeta_3+26827200 \zeta_4+13512960 \zeta_5) \eps^4}{316406250}+ \Ord(\eps^5)\, ,  \nonum 
\ea
with each term in the series getting divided by an increasing powers of $1/5$. Numerically, this series evaluates to
\be
\eta(\text{P}'_2) = 0.8000 \eps- 0.005333 \eps^2 + 0.01102 \eps^3+ 0.001369\eps^4
+\Ord(\eps^5)\,,
\label{eq:pertseriesp2}
\ee
with surprisingly small coefficients as already noticed in \cite{Coquand:2020a,Metayer:2021kxm}. As discussed in these papers, perturbative series are asymptotic in nature but the case of polymerized membranes is quite peculiar in the sense that various factors (such as increasing powers of $1/(d_c+4)$ in the case of P$'_2$) conspire to numerically reduce the coefficient of the epsilon-series over several orders. The series therefore effectively looks convergent even in the case of interest $\eps=1$. As a matter of fact, note that the the two and four loop contributions 
in \eqref{eq:pertseriesp2} are not only numerically small but also smaller than the one-loop coefficient by two orders of magnitude. We still observe an increase upon going to three loops as the third order coefficient is twice the second order one. Therefore, we do expect the asymptotic nature of the series to manifest at higher orders. But up to four loops the series behave remarkably well and no resummation is needed.\footnote{By this we mean that, up to four loops, resummations of the series \eqref{eq:pertseriesp2} give results close to the raw ones in the limit $\eps=1$. As an example, a simple three-loop Padé approximant yields either $\eta^{[2/1]}(\text{P}'_2)=0.7983$ or $\eta^{[1/2]}(\text{P}'_2)=0.8057$ and the four-loop one $\eta^{[2/2]}=0.8074$, which are all very close if not indistinguishable from the 
non-resummed results of \eqref{eq:P2primeres}. 
} 
Hence, a direct substitution of $\eps=1$ in \eqref{eq:pertseriesp2} yields successively
\begin{empheq}[box=\widefbox]{equation}
\label{eq:P2primeres}
\eta_{\text{1-loop}}(\text{P}'_2)=0.8000, ~~~
\eta_{\text{2-loop}}(\text{P}'_2)=0.7947, ~~~
\eta_{\text{3-loop}}(\text{P}'_2)=0.8057, ~~~
\eta_{\text{4-loop}}(\text{P}'_2)=0.8071.
\end{empheq}
The one-loop result has been first obtained in \cite{Aronovitz:1988}, the two-loop result (32 year later) in \cite{Mauri:2020,Coquand:2020a}, the three-loop result in \cite{Metayer:2021kxm} and the four-loop one in \cite{Metayer:2024}.

\subsubsection{Infinitely compressible fixed point P\titlemath{_3}}

The infinitely compressible fixed point P$_3$ is characterized by 
\bs
\ba
\text{P}_3 : \quad 
&\mu_3^* = (4\pi)^2 \bigg[\frac{12 \eps}{d_c+20}+\bigg(\frac{1680}{(d_c+20)^3}-\frac{260}{3 (d_c+20)^2}\bigg) \eps^2 + \bigg(
\frac{470400}{(d_c+20)^5}
\label{mem:FP:3l:P3} \\ 
& \hspace{1cm}+ \frac{8(591624 \zeta_3-709633)}{9(d_c+20)^4}
- \frac{4(144504\zeta_3-171025)}{27(d_c+20)^3}
+ \frac{263}{27(d_c+20)^2} \bigg)\eps^3+ \Ord(\eps^4)\bigg],
\nonum \\
&b_3^*= 0+ \Ord(\eps^4) \text{~~(Infinitely compressible)} 
\ea
\es
with a non-trivial value for the shear modulus (the four-loop contribution of which is too lengthy to be displayed) but a vanishing bulk modulus (including at four-loops). 
It is once unstable, \ie, attractive in the $\mu_r$ direction but repulsive in the $b_r$ direction (which is exactly the opposite of P$'_2$). 
%
This fixed point leads to a non-trivial value for the field anomalous dimension of the flexuron, reading 
\ba
\label{eq:mem:etap3}
&\eta(\text{P}_3) = \frac{20 \eps}{d_c+20}
+\bigg(
\frac{2800}{(d_c+20)^3}
+\frac{1060}{3 (d_c+20)^2}
-\frac{74}{3 (d_c+20)} \bigg) \eps^2
+ \bigg(
\frac{784000}{(d_c+20)^5} \\ 
& + \frac{40(591624 \zeta_3-615553)}{27(d_c+20)^4}
- \frac{2(1006344 \zeta_3-1024193)}{27(d_c+20)^3}
+ \frac{2(20736 \zeta_3-17105)}{27(d_c+20)^2}
- \frac{155}{9(d_c+20)}\bigg)\eps^3+ \Ord(\eps^4), \nonum
\ea
where the structure in $d_c$ now displays denominators in powers of $d_c+20$. We therefore expect the first terms of the series to be even more convergent than for P$'_2$. Indeed, in the physical case $d_c=1$, including explicitly the four-loop contribution, the coefficients simplify as
\ba
\label{eq:mem:etap3:num}
\eta(\text{P}_3) & = \frac{20 \eps}{21} - \frac{94 \eps^2}{1323} - \frac{(312336 \zeta_3-9011) \eps^3}{5250987}  \\
& - \frac{(14383003505+36705338304 \zeta_3+59031504 \zeta_4-56435313600 \zeta_5) \eps ^4}{661624362} + \Ord(\eps^5), \nonum
\ea
and numerically evaluate to
\be
\eta(\text{P}_3) = 0.9524 \eps- 0.07105 \eps^2 - 0.06978 \eps^3- 0.07475 \eps^4 +\Ord(\eps^5).
\label{eq:pertseriesp3}
\ee
%
%
Remarkably, all the coefficients appearing in (\ref{eq:pertseriesp3}) are small and even decreasing up to three loops. We observe a slight increase of the four-loop correction, which, as for P$_2'$, indicates the asymptotic nature of the series. However, this increase is so small and all corrections are so close to each other that resummations of this series behave badly.
\footnote{Indeed, a resummation of the series \eqref{eq:mem:etap3:num} with a simple $[1/2]$ Padé approximant yields $\eta^{[1/2]}(\text{P}_3)=0.8257$ while the $[2/1]$ case is negative hence unphysical. Similarly, at four loops, the Padé approximant behaves badly with a value larger than one $\eta^{[2/2]}(\text{P}_3)=1.8695$, \ie, mechanically unstable.} Hence, a direct substitution of $\eps=1$ in \eqref{eq:pertseriesp3} yields, successively
\begin{empheq}[box=\widefbox]{equation}
\eta_{\text{1-loop}}(\text{P}_3) = 0.9524, ~~~
\eta_{\text{2-loop}}(\text{P}_3) = 0.8813, ~~~
\eta_{\text{3-loop}}(\text{P}_3) = 0.8115, ~~~
\eta_{\text{4-loop}}(\text{P}_3) = 0.7368.
\end{empheq}
The one-loop result has been first obtained in \cite{Aronovitz:1988}, the two-loop result (32 year later) in \cite{Coquand:2020a} and the three-loop result in \cite{Metayer:2021kxm}. The four-loop result has been first computed in \cite{Pikelner:2021} in the equivalent two-field model, recently confirmed by \cite{Metayer:2024} in the effective flexural model. 

\subsubsection{Flat phase fixed point P\titlemath{_4}}

The most important fixed point is P$_4$ which is characterized by
\bs
\label{eq:mem:fp4coord}
\ba
\text{P}_4 : \quad 
&\mu_4^* = (4\pi)^2 \bigg[\frac{12 \eps}{d_c+24}+\bigg(\frac{1440}{(d_c+24)^3}-\frac{616}{5 (d_c+24)^2}\bigg) \eps^2 
+ \bigg(
\frac{345600}{(d_c+24)^5}
\label{mem:FP:3l:P4} \\ 
& \hspace{0.8cm}
+\frac{96 (576288 \zeta_3-812161)}{125 (d_c+24)^4}
-\frac{144 (12288 \zeta_3-20401)}{125 (d_c+24)^3}
-\frac{8168}{75(d_c+24)^2}
\bigg)\eps^3+ \Ord(\eps^4)\bigg],
\nonum \\
&b_4^*= (4\pi)^2 \bigg[\frac{24 \eps}{5 (d_c+24)}+\bigg(\frac{576}{(d_c+24)^3}+\frac{1936}{25 (d_c+24)^2}\bigg) \eps^2 
+ \bigg(
\frac{138240}{(d_c+24)^5} \\
& \hspace{0.8cm}
+\frac{4416 (25056 \zeta_3-31007)}{625 (d_c+24)^4}
-\frac{48 (71136 \zeta_3-163967)}{625 (d_c+24)^3}
-\frac{77512}{375(d_c+24)^2}
\bigg)\eps^3+ \Ord(\eps^4)\bigg], \nonum
\nonum
\ea
\es
where both couplings are non-vanishing (their four-loop contributions are too lengthy to be displayed). This non-trivial fixed point is fully stable, \ie, attractive in both $\mu_r$ and $b_r$ directions. In that sense, it fully controls the flat phase of the membrane in the long-distance limit. 


At P$_4$, the anomalous dimension of the flexuron reads
\ba
\label{eq:mem:etap4}
&\eta(\text{P}_4) = \frac{24 \eps}{d_c+24}+\bigg(
\frac{2880}{(d_c+24)^3}
+\frac{456}{(d_c+24)^2}
-\frac{24}{(d_c+24)} \bigg) \eps^2
+ \bigg(
\frac{691200}{(d_c+24)^5} \\
& +\frac{576 (192096 \zeta_3-234137)}{125 (d_c+24)^4}
-\frac{8 (923616 \zeta_3-1031777)}{125 (d_c+24)^3}
+\frac{4 (86832 \zeta_3-39029)}{375 (d_c+24)^2}
-\frac{64}{3(d_c+24)}
\bigg)\eps^3+ \Ord(\eps^4), \nonum 
\ea
where the structure in $d_c$ now has denominators in powers of $d_c+24$ that further improves the behavior of the series with respect to the previous fixed points. Indeed, in the case of interest, $d_c=1$, including explicitly the four-loop contribution, the series simplifies to
\ba
\eta(\text{P}_4) & = \frac{24 \eps}{25} - \frac{144 \eps^2}{3125} - \frac{4(1286928 \zeta_3-568241) \eps^3}{146484375} \\
& - \frac{4 (139409079893+355002697944 \zeta_3+723897000 \zeta_4-546469130880 \zeta_5) \eps ^4}{54931640625} + \Ord(\eps^5), \nonum
\ea
and the coefficients numerically evaluate to
\be
\eta(\text{P}_4) = 0.9600 \eps- 0.04608 \eps^2 - 0.02673 \eps^3-0.02017\eps^4+ \Ord(\eps^5)\,.
\label{eq:pertseriesp4}
\ee
From \eqref{eq:pertseriesp4}, we see that all the coefficients of the $\eps$-series are small and successively decreasing. Up to four-loops there is no sign of the asymptotic nature of the series. It may therefore straightforwardly be evaluated in the limit $\eps=1$ without any resummation.\footnote{For completeness, let's indicate that resummations of the series with simple Padé approximants lead to the three-loop values $\eta^{[1/2]}(\text{P}_4)=0.8904$, $\eta^{[2/1]}(\text{P}_4)=0.8503$ and the four-loop result $\eta^{[2/2]}=0.8060$.} Hence, a direct substitution of $\eps=1$ in \eqref{eq:pertseriesp3} yields, successively
\begin{empheq}[box=\widefbox]{equation}
\eta_{\text{1-loop}}(\text{P}_4) = 0.9600, ~~~
\eta_{\text{2-loop}}(\text{P}_4) = 0.9139, ~~~
\eta_{\text{3-loop}}(\text{P}_4) = 0.8872, ~~~
\eta_{\text{4-loop}}(\text{P}_4) = 0.8670.
\end{empheq}
Similarly to P$_3$, the one-loop result has been first obtained in \cite{Aronovitz:1988}, the two-loop result (32 year later) in \cite{Coquand:2020a}, the three-loop result in \cite{Metayer:2021kxm} and a few months after, the four-loop computation has been achieved in the two-field model in \cite{Pikelner:2021}. This four-loop result has recently been confirmed by \cite{Metayer:2024} in the effective flexural model. These results show that the perturbative value for the anomalous stiffness slowly decreases with the loop order seemingly converging to a value which is well within the generally accepted range $[0.7,0.9]$, see discussion in the Introduction.

\begin{figure}[h!]
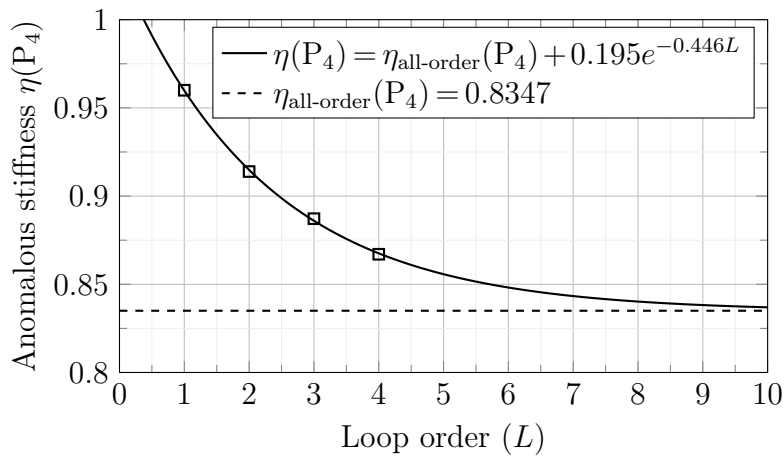

\centering
\Plotetafit
\caption{Exponential fit of the results found for $\eta(\text{P}_4)$ from 1 to 4-loop.}
\label{fig:rqed:expfiteta}
\end{figure}

Following \cite{Metayer:2024}, it is tempting at this point to plot the values obtained for the anomalous dimension as a function of the loop order. As can be seen from figure \ref{fig:rqed:expfiteta}, a simple exponential fits the 4 data points. Its extrapolation to infinite order yields 
\begin{empheq}{equation}
\eta_{\text{all-order}}(\text{P}_4)=0.8347\, , 
\label{eq:infloopfit}
\end{empheq}
which is again within the generally accepted range $[0.7,0.9]$.

\subsection{Flow diagram}

\begin{figure}[h!]
\centering
\begin{picture}(75,75)
\put(0,0){\includegraphics[scale=0.7]{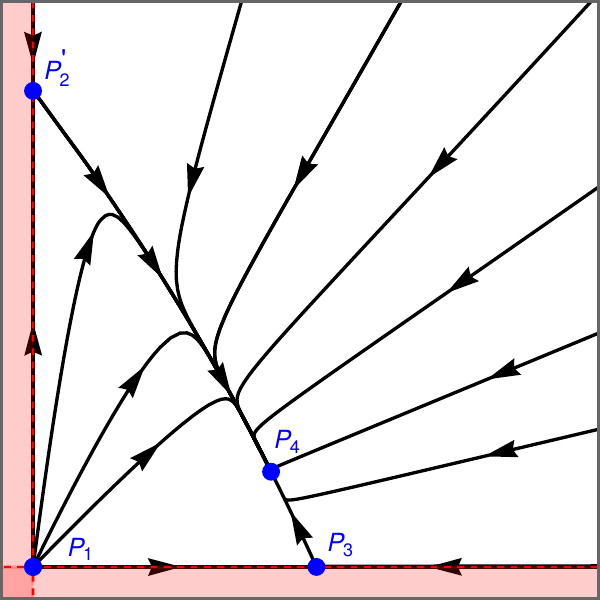}}
\put(73,3.5){$\mu_r$}
\put(2.5,73){$b_r$}
\end{picture}
\caption{RG-flow diagram in the ($\mu_r,b_r$) plane. The mechanical stability of the model imposes $\mu_r>0$ and $b_r>0$. The corresponding non-physical regions are indicated in red and delimited by the red dashed lines $\mu_r=0$, $b_r=0$, on which lie the fixed points P$_1$, P$'_2$ and P$_3$ at all loop orders. This plot has been obtained from the beta functions \eqref{eq:mem:betafuncs} and remains qualitatively the same for all values $0.001<\eps<0.8$.}
\label{fig:mem:phasediag}
\end{figure}

We close this section by displaying on figure \ref{fig:mem:phasediag} the flow diagram obtained from all the previous results. The coordinates of the fixed points do vary slightly with the loop order but the diagram remains essentially the same. In particular, the fixed points P$_1$, P$'_2$ and P$_3$ stay exactly on the mechanical stability lines imposed by positive shear modulus $\mu_r>0$ and positive bulk modulus $b_r>0$. The fixed point P$_4$ is clearly seen to be attractive and it is this fixed point that controls the flat phase. 

\section{Comparisons with other approaches}
\label{sec:benchmarks}

The results that we have obtained in the previous section were limited to a given order in the loop expansion but are otherwise exact. In this section, we will use them to benchmark results obtained in the literature using other methods that work directly in the dimension of interest $d=2$ and resum part of the perturbative series but are otherwise subject to their own sets of approximations.

\subsection[SCSA]{Self Consistent Screening Approximation}

The SCSA was originally introduced by Bray \cite{Bray:1974} for the $O(N)$ model as a self-consistent version of the $1/N$ expansion. An early application of this technique to membranes has been carried out in \cite{LeDoussal:1992} 
for the effective flexural model \eqref{mem:SEFT}. Though approximate (see more below), the method has the advantage of being exact both at one-loop in the $\eps$-expansion and at leading order in the $1/d$-expansion. It is regarded as being quantitatively successful in application to membranes, see the review \cite{LeDoussal:2018}. 

Following \cite{LeDoussal:1992,LeDoussal:2018}, the SCSA equations, written in the conventions of the present manuscript and for each fixed point, read
\bs
\ba
& \text{P}_1: ~~ \eta=0, \\
& \text{P}'_2: ~~ 2F(d,\eta)-d_c=0, \\
& \text{P}_3: ~~ (d+1)(d-2)F(d,\eta)-d_c=0, \\
& \text{P}_4: ~~ d(d-1)F(d,\eta)-d_c=0,
\ea
\es
where
\be
F(d,\eta)=\frac{\Gamma (2-\eta) \Gamma \left(2-\eta/2\right) \Gamma \left(\eta/2\right) \Gamma (\eta+d)}{\Gamma \left(2-\eta-d/2\right) \Gamma \left((4-\eta+d)/2\right) \Gamma \left((\eta+d)/2\right) \Gamma \left(\eta+d/2\right)}.
\ee
Solving non-perturbatively these equations for $\eta$ in $d=2$, leads to the well-known value $\eta_{\text{SCSA}}=0.8209$ for $d_c=1$ at the fixed point P$_4$ relevant to the flat phase. In order to compare the SCSA results to ours, we now set $d=4-2\eps$ in the above SCSA equations and solve them perturbatively at each fixed point. In expanded form, the SCSA results read
\bs
\label{eq:SCSA:perturbative}
\ba
\eta_{\text{SCSA}}(\text{P}_1) & = 0+\Ord(\eps^4),\\
\eta_{\text{SCSA}}(\text{P}'_2) & = \frac{4 \eps }{d_c+4} + \left(\frac{16}{(d_c+4)^3}-\frac{20}{3 (d_c+4)^2}+\frac{2}{3 (d_c+4)}\right) \eps^2 \\ 
& +\left(\frac{128}{(d_c+4)^5}+\frac{128}{3 (d_c+4)^4}-\frac{400}{3 (d_c+4)^3}+\frac{406}{9 (d_c+4)^2}-\frac{37}{9 (d_c+4)}\right) \eps^3 +\Ord(\eps^4),\nonum \\
\eta_{\text{SCSA}}(\text{P}_3) & = \frac{20 \eps }{d_c+20}+ \left(\frac{2000}{(d_c+20)^3}+\frac{1180}{3 (d_c+20)^2}-\frac{74}{3 (d_c+20)}\right) \eps^2 \\ 
& +\left(\frac{400000}{(d_c+20)^5}+\frac{584000}{3 (d_c+20)^4}-\frac{18720}{(d_c+20)^3}+\frac{6694}{9 (d_c+20)^2}-\frac{155}{9 (d_c+20)}\right) \eps^3 +\Ord(\eps^4), \nonum \\
\eta_{\text{SCSA}}(\text{P}_4) & = \frac{24 \eps }{d_c+24} + \left(\frac{3456}{(d_c+24)^3}+\frac{432}{(d_c+24)^2}-\frac{24}{(d_c+24)}\right) \eps^2 \\
& +\left(\frac{995328}{(d_c+24)^5}+\frac{345600}{(d_c+24)^4}-\frac{35520}{(d_c+24)^3}+\frac{1320}{(d_c+24)^2}-\frac{64}{3 (d_c+24)}\right) \eps^3 +\Ord(\eps^4),\nonum 
\ea
\es
where the four-loop contributions are too lengthy to be displayed.
Clearly, the SCSA results miss a great part of the transcendental structure of the series ($\zeta_3$ terms start to appear only at four loops in SCSA) but reproduce well the $1/(d_c+n)$ structure previously observed with $n=4,20,24$. The quantitative agreement is even more apparent numerically in the case $d_c=1$: 
\bs
\label{eq:SCSA:perturbative:num}
\ba
\eta_{\text{SCSA}}(\text{P}_1)& = 0+\Ord(\eps^5),\\ 
\eta(\text{P}_1)& = 0+\Ord(\eps^5), \nonum \\[0.1cm]
\eta_{\text{SCSA}}(\text{P}'_2)& = 0.800 \eps -0.00533 \eps^2+0.0248 \eps^3 - 0.00339\eps^4 +\Ord(\eps^5), \\ 
\eta(\text{P}'_2)& = 0.800 \eps- 0.00533 \eps^2 + 0.0110 \eps^3 + 0.00137\eps^4 +\Ord(\eps^5), \nonum \\[0.1cm]
\eta_{\text{SCSA}}(\text{P}_3)& = 0.952 \eps-0.0667 \eps^2 -0.0560 \eps^3 -0.0519\eps^4 +\Ord(\eps^5),\\ 
\eta(\text{P}_3)& = 0.952 \eps- 0.0711 \eps^2 - 0.0698 \eps^3 - 0.0748\eps^4 +\Ord(\eps^5),\nonum \\[0.1cm]
\eta_{\text{SCSA}}(\text{P}_4)& = 0.960 \eps-0.0476 \eps^2 -0.0280 \eps^3 - 0.0177\eps^4 +\Ord(\eps^5),\\ 
\eta(\text{P}_4)& = 0.960 \eps- 0.0461 \eps^2 - 0.0267 \eps^3 - 0.0202\eps^4 +\Ord(\eps^5).
\nonum
\ea
\es
%
%
%
The first order is exact as expected from such a technique. The second order is very close for both P$_3$ and P$_4$ and even exact for P$'_2$. The three-loop order is very close for both P$_3$ and P$_4$ but differs by a factor of two for P$'_2$. Interestingly, at four loops, the SCSA is very close numerically for both P$_3$ and P$_4$ but gives the wrong sign and misses a factor of 3 for P$_2'$ indicating the limits of the SCSA approximation.

In order to clarify the approximation involved in the SCSA calculations, we have recomputed all of our result adding a constant $V$ (such that $V^2=V$) in factor of each diagram containing a vertex correction. To be more precise, we add a factor $V$ in front of all the diagrams that cannot be reduced to a line by successively removing two-lines bubbles. We therefore add a factor $V$ to the two-loop diagram $c$ of figure \ref{mem:fig:two-loop:Sigma}, the diagram $b$ of figure \ref{mem:fig:two-loop:vertex}, the three-loop diagrams $f, g, h, i, j, k, l, m$, and $o$ of figure \ref{mem:fig:three-loop:sigma}, the three-loop  diagrams $d, e, f, g, h, i$, and $k$ of figure \ref{mem:fig:three-loop:V}, as well as in front of 209 of the 246 four-loop diagrams. With this factor included, our perturbative computations read (in numerical form)
\ba
\label{eq:generalizedscsares}
& \eta(\text{P}_1)=0+\Ord(\eps^5), \\
& \eta(\text{P}'_2)=0.800 \eps -0.0053 \eps^2+(0.0248 -0.0138 V)\eps ^3 -(0.00339-0.00476 V)\eps^4 +\Ord(\eps^5), \nonum \\
& \eta(\text{P}_3)=0.952 \eps -(0.0667+0.0043 V) \eps ^2-(0.0560+0.0138 V) \eps ^3-(0.0519+0.0228 V)\eps^4 +\Ord(\eps^5), \nonum \\
& \eta(\text{P}_4)=0.960 \eps -(0.0476-0.0015 V) \eps ^2 - (0.0280-0.0012 V) \eps ^3-(0.0177+0.0025 V)\eps^4 + \Ord(\eps^5), \nonum 
\ea
and are such that vertex corrections are completely neglected in the case $V=0$ and fully included in the limit $V=1$.
A comparison with \eqref{eq:SCSA:perturbative:num} then reveals that 
the approximation made in the SCSA is to neglect all vertex corrections ($V=0$). Certainly such an approximation allows for a drastic simplification of the calculations. In the present case, it seems reliable for fixed points P$_3$ and P$_4$ due to the fact that vertex corrections are remarkably small up to four loops, as can be seen from \eqref{eq:generalizedscsares}. Just as the overall smallness of the coefficients of the perturbative series, this arises from strong cancellations among the diagrams and may explain why the SCSA approximation is so successful at fixed points P$_3$ and P$_4$. This is to be contrasted with the fixed point P$'_2$, for which vertex corrections are of the same order as non-vertex corrections. It is interesting at this point to evaluate \eqref{eq:generalizedscsares} at $\eps=1$. This yields:
\bs
\ba
& \eta(\text{P}_1)=0 \,, \\
& \eta(\text{P}'_2) = 0.8161 - 0.0090 V \,,\\
& \eta(\text{P}_3) = 0.7777 - 0.0409 V \,,\\
& \eta(\text{P}_4) =0.8667 + 0.0003 V \,,
\ea
\es
and shows that vertex corrections are surprisingly small overall, e.g., of the order of $1\%$ for P$'_2$, $5$\% for P$_3$ and even $0.03\%$ for the physical stable fixed point P$_4$. The renormalization-group functions generalized for all $V$ are available in computer readable files as ancillary files to the arXiv version of the letter \cite{Metayer:2024}.

\subsection[NPRG]{Non-Perturbative Renormalization Group}

We proceed along the same lines for the NPRG technique. The latter has been introduced in \cite{Wetterich:1993} and is based on a truncated solution of exact RG equations. The NPRG approach therefore resums an infinite number of terms of the perturbative series though the approximation involved does not have any diagrammatic interpretation contrary to the SCSA one and, hence, appears to be less controlled. Nevertheless, it is exact at one-loop and allows to work directly in $d=2$. It has been applied to the flat phase of polymerized membranes in, \eg, \cite{Kownacki:2009,Braghin:2010,Hasselmann:2011}.

Following \cite{Kownacki:2009}, the NPRG equations, written in the conventions of the present manuscript and for each fixed point, read
\be
\beta^{\text{NPRG}}_\mu= \mu F\left(\frac{2 \mu }{d+1}\right), \qquad
\beta^{\text{NPRG}}_b = b F(b), \qquad
\eta_{\text{NPRG}} =\frac{(d+4) A_d}{(d+2) (d+4) d^2+A_d},
\ee
where 
\be
F(x)=d-4+2 \eta +\frac{d(d-\eta +8)d_c\eta x}{(d+8) (d-\eta +4) (b+(d-2) \mu)}, \quad 
A_d=\frac{16 (d+1) (b+(d-2) \mu)}{(4 \pi)^{d/2} \Gamma \left(d/2\right)}.
\ee
Solving non-perturbatively these equations in $d=2$ for $\eta$ leads to the value $\eta_{\text{NPRG}}=0.8491$ for $d_c=1$ at the fixed point P$_4$ relevant to the flat phase. In order to compare the NPRG results to ours, we set $d=4-2\eps$ in the above NPRG equations and solve them perturbatively at each fixed point. In expanded form, the NPRG results read
\bs
\ba
\eta_{\text{NPRG}}(\text{P}_1) & = 0+\Ord(\eps^4), \\
\eta_{\text{NPRG}}(\text{P}'_2) & = \frac{4 \eps }{d_c+4}+ \bigg(\frac{8}{3(d_c+4)^3}-\frac{14}{3(d_c+4)^2}+\frac{1}{d_c+4}\bigg) \eps^2 \\ 
& +\bigg(\frac{32}{9(d_c+4)^5}-\frac{76}{9(d_c+4)^4}+\frac{25}{3(d_c+4)^3}-\frac{65}{18(d_c+4)^2}+\frac{1}{2(d_c+4)}\bigg) \eps^3 + \Ord(\eps^4), \nonum
\\
\eta_{\text{NPRG}}(\text{P}_3) & = \frac{20 \eps }{d_c+20}+ \bigg(\frac{1000}{3(d_c+20)^3}+\frac{1330}{3(d_c+20)^2} -\frac{23}{d_c+20}\bigg) \eps^2\\ 
& +\bigg(\frac{100000}{9(d_c+20)^5}+\frac{204500}{9(d_c+20)^4}+\frac{26365}{3(d_c+20)^3}-\frac{10217}{18(d_c+20)^2}+\frac{7}{2(d_c+20)}\bigg) \eps^3 + \Ord(\eps^4), \nonum 
\\
\eta_{\text{NPRG}}(\text{P}_4) & =\frac{24 \eps }{d_c+24}+ \bigg(\frac{576}{(d_c+24)^3}+\frac{504}{(d_c+24)^2}-\frac{22}{d_c+24}\bigg) \eps^2 \\
& +\bigg(\frac{27648}{(d_c+24)^5}+\frac{37440}{(d_c+24)^4}+\frac{9192}{(d_c+24)^3}-\frac{546}{(d_c+24)^2}+\frac{4}{d_c+24}\bigg) \eps^3 + \Ord(\eps^4)\,, \nonum 
\ea
\es
where the four-loop contributions are too lengthy to be displayed.
As for the SCSA approach, the NPRG reproduce well the $1/(d_c+n)$ structure previously observed with $n=4,20,24$. It is however completely missing the transcendental structure as there is no $\zeta_n$ in the result, at all orders. Despite this, the quantitative agreement is still apparent numerically in the case $d_c=1$:
%
\bs
\ba
\eta_{\text{NPRG}}(\text{P}_1) & = 0+\Ord(\eps^5), \nonum \\
\eta(\text{P}_1) & = 0+\Ord(\eps^5), \\[0.1cm]
\eta_{\text{NPRG}}(\text{P}'_2) & = 0.800 \eps +0.0347 \eps^2 +0.0099 \eps^3 +0.00305\eps^4+\Ord(\eps^5) , \nonum \\
\eta(\text{P}'_2) & = 0.800 \eps- 0.0053 \eps^2 + 0.0110 \eps^3+0.00137\eps^4+\Ord(\eps^5), \\[0.1cm]
\eta_{\text{NPRG}}(\text{P}_3) & = 0.952 \eps -0.0540 \eps^2 -0.0519 \eps^3-0.0485\eps^4+\Ord(\eps^5), \nonum \\
\eta(\text{P}_3) & = 0.952 \eps- 0.0711 \eps^2 - 0.0698 \eps^3-0.0748\eps^4+\Ord(\eps^5), \\[0.1cm]
\eta_{\text{NPRG}}(\text{P}_4) & = 0.960 \eps -0.0367 \eps^2 -0.0266 \eps^3-0.0178\eps^4+\Ord(\eps^5), \nonum \\
\eta(\text{P}_4) & = 0.960 \eps- 0.0461 \eps^2 - 0.0267 \eps^3-0.0202\eps^4+\Ord(\eps^5), 
\ea
\es
The NPRG is successful in reproducing numerically the loop expansion for both P$_3$ and P$_4$ up to four loops, with striking quantitative agreement. Just as for the SCSA, this might be due to the fact that the perturbative series behaves remarkably well for the present problem. However, similarly to the SCSA, the NPRG is less accurate in reproducing the anomalous dimension of the flexuron field at P$'_2$, with a wrong sign already at two-loop as well as a four-loop contribution 
which is the correct sign (contrary to SCSA) but is too large by a factor of three.

\subsection{Large-\titlemath{d_c} approaches}
\label{sec:mem:largedc}

As advertised in the introduction, another approach to studying polymerized membranes is based on an expansion in large codimension $d_c$, the case of interest being $d_c=1$. Despite the fact that a $1/d_c$-expansion does not seem to be (quantitatively) reliable for the present problem, it is a very useful check to compare it with our computations properly re-expanded in large-$d_c$.

By construction, the SCSA is exact at leading order in $1/d_c$ with the corresponding field anomalous dimension (valid for any $d$) reading \cite{LeDoussal:1992,LeDoussal:2018}
\be
\eta_{\text{SCSA}}(d,d_c)=\frac{8}{d_c}\frac{d-1}{d+2}\frac{\Gamma(d)}{\Gamma^3(d/2)\Gamma(2-d/2)}+\Ord\left(1/d_c^2\right) \, .
\label{eq:mem:scsaLO}
\ee
In the case $d=2$, this results in, $\eta_{\text{SCSA}}(\text{P}_4)=2/d_c+\Ord(1/d_c^2)$, which agrees with the early finding of \cite{Aronovitz:1988,David:1988}. 
For each fixed point and in the conventions of the present manuscript, \eqref{eq:mem:scsaLO} leads to
\bs
\ba
\eta_{\text{SCSA}}(\text{P}_1) & = 0, \\
\eta_{\text{SCSA}}(\text{P}'_2) & =\eta_{\text{SCSA}}\left(d,\frac{d_c d(d-1)}{2}\right), \\
\eta_{\text{SCSA}}(\text{P}_3) & =\eta_{\text{SCSA}}\left(d,\frac{d_c d(d-1)}{(d-2)(d+1)}\right), \\
\eta_{\text{SCSA}}(\text{P}_4) & =\eta_{\text{SCSA}}(d,d_c).
\ea
\es
For our purposes, we may now set $d=4-2\eps$ in the above equations and expand them up to $\Ord(\eps^4/d_c)$, which yields 
\bs
\ba
\eta_{\text{SCSA}}(\text{P}_1) & = 0+\Ord(\eps^5/d_c^2), \\
\eta_{\text{SCSA}}(\text{P}'_2) & = \frac{1}{d_c}\left(4\eps+\frac{2 \eps^2}{3}-\frac{37 \eps^3}{9}+4\left(\frac{479}{216}-2\zeta_3\right)\eps^4+\Ord(\eps^5)\right)+\Ord\left(1/d_c^2\right), \\
\eta_{\text{SCSA}}(\text{P}_3) & = \frac{1}{d_c}\left(20 \eps-\frac{74 \eps^2}{3}-\frac{155 \eps^3}{9}+20\left(\frac{769}{1080}-2\zeta_3\right)\eps^4+\Ord(\eps^5)\right)+\Ord\left(1/d_c^2\right), \\
\eta_{\text{SCSA}}(\text{P}_4) & = \frac{1}{d_c}\left(24 \eps-24 \eps^2-\frac{64 \eps^3}{3}+24\left(\frac{26}{27}-2\zeta_3\right)\eps^4+\Ord(\eps^5)\right)+\Ord\left(1/d_c^2\right).
\ea
\es
These results are in perfect accordance with our results, see \eqref{eq:mem:etap2}, \eqref{eq:mem:etap3}, \eqref{eq:mem:etap4}, re-expanded to the first large-$d_c$ order. For example, for P$_4$, it is very easy to see that the series of numbers $\{+24,-24,-64/3,...\}$ exactly corresponds to the last term in $1/(d_c+24)$ at each order of the $\eps$ expansion in our result \eqref{eq:mem:etap4}. This correspondence is also exact for the four fixed points at four-loop order, whose terms are not displayed in \eqref{eq:mem:etap2}, \eqref{eq:mem:etap3} and \eqref{eq:mem:etap4} because they are too lengthy. This comparison has been first done up to four loops in \cite{Pikelner:2021} for P$_3$ and P$_4$, and in \cite{Metayer:2024} for P$'_2$.


At the next-to-leading order (NLO) in the $1/d_c$-expansion there are no available results valid for arbitrary $d$. There has been however a recent computation of $\eta$ at the fixed point P$_4$ in $d=2$ \cite{Saykin:2020} that reads
\be
\eta_{\text{large-}d_c}(\text{P}_4)=\frac{2}{d_c}+\frac{73-68\zeta_3}{27d_c^2}+\Ord(1/d_c^2).
\ee
Surprisingly, the series yields the numerical value $\eta_{\text{large-}d_c}(\text{P}_4)=2/d_c-0.32/d_c^2+\Ord(1/d_c^3)$, which, for $d_c=1$, evaluates to $\eta_{\text{large-}d_c}(\text{P}_4)=1.68$, well above unity. It therefore seems that, for the present problem, the $1/d_c$-expansion behaves rather badly with respect to the loop expansion. Let's also note that the NLO (in the $1/d_c$-expansion) SCSA result has been obtained semi-analytically in \cite{Gazit:2009} and, according to this publication, yields a value of $\eta^{\text{NLO}}_{\text{SCSA}}(\text{P}_4)=0.78922(5)$. However, to our knowledge, this result is not available analytically for arbitrary dimension $d$ so we cannot further compare it with our results.

As a concluding remark, let us recall that early computations \cite{Guitter:1988,Guitter:1989} (see also \cite{LeDoussal:1992b}) considered a large-$D$ approach which is strictly equivalent to large $d_c$ at leading order since $d_c=D-d$ and hence
\be
\eta_{\text{large}-D}(\text{P}_4)=\frac{2}{D}+\Ord(1/D^2).
\ee
The key difference is that $D=3$ providing a reasonably small $1/D$ expansion parameter; this leads to $\eta_{\text{large}-D}(\text{P}_4)=2/3=0.667$ which is a much better result than the one obtained with the corresponding large-$d_c$ approach. 
To our knowledge, there is presently no available NLO result in the $1/D$-expansion.

\newpage

\subsection{Table of results for \titlemath{\eta(\text{P}_4)} }


As a summary, we list in table \ref{tab:mem:literatureeta} values obtained in the literature during the last three decades for the flexuron-field anomalous dimension at the flat fixed point P$_4$ in $d=2$ and $d_c=1$. 

\begin{table}[h!]
\renewcommand{\arraystretch}{1.3}
\centering
\begin{tabular}{|c|c|l|}
\hline
$\eta(\text{P}_4)$ & Method & Year/ref \\
\hline\hline
$\approx0.66$ & Monte Carlo (membrane) & 1990 \cite{Abraham:1990} Abraham, Nelson \\
$0.667$ & Large $D$ (LO) & 1988 \cite{Guitter:1988,Guitter:1989} Guitter, \etal \\
$\approx0.7$ & Monte Carlo (vesicles) & 1991 \cite{Komura:1991} Komura, Baumgärtner \\
$0.72(4)$ & Monte Carlo (membrane) & 1989 \cite{Leibler:1989} Leibler, Maggs \\
$0.75(5)$ & Monte Carlo (membrane) & 1990 \cite{Guitter:1990b} Guitter \etal \\
$0.750(5)$ & Monte Carlo (membrane) & 1996 \cite{Bowick:1996} Bowick \etal \\
$0.789$ & SCSA (large-$d_c$ NLO, semi-numerical) & 2009 \cite{Gazit:2009} Gazit \\
$0.795(10)$ & Monte Carlo (graphene) & 2013 \cite{Troster:2013} Tröster \\
$0.81(3)$ & Monte Carlo (membrane) & 1993 \cite{Zhang:1993} Zhang \etal \\
$\approx 0.82$ & Molecular dynamics simulations & 1996 \cite{Zhang:1996} Zhang \etal \\
$\approx0.82$ & SCSA (LO, semi-numerical) & 2010 \cite{Zakharchenko:2010,Roldan:2011} Zakharchenko \etal \\
$0.821$ & SCSA (LO, analytical) & 1992 \cite{LeDoussal:1992,LeDoussal:2018} Le Doussal, Radzihovsky \\
$0.849$ & NPRG (analytical) & 2009 \cite{Kownacki:2009} Kownacki, Mouhanna \\
$\approx0.85$ & NPRG (semi-numerical) & 2009 \cite{Braghin:2010,Hasselmann:2011} Braghin, Hasselmann \\
$\approx 0.85$ & Monte Carlo (graphene) & 2009 \cite{Los:2009} Los \etal \\
$0.867$ & 4-loop (2-field) & 2021 \cite{Pikelner:2021} Pikelner \\
\rowcolor{mygray} $0.867$ & 4-loop (effective flexural) & 2024 \cite{Metayer:2024} Metayer \\
\rowcolor{mygray} $0.887$ & 3-loop & 2021 \cite{Metayer:2021kxm} Metayer \etal \\
$0.90(4)$ & Molecular dynamics simulations & 1993 \cite{Petsche:1993} Petsche, Grest \\
\rowcolor{mygray} $0.914$ & 2-loop & 2020 \cite{Coquand:2020a} Coquand \etal \\
\rowcolor{mygray} $0.960$ & 1-loop & 1988 \cite{Aronovitz:1988,Aronovitz:1989} Aronovitz, Lubensky \\
$1$ & Self-consistent & 1987 \cite{Nelson:1987} Nelson, Peliti \\
\hline
\end{tabular}
\caption{Results for the anomalous stiffness of a flat two-dimensional membrane embedded in three dimensions obtained from 1987 to 2024. 
Shaded lines are the multi-loop results explicitly reviewed in this paper. Note that some references evaluated the roughness exponent $(\zeta)$ which we converted to the anomalous stiffness using $\eta=2(1-\zeta)$, see \eqref{eq:mem:roughness}. 
}
\label{tab:mem:literatureeta}
\end{table}

\newpage

\section{Conclusion}
\label{sec:conclusion}

We have reviewed the field-theoretic approach to the flat phase of polymerized membranes within the flexural effective model. Our general formalism together with fully automated computations allowed us to reach a four-loop accuracy in the computation of the renormalization-group functions and corresponding anomalous dimensions of the model. Our results complete the 
four-loop results obtained in \cite{Pikelner:2021} for the equivalent two-field model. In particular, the anomalous stiffness of a flat two-dimensional membrane embedded in three dimensions is found to be $\eta(\text{P}_4) = 0.867$ (and an exponential fit slightly lowers this value to $0.8347$) which is in very good agreement with results obtained by numerical simulations and non-perturbative techniques. In addition, a four-loop accuracy was reached for the peculiar P$_2'$ fixed point that is absent in the two-field model. We also used our (exact order-by-order) results to benchmark several non-perturbative approaches, such as NPRG and SCSA. Up to four loops, a rather impressive quantitative agreement is found. Its origin maybe traced to the very peculiar structure of the perturbative series associated to the present problem that is characterized, up to four loops, by surprisingly small vertex corrections together with overall small and even decreasing coefficients for the important fully stable fixed point P$_4$. In closing, we have provided a detailed and hopefully pedagogical review of the field-theoretic formalism useful to study critical properties of polymerized membranes (see \cite{Metayer:2023tii} for more). We hope that it will benefit (young) researchers interested in the field and related issues.

\section*{Acknowledgements}
 This work benefited from access to the HPC resources of the MeSU platform at Sorbonne University. S.M.\ would like to thank N.~Benoit, head of the MeSU SACADO team, for excellent technical support as well as P.~Descombes for useful discussion on optimal diagram parametrization and finally D.~Mouhanna for fruitful discussions on the physics of the model. S.T.\ would like to thank M.~Kompaniets for fruitful discussions and advice at the beginning of this project as well as O.~Coquand and D.~Mouhanna for initial collaboration.

\begin{appendix}

\section{Multiloop Feynman integrals}
\label{chap:appendix}


We consider an Euclidean space of dimension $d$ and follow most of the notations of the review \cite{Kotikov:2018wxe}. In momentum space, the one-loop massless propagator-type master integral
corresponds to a simple loop that is given by
\be
J(d,\vec p,\al,\beta) 
= \generaloneloop{\al}{\beta}
= \int \frac{[\D^d k]}{k^{2\al} (\vec p- \vec k)^{2\beta}}
= \frac{(p^2)^{d/2-\al-\beta}}{(4\pi)^{d/2}} G(d,\al,\beta)\, ,
\label{def:one-loop:2-pt}
\ee
where $[\D^dk]= \D^dk /(2\pi)^d$ and $\al$, $\beta$ are the so-called indices of the propagators. The dimensionless function $G$ is well-known, see, \eg, \cite{Kotikov:2018wxe}, and has a simple expression in terms of Euler $\Gamma$-functions
\ba
G(d,\al,\beta)= \frac{a(d,\al) a(d,\beta)}{a(d,\al+\beta-d/2)}, 
\qquad a(d,\al)= \frac{\Gamma(d/2-\al)}{\Gamma(\al)}\, .
\label{eq:multiloop:Gres}
\ea
Note that this function vanishes if at least one integer index is negative or zero and is also symmetrical under the exchange of indices, \ie, 
\be
G(d,\al,\beta)=0, \text{ if } \al\leq0 \text{ or } \beta\leq0, \qquad \text{and} \qquad
G(d,\al,\beta)=G(d,\beta,\al).
\label{eq:multiloop:Grules}
\ee

At two-loop, the massless propagator-type master integral (the so called \textit{diamond} diagram) is given by
\ba
J(d,\vec p,\al_i) =
\generaltwoloop{\al_1}{\al_2}{\al_3}{\al_4}{\al_5}
& = \int \frac{[\D^d k_1 ] [\D^d k_2]}{
k_1^{2\al_1} 
k_2^{2\al_2} 
(\vec p - \vec k_2)^{2\al_3} 
(\vec p- \vec k_1)^{2\al_4} 
(\vec k_{12})^{2\al_5}} =\frac{(p^2)^{d-\sum\al_i}}{(4\pi)^d} G(d,\al_i), \nonum \\[-0.5cm]
\label{eq:multiloopdiamond}
\ea
where $\vec k_{12}=\vec k_1-\vec k_2$ and $G(d,\al_1,...,\al_5)$ is dimensionless and unknown for arbitrary indices $\{\al_i\}_{i=1-5}$. However, for the set of indices relevant to our study, this function is known exactly and can be expressed using the one-loop result \eqref{eq:multiloop:Gres}.

At three loops, there are three different topologies for the massless propagator-type master integral, namely Ladder (L3), Benz (B3) and non-planar (N3). These are respectively defined as
\bs
\label{def:three-loop:2-pt}
\ba
& J_{\text{L3}}
=
\!\GraphGeneraThreeLoopLA\!
= \int \! \frac{[\D^d k_1 ][\D^d k_2][\D^d k_3]}{
k_1^{2\al_1} 
k_2^{2\al_2} 
k_3^{2\al_3} 
(\vec p - \vec k_3)^{2\al_4} 
(\vec p - \vec k_2)^{2\al_5} 
(\vec p - \vec k_1)^{2\al_6} 
(\vec k_{12})^{2\al_7} 
(\vec k_{23})^{2\al_8}},
\\
& J_{\text{B3}}
=
\!\GraphGeneraThreeLoopBE\!
= \int \! \frac{[\D^d k_1 ][\D^d k_2][\D^d k_3]}{
k_1^{2\al_1} 
k_2^{2\al_2} 
k_3^{2\al_3} 
(\vec p - \vec k_3)^{2\al_4} 
(\vec p - \vec k_1)^{2\al_5} 
(\vec k_{12})^{2\al_6} 
(\vec k_{23})^{2\al_7} 
(\vec k_{13})^{2\al_8}},
\\
& J_{\text{N3}}
=
\!\GraphGeneraThreeLoopNP\! 
= \int \! \frac{[\D^d k_1 ][\D^d k_2][\D^d k_3]}{
k_1^{2\al_1} 
k_2^{2\al_2} 
k_3^{2\al_3} 
(\vec p - \vec k_3)^{2\al_4} 
(\vec p_{123})^{2\al_5} 
(\vec p-\vec k_1)^{2\al_6} 
(\vec k_{32})^{2\al_7} 
(\vec k_{12})^{2\al_8}},
\ea
\es
with $J_{\text{X}}=J_{\text{X}}(d,\vec p,\al_i)$ for $\text{X}\in\{{\text{L3},\text{B3},\text{N3}}\}$ and $\vec k_{ij}=\vec k_i- \vec k_j$ as well as $\vec p_{123}=\vec p-\vec k_1+\vec k_2-\vec k_3$. Similarly to the one- and two-loop cases, the external momentum ($p$) dependence is easily extracted from dimensional analysis which allows us to write the diagrams in the following form
\ba
J_\text{X}(d,\vec p,\al_1,\dots,\al_8) = \frac{(p^2)^{3d/2 - \sum_{i=1}^8 \al_i}}{(4\pi)^{3d/2}} G_\text{X}(D,\al_1,\dots,\al_8), ~~ \text{X}\in\{\text{L3}, \text{B3}, \text{N3}\},
\label{def:three-loop-G-func}
\ea
where $G_{\text{X}}(D,\al_1,\dots,\al_8)$ is the (dimensionless) coefficient function of the diagram with topology X. 

In this review, at three loops, we consider a theory where all indices $\al_i$ ($i=1,\dots,8$) are integers, in which case the integration by parts (IBP) reduction technique \cite{Vasiliev:1981,Tkachov:1981,Chetyrkin:1981} is very powerful. In the following, we choose the ladder (L3) topology to be the default one. This implies that if a diagram is sufficiently trivial to be issued from several topologies, we shall choose the ladder (L3) one. Using IBP-reduction techniques, all possible three-loop diagrams can be expressed on three different basis corresponding to the three topologies
\begin{itemize}
\item The Ladder (L3) master integral basis:
\bs
\label{eq:examplenontrivial}
\ba
& \hspace{-0.5cm} J_{\text{L3}}(d,\vec p, 0, 0, 1, 0, 0, 1, 1, 1)=\MasterThreeA=\frac{(p^2)^{3d/2-4}}{(4\pi)^{3d/2}} G(d, 1, 1) G(d, 1, 2-d/2) G(d, 1, 3-d), \\[-0.5cm]
& \hspace{-0.5cm} J_{\text{L3}}(d,\vec p, 0, 1, 0, 1, 0, 1, 1, 1)=\MasterThreeB=\frac{(p^2)^{3d/2-5}}{(4\pi)^{3d/2}}G^2(d, 1, 1) G(d, 1, 4-d), \\
& \hspace{-0.5cm} J_{\text{L3}}(d,\vec p, 0, 1, 1, 1, 0, 1, 1, 0)=\MasterThreeC=\frac{(p^2)^{3d/2-5}}{(4\pi)^{3d/2}}G^2(d, 1, 1) G(d, 1, 2-d/2), \\
& \hspace{-0.5cm} J_{\text{L3}}(d,\vec p, 1, 0, 1, 1, 0, 1, 1, 1)=\MasterThreeD=\frac{(p^2)^{3d/2-6}}{(4\pi)^{3d/2}}G(d, 1, 1) G(d, 1, 1, 1, 1, 2-d/2), \\
& \hspace{-0.5cm} J_{\text{L3}}(d,\vec p, 1, 1, 1, 1, 1, 1, 0, 0)=\MasterThreeE=\frac{(p^2)^{3d/2-6}}{(4\pi)^{3d/2}}G^3(d, 1, 1).
\ea
\es
\item The Benz (B3) master integral basis:
\bs
\ba
& \hspace{-0.5cm} J_{\text{B3}}(d,\vec p, 0, 0, 1, 0, 1, 1, 1, 0)=\MasterThreeA=\frac{(p^2)^{3d/2-4}}{(4\pi)^{3d/2}}G(d, 1, 1) G(d, 1, 2-d/2) G(d, 1, 3-d), \\[-0.5cm]
& \hspace{-0.5cm} J_{\text{B3}}(d,\vec p, 0, 1, 0, 1, 1, 1, 1, 0)=\MasterThreeB=\frac{(p^2)^{3d/2-5}}{(4\pi)^{3d/2}}G^2(d, 1, 1) G(d, 1, 4-d), \\
& \hspace{-0.5cm} J_{\text{B3}}(d,\vec p, 0, 1, 1, 1, 1, 1, 0, 0)=\MasterThreeC=\frac{(p^2)^{3d/2-5}}{(4\pi)^{3d/2}}G^2(d, 1, 1) G(d, 1, 2-d/2), \\
& \hspace{-0.5cm} J_{\text{B3}}(d,\vec p, 1, 0, 1, 1, 1, 1, 1, 0)=\MasterThreeD=\frac{(p^2)^{3d/2-6}}{(4\pi)^{3d/2}}G(d, 1, 1) G(d, 1, 1, 1, 1, 2-d/2).
\ea
\es
\item The Non-planar (N3) master integral basis:
\bs
\ba
& \hspace{-0.5cm} J_{\text{N3}}(d,\vec p, 0, 0, 1, 0, 0, 1, 1, 1)=\MasterThreeA=\frac{(p^2)^{3d/2-4}}{(4\pi)^{3d/2}}G(d, 1, 1) G(d, 1, 2-d/2) G(d, 1, 3-d), \\[-0.5cm]
& \hspace{-0.5cm} J_{\text{N3}}(d,\vec p, 0, 1, 0, 1, 0, 1, 1, 1)=\MasterThreeB=\frac{(p^2)^{3d/2-5}}{(4\pi)^{3d/2}}G^2(d, 1, 1) G(d, 1, 4-d), \\
& \hspace{-0.5cm} J_{\text{N3}}(d,\vec p, 0, 1, 1, 1, 0, 1, 0, 1)=\MasterThreeC=\frac{(p^2)^{3d/2-5}}{(4\pi)^{3d/2}}G^2(d, 1, 1) G(d, 1, 2-d/2), \\
& \hspace{-0.5cm} J_{\text{N3}}(d,\vec p, 1, 0, 1, 1, 0, 1, 1, 1)=\MasterThreeD=\frac{(p^2)^{3d/2-6}}{(4\pi)^{3d/2}}G(d, 1, 1) G(d, 1, 1, 1, 1, 2-d/2), \\
& \hspace{-0.5cm}J_{\text{N3}}(d,\vec p, 1, 1, 1, 1, 1, 1, 1, 1)=\MasterThreeF=\frac{(p^2)^{3d/2-8}}{(4\pi)^{3d/2}} G_{\text{N3}}(d, 1, 1, 1, 1, 1, 1, 1, 1).
\ea
\es
\end{itemize}
We therefore see that these master integrals can be expressed via the one-loop integral $G(d,\al,\beta)$ (they are said to be primitively one-loop) and two non-trivial integrals: the two-loop coefficient function $G(d, 1, 1, 1, 1, 2-d/2)$ and the three-loop $G_{\text{N3}}(d, 1, 1, 1, 1, 1, 1, 1, 1)$, that are not reducible to simpler integrals. 

Let us first focus on $G_{\text{N3}}(d, 1, 1, 1, 1, 1, 1, 1, 1)$ that is the only truely three-loop integral entering the above basis. The result of this integral in $d=4-2\eps$ can be found, \eg, in the work \cite{Baikov:2010} and reads
\be
G_{\text{N3}}(4-2\eps, 1, 1, 1, 1, 1, 1, 1, 1)= e^{-3 \eps \gamma_E} \Big[ 20 \zeta_5+\Ord(\eps) \Big] \, ,
\label{eq:GNP(1,1,1,1,1,1,1,1)}
\ee
where $\zeta_n$ is the Riemann zeta function. 

We now focus on the second non-trivial integral, $G(d, 1, 1, 1, 1, 2-d/2)$, that contributes to many of our diagrams. It corresponds to $G(d, 1, 1, 1, 1, \eps)$ in $d=4-2\eps$, a two-loop diamond integral with a non-integer index on the central line. A more general integral, $G(d,1,1,1,1,\alpha)$, has been evaluated exactly in \cite{Kazakov:1983,Kotikov:1996}. From \cite{Kotikov:1996} its expression is given by
\ba
\label{eq:multiloop:G(1111a)}
G(d,1,1,1,1,\al) & =
-2 \Gamma(\lambda)\Gamma(\lambda-\al) \Gamma(1-2\lambda+\al) \\
&\times \left [ \frac{\Gamma(\lambda)}{\Gamma(2\lambda)\Gamma(3\lambda-\al-1)} 
\sum_{n=0}^{\infty} \frac{\Gamma(n+2\lambda)\Gamma(n+1)}{n! \Gamma(n+1+\al)} \frac{1}{n+1-\lambda+\al}
+\frac{\pi \cot \pi (2\lambda-\al)}{\Gamma(2\lambda)} \right ],
\nonum
\ea
where $\lambda=(d-2)/2=(1-2\eps)/2$. Note that \eqref{eq:multiloop:G(1111a)} may be written with a generalized hypergeometric function ${}_3F_2$ of argument $1$, since
\be
\setlength\arraycolsep{1pt}
{}_3 F_2\left(\begin{matrix} 1 ~~~ \alpha -\lambda +1 ~~~ 2 \lambda \\ ~~ \alpha +1 ~~~ \alpha -\lambda +2 ~~ \end{matrix}\bigg|~1\right)=
\frac{(\alpha -\lambda +1) \Gamma(\alpha+1)}{\Gamma (2 \lambda)} \sum_{n=0}^{\infty } \frac{\Gamma(n+2\lambda)\Gamma (n+1)}{n! \Gamma (n+\alpha +1)}\frac{1}{n+1-\lambda+\alpha}.
\label{res:I(al):Kotikov}
\ee
We can then set $\al=\eps$ in (\ref{res:I(al):Kotikov}) and expand it in series using, \eg, the \textsc{Mathematica} package \textsc{HypExp} \cite{Huber:2006,Huber:2008}. This yields
\ba
& G(4-2\eps,1,1,1,1,\eps) =e^{-2 \eps \gamma_E}\bigg[\frac{1}{3 \eps^2}+ \frac{5}{3 \eps} +\frac{17-\zeta_2}{3}+\frac{\eps}{9}\left(123-15 \zeta_2+28 \zeta_3\right)+ \Ord(\eps^2) \bigg].
\label{eq:G(1,1,1,1,eps)}
\ea
The non-trivial master integrals series expansions \eqref{eq:GNP(1,1,1,1,1,1,1,1)} and \eqref{eq:G(1,1,1,1,eps)} have also been checked numerically using sector decomposition Monte-Carlo technique with the \textsc{Mathematica} package \textsc{FIESTA} \cite{Smirnov:2009,Smirnov:2011,Smirnov:2014}.

\begin{figure}[h!]
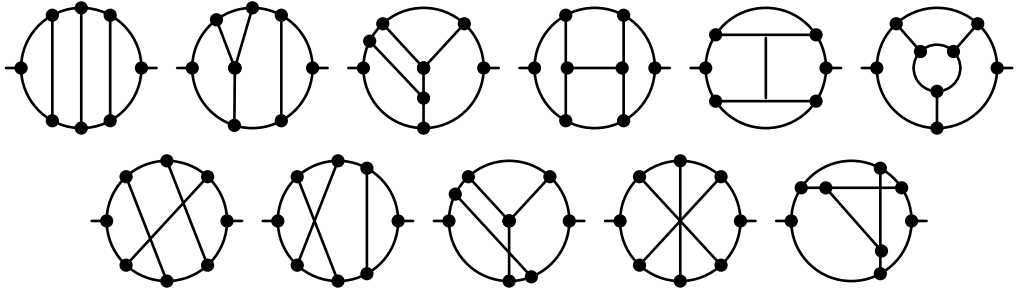

\centering
\GraphTolopogyLfour \GraphTolopogyBfour \GraphTolopogyFfour \GraphTolopogyHfour \GraphTolopogyEfour \GraphTolopogyDfour \\
\GraphTolopogyNfour \GraphTolopogyJfour \GraphTolopogyCfour \GraphTolopogyMfour \GraphTolopogyGfour \\
\caption{Propagator-type topologies at four loops.}
\label{fig:topologies}
\end{figure}

At four loops, there is a total of 11 topologies, 6 planar and 5 non-planar, see figure \ref{fig:topologies}

Similarly to the three-loop case, after IBP-reduction of the maximum topologies, most of the 39 obtained masters can be partially computed with techniques of massless integrals \cite{Kotikov:2018wxe}. We are left with a total of $16$ non-trivial integrals to compute. Eight of them are primitively two or three loop, \ie, are of the topologies \eqref{eq:multiloopdiamond} or \eqref{def:three-loop:2-pt} with generalized indices and can be found in \cite{Kazakov:1983,Kotikov:1996}. The remaining eight non-trivial four-loop masters can be found in \cite{Baikov:2010}. Finally, all master integrals, including the previous results from one to three loop, should be expanded up to transcendental weight five, \ie, series terms including $\zeta_5$ coefficient. We double-checked masters numerically using the sector decomposition tool \textsc{Fiesta} \cite{Smirnov:2009,Smirnov:2011,Smirnov:2014,Smirnov:2015mct,Smirnov:2021rhf} with the Monte-Carlo integrator \textsc{Vegas} from the \textsc{Cuba} library \cite{Hahn:2004fe,Hahn:2014fua}.

\end{appendix}

\newpage

\small

\bibliographystyle{unsrt.bst}
\bibliography{main}

\end{fmffile}
\end{document}

%% file: texcode/packages.tex
\usepackage[utf8]{inputenc}
\usepackage[T1]{fontenc}

\usepackage[colorlinks]{hyperref}
\usepackage{
color,
colortbl,
titlesec,
caption,
subcaption,
listings,
appendix,
titlesec,
enumitem
}

\usepackage[makeroom]{cancel}
\usepackage{
slashed,
mathtools,
centernot,
empheq,
array
}

\usepackage{
feynmp-auto,
graphics,
graphicx,
tikz,
pgfplots,
pdfpages
}

%% file: texcode/packages_settings.tex
\allowdisplaybreaks

\usetikzlibrary{patterns}
\usetikzlibrary{calc}
\usetikzlibrary{tikzmark}

\definecolor{mygray}{gray}{0.8}

%% file: graphs/graphs_memeft.tex
\newcommand{\flexuronprop}{
\parbox[c][10mm][c]{25mm}{
\centering
\begin{fmfgraph*}(15,10)
\fmfleft{in}
\fmflabel{$\al$}{in}
\fmfright{out}
\fmflabel{$\beta$}{out}
\fmf{plain,label=$\vec p$}{in,out}
\end{fmfgraph*}
}
}

\newcommand{\effprop}{
\parbox[c][10mm][c]{30mm}{
\centering
\begin{fmfgraph*}(15,10)
\fmfleft{in}
\fmflabel{$ab$}{in}
\fmfright{out}
\fmflabel{$cd$}{out}
\fmf{dashes,label=$\vec p$}{in,out}
\end{fmfgraph*}
}
}

\newcommand{\fourpointvertex}{
\parbox[c][20mm][c]{35mm}{
\centering
\begin{fmfgraph*}(15,10)
\fmfleft{i1,i2}
\fmfright{o1,o2}
\fmflabel{$\beta~\vec p_2$}{i1}
\fmflabel{$\al~\vec p_1$}{i2}
\fmflabel{$\gamma~\vec p_3$}{o1}
\fmflabel{$\delta~\vec p_4$}{o2}
\fmf{plain}{i1,v1,i2}
\fmf{plain}{o1,v2,o2}
\fmf{plain}{i1,v1}
\fmf{plain}{i2,v1}
\fmf{plain}{o1,v2}
\fmf{plain}{o2,v2}
\fmf{dashes,label=$\vec p$}{v1,v2}
\fmfdot{v1,v2}
\end{fmfgraph*}
}
}

\newcommand{\threepointvertex}{
\parbox[c][20mm][c]{35mm}{
\centering
\begin{fmfgraph*}(12,10)
\fmfleft{i1,i2}
\fmfright{o}
\fmflabel{$\beta~\vec p_2$}{i1}
\fmflabel{$\al~\vec p_1$}{i2}
\fmflabel{$ab~\vec p_3$}{o}
\fmf{plain}{i1,v}
\fmf{plain}{i2,v}
\fmf{dashes}{v,o}
\fmfdot{v}
\end{fmfgraph*}
}
}

\newcommand{\GraphSigmaOne}{
\parbox[c][30mm][c]{40mm}{
\centering
\begin{fmfgraph*}(25,25)
\fmfleft{in}
\fmflabel{$\al$}{in}
\fmfright{out}
\fmflabel{$\beta$}{out}
\fmf{plain,label=$\vec p$}{in,vi}
\fmf{plain,tension=0.2,label=$\vec k$}{vi,vo}
\fmf{dashes,left,tension=0.2,label=$\vec p -\vec k$}{vi,vo}
\fmf{plain}{vo,out}
\fmfdot{vi,vo}
\end{fmfgraph*}
}
}

\newcommand{\GraphVOne}{
\parbox[c][30mm][c]{40mm}{
\centering
\begin{fmfgraph*}(30,15)
\fmfleft{i1,i2}
\fmfright{o1,o2}
\fmflabel{$\beta~\vec p_2$}{i1}
\fmflabel{$\al~\vec p_1$}{i2}
\fmflabel{$\gamma~\vec p_3$}{o1}
\fmflabel{$\delta~\vec p_4$}{o2}
\fmf{plain,tension=0.7}{i1,v1,i2}
\fmf{plain,tension=0.7}{o1,v2,o2}
\fmf{plain}{i1,v1}
\fmf{plain}{i2,v1}
\fmf{plain}{o1,v2}
\fmf{plain}{o2,v2}
\fmf{dashes,label=$\vec p$}{v1,v3}
\fmf{dashes,label=$\vec p$}{v4,v2}
\fmf{plain,left,tension=0.2,label=$\vec p - \vec k$}{v3,v4}
\fmf{plain,right,tension=0.2,label=$\vec k$}{v3,v4}
\fmfdot{v1,v2,v3,v4}
\end{fmfgraph*}
}
}

\newcommand{\GraphSigmaTwoA}{
\parbox[c][30mm][c]{45mm}{
\centering
\begin{fmfgraph*}(35,30)
\fmfleft{i1}
\fmflabel{$\al$}{i1}
\fmfright{o1}
\fmflabel{$\beta$}{o1}
\fmf{plain,label=$\vec p$}{i1,i2}
\fmf{plain,tension=1.4}{i2,i3}
\fmf{plain}{i3,i4}
\fmf{plain,tension=0.4,label=$\vec p- \vec k_2$}{i4,o4}
\fmf{plain}{o4,o3}
\fmf{plain,tension=1.4}{o3,o2}
\fmf{plain}{o2,o1}
\fmffreeze
\fmf{phantom,left,tension=0.1,tag=2}{i2,o2}
\fmf{phantom,left,tension=0.1,tag=3}{i3,o3}
\fmf{phantom,left,tension=0.1,tag=4}{i4,o4}
\fmfdot{i3,o3}
\fmfposition
\fmfipath{p[]}
\fmfiset{p2}{vpath2(__i2,__o2)}
\fmfiset{p3}{vpath3(__i3,__o3)}
\fmfiset{p4}{vpath4(__i4,__o4)}
\fmfi{dashes}{subpath (0,3length(p3)/8) of p3}
\fmfi{dashes,label=$\vec k_2$}{subpath (5length(p3)/8,length(p3)) of p3}
\fmfi{plain}{point 3length(p3)/8 of p3 .. point length(p2)/2 of p2 .. point 5length(p3)/8 of p3}
\fmfi{plain}{point 5length(p3)/8 of p3 .. point length(p4)/2 of p4 .. point 3length(p3)/8 of p3}
\myvert{point 3length(p3)/8 of p3}
\myvert{point 5length(p3)/8 of p3}
\end{fmfgraph*} \\
\vspace*{-1cm}
}
}

\newcommand{\GraphSigmaTwoB}{
\parbox[c][30mm][c]{45mm}{
\centering
\begin{fmfgraph*}(35,30)
\fmfleft{i}
\fmfright{o}
\fmflabel{$\al$}{i}
\fmflabel{$\beta$}{o}
\fmf{plain,label=$\vec p$}{i,v1}
\fmf{plain}{o,v5}
\fmfn{phantom}{v}{5}
\fmffreeze
\fmf{plain,label=$\vec k_2$}{v1,v2}
\fmf{plain}{v2,v4}
\fmf{plain}{v4,v5}
\fmf{dashes,left,label=$\vec p-\vec k_2$}{v1,v5}
\fmf{dashes,left}{v2,v4}
\fmfdot{v1,v2,v4,v5}
\end{fmfgraph*} \\
\vspace*{-1cm}
}
}

\newcommand{\GraphSigmaTwoC}{
\parbox[c][30mm][c]{45mm}{
\centering
\begin{fmfgraph*}(35,30)
\fmfleft{i1}
\fmflabel{$\al$}{i1}
\fmfright{o1}
\fmflabel{$\beta$}{o1}
\fmf{plain,tension=1.7,label=$\vec p$}{i1,i2}
\fmf{plain}{i2,i3}
\fmf{plain,label=\phantom{a}$\vec k_{21}$,label.side=right}{i3,o3}
\fmf{plain,label=$\vec k_2$,label.side=left}{o3,o2}
\fmf{plain,tension=1.7}{o2,o1}
\fmffreeze
\fmf{dashes,left,tension=0.1,label=$\vec k_1$,label.side=left}{i2,o3}
\fmf{dashes,right,tension=0.1,label.side=right}{i3,o2}
\fmfdot{i2,i3,o2,o3}
\end{fmfgraph*}
}
}

\newcommand{\GraphVTwoA}{
\parbox[c][30mm][c]{50mm}{
\centering
\begin{fmfgraph*}(45,20)
\fmfleft{i1,i2}
\fmfright{o1,o2}
\fmflabel{$\beta~\vec p_2$}{i1}
\fmflabel{$\al~\vec p_1$}{i2}
\fmflabel{$\gamma~\vec p_3$}{o1}
\fmflabel{$\delta~\vec p_4$}{o2}
\fmf{plain,tension=0.7}{i1,vi,i2}
\fmf{plain,tension=0.7}{o1,vo,o2}
\fmf{plain}{i1,vi}
\fmf{plain}{i2,vi}
\fmf{plain}{o1,vo}
\fmf{plain}{o2,vo}
\fmffreeze
\fmfforce{(0.2w,0.5h)}{vi}
\fmfforce{(0.8w,0.5h)}{vo}
\fmfforce{(0.35w,0.5h)}{v1}
\fmfforce{(0.65w,0.5h)}{v2}
\fmfforce{(.5w,0.5h)}{v}
\fmffreeze
\fmf{dashes,l.d=0.1cm,label=$\vec p$}{vi,v1}
\fmf{dashes}{v2,vo}
\fmf{phantom,left,tension=0.2,tag=1}{v1,v2}
\fmf{phantom,right,tension=0.2,tag=2}{v1,v2}
\fmf{phantom,tension=0.1,tag=3}{v1,v2}
\fmfdot{v1,v2,vi,vo}
\fmfposition
\fmfipath{p[]}
\fmfiset{p1}{vpath1(__v1,__v2)}
\fmfiset{p2}{vpath2(__v2,__v1)}
\fmfiset{p3}{vpath3(__v1,__v2)}
\fmfi{plain,l.d=0.1cm,label=$\vec p-\vec k_2$}{subpath (0,length(p1)) of p1}
\fmfi{plain}{subpath (0,length(p2)/4) of p2}
\fmfi{plain}{subpath (length(p2)/4,3length(p2)/4) of p2}
\fmfi{plain,l.d=0.1cm,label=$\vec k_2$}{subpath (3length(p2)/4,length(p2)) of p2}
\fmfi{dashes}{point length(p2)/4 of p2 .. point length(p3)/2 of p3 .. point 3length(p2)/4 of p2}
\myvert{point length(p2)/4 of p2}
\myvert{point 3length(p2)/4 of p2}
\end{fmfgraph*} 
}
}

\newcommand{\GraphVTwoB}{
\parbox[c][30mm][c]{50mm}{
\centering
\begin{fmfgraph*}(45,20)
\fmfleft{i1,i2}
\fmfright{o1,o2}
\fmflabel{$\beta~\vec p_2$}{i1}
\fmflabel{$\al~\vec p_1$}{i2}
\fmflabel{$\gamma~\vec p_3$}{o1}
\fmflabel{$\delta~\vec p_4$}{o2}
\fmf{plain}{i1,vi,i2}
\fmf{plain}{o1,vo,o2}
\fmf{plain}{i1,vi}
\fmf{plain}{i2,vi}
\fmf{plain}{o1,vo}
\fmf{plain}{o2,vo}
\fmffreeze
\fmfforce{(0.2w,0.5h)}{vi}
\fmfforce{(0.8w,0.5h)}{vo}
\fmfforce{(0.35w,0.5h)}{v1}
\fmfforce{(0.65w,0.5h)}{v2}
\fmfforce{(.5w,0.5h)}{v}
\fmffreeze
\fmf{dashes,l.d=0.1cm,label=$\vec p$}{vi,v1}
\fmf{dashes}{v2,vo}
\fmf{phantom,tension=0.2,tag=3}{v1,v2}
\fmf{phantom,left,tension=0.5,tag=1}{v1,v2}
\fmf{phantom,right,tension=0.5,tag=2}{v1,v2}
\fmfdot{v1,v2,vi,vo}
\fmfposition
\fmfipath{p[]}
\fmfiset{p1}{vpath1(__v1,__v2)}
\fmfiset{p2}{vpath2(__v2,__v1)}
\fmfi{plain,l.d=0.1cm,label=$\vec k_1$}{subpath (0,length(p1)/2) of p1}
\fmfi{plain,l.d=0.1cm,label=$\vec k_2$}{subpath (length(p1)/2,length(p1)) of p1}
\fmfi{plain}{subpath (0,length(p2)/2) of p2}
\fmfi{plain}{subpath (length(p2)/2,length(p2)) of p2}
\fmfi{dashes,label=$\vec k_{12}$,label.side=left,l.d=0.01w}{point length(p1)/2 of p1 -- point length(p2)/2 of p2}
\myvert{point length(p1)/2 of p1}
\myvert{point length(p2)/2 of p2}
\end{fmfgraph*}
}
}

\newcommand{\GraphSigmaThreeA}{
\parbox[c][30mm][c]{40mm}{
\centering
\vspace*{1cm}
\begin{fmfgraph*}(40,25)
\fmfleft{i}
\fmfright{o}
\fmf{plain}{i,v1}
\fmf{plain}{o,v4}
\fmfn{plain}{v}{4}
\fmf{dashes,left}{w1,w2}
\fmffreeze
\fmf{phantom,left,tag=1}{v1,v4}
\fmfposition
\fmfipath{p[]}
\fmfiset{p1}{vpath1(__v1,__v4)}
\fmfi{dashes}{subpath (0,1length(p1)/3) of p1}
\fmfi{plain}{point 1length(p1)/3 of p1 .. point 2length(p1)/3 of p1 .. point 1length(p1)/3 of p1}
\fmfi{dashes}{subpath (2length(p1)/3,3length(p1)/3) of p1}
\fmfdot{v1,v4}
\myvert{point 1length(p1)/3 of p1}
\myvert{point 2length(p1)/3 of p1}
\fmfdot{w1}
\fmfdot{w2}
\fmfforce{(0.4w,0.73h)}{w1}
\fmfforce{(0.6w,0.73h)}{w2}
\end{fmfgraph*}   \\
\vspace*{-1cm}
}
}

\newcommand{\GraphSigmaThreeB}{
\parbox[c][30mm][c]{40mm}{
\centering
\vspace*{1cm}
\begin{fmfgraph*}(40,25)
\fmfleft{i}
\fmfright{o}
\fmf{plain}{i,v1}
\fmf{plain}{o,v6}
\fmfn{plain}{v}{6}
\fmffreeze
\fmf{phantom,left,tag=1}{v2,v5}
\fmf{dashes,left}{v1,v6}
\fmfposition
\fmfipath{p[]}
\fmfiset{p1}{vpath1(__v2,__v5)}
\fmfi{dashes}{subpath (0,1length(p1)/3) of p1}
\fmfi{plain}{point 1length(p1)/3 of p1 .. point 2length(p1)/3 of p1 .. point 1length(p1)/3 of p1}
\fmfi{dashes}{subpath (2length(p1)/3,3length(p1)/3) of p1}
\fmfdot{v1,v2,v5,v6}
\myvert{point 1length(p1)/3 of p1}
\myvert{point 2length(p1)/3 of p1}
\end{fmfgraph*} \\
\vspace*{-1cm}

}
}

\newcommand{\GraphSigmaThreeC}{
\parbox[c][30mm][c]{40mm}{
\centering
\vspace*{1cm}
\begin{fmfgraph*}(40,25)
\fmfleft{i}
\fmfright{o}
\fmf{plain}{i,v1}
\fmf{plain}{o,v7}
\fmfn{plain}{v}{7}
\fmffreeze
\fmf{dashes,left}{v1,v7}
\fmf{dashes,left}{v2,v6}
\fmf{dashes,left}{v3,v5}
\fmfdot{v1,v2,v3,v5,v6,v7,v7}
\end{fmfgraph*} \\
\vspace*{-1cm}
}
}

\newcommand{\GraphSigmaThreeD}{
\parbox[c][30mm][c]{40mm}{
\centering
\vspace*{1cm}
\begin{fmfgraph*}(40,25)
\fmfleft{i}
\fmfright{o}
\fmf{plain}{i,v1}
\fmf{plain}{o,v7}
\fmfn{plain}{v}{7}
\fmffreeze
\fmf{phantom,left,tag=1}{v1,v7}
\fmfposition
\fmfipath{p[]}
\fmfiset{p1}{vpath1(__v1,__v7)}
\fmfi{dashes}{subpath (0,1length(p1)/5) of p1}
\fmfi{plain}{point 1length(p1)/5 of p1 .. point 2length(p1)/5 of p1 .. point 1length(p1)/5 of p1}
\fmfi{dashes}{subpath (2length(p1)/5,3length(p1)/5) of p1}
\fmfi{plain}{point 3length(p1)/5 of p1 .. point 4length(p1)/5 of p1 .. point 3length(p1)/5 of p1}
\fmfi{dashes}{subpath (4length(p1)/5,5length(p1)/5) of p1}
\fmfdot{v1,v7}
\myvert{point 1length(p1)/5 of p1}
\myvert{point 2length(p1)/5 of p1}
\myvert{point 3length(p1)/5 of p1}
\myvert{point 4length(p1)/5 of p1}
\end{fmfgraph*} \\
\vspace*{-1cm}
}
}

\newcommand{\GraphSigmaThreeE}{
\parbox[c][30mm][c]{40mm}{
\centering
\vspace*{1cm}
\begin{fmfgraph*}(40,25)
\fmfleft{i}
\fmfright{o}
\fmf{plain}{i,v1}
\fmf{plain}{o,v8}
\fmfn{plain}{v}{8}
\fmffreeze
\fmf{dashes,left}{v1,v8}
\fmf{dashes,left}{v2,v4}
\fmf{dashes,left}{v5,v7}
\fmfdot{v1,v2,v4,v5,v7,v8}
\end{fmfgraph*}\\
\vspace*{-1cm}
}
}

\newcommand{\GraphSigmaThreeF}{
\parbox[c][30mm][c]{40mm}{
\centering
\vspace*{1cm}
\begin{fmfgraph*}(40,25)
\fmfleft{i}
\fmfright{o}
\fmf{plain}{i,v1}
\fmf{plain}{o,v4}
\fmfn{plain}{v}{4}
\fmf{dashes}{w1,w2}
\fmffreeze
\fmf{phantom,left,tag=1}{v1,v4}
\fmfposition
\fmfipath{p[]}
\fmfiset{p1}{vpath1(__v1,__v4)}
\fmfi{dashes}{subpath (0,1length(p1)/3) of p1}
\fmfi{plain}{point 1length(p1)/3 of p1 .. point 2length(p1)/3 of p1 .. point 1length(p1)/3 of p1}
\fmfi{dashes}{subpath (2length(p1)/3,3length(p1)/3) of p1}
\fmfdot{v1,v4}
\myvert{point 1length(p1)/3 of p1}
\myvert{point 2length(p1)/3 of p1}
\fmfdot{w1}
\fmfdot{w2}
\fmfforce{(0.5w,1.15h)}{w1}
\fmfforce{(0.5w,0.68h)}{w2}
\end{fmfgraph*}   \\
\vspace*{-1cm}
}
}

\newcommand{\GraphSigmaThreeG}{
\parbox[c][30mm][c]{40mm}{
\centering
\vspace*{1cm}
\begin{fmfgraph*}(40,25)
\fmfleft{i}
\fmfright{o}
\fmf{plain}{i,v1}
\fmf{plain}{o,v6}
\fmfn{plain}{v}{6}
\fmffreeze
\fmf{dashes,left}{v1,v6}
\fmf{dashes,left}{v2,v4}
\fmf{dashes,right}{v3,v5}
\fmfdotn{v}{6}
\end{fmfgraph*} \\
\vspace*{-0.5cm}
}
}

\newcommand{\GraphSigmaThreeH}{
\parbox[c][30mm][c]{40mm}{
\centering
\vspace*{0.5cm}
\begin{fmfgraph*}(40,20)
\fmfleft{i}
\fmfright{o}
\fmf{plain}{i,v1}
\fmf{plain}{o,v7}
\fmfn{plain}{v}{7}
\fmffreeze
\fmf{dashes,left}{v1,v6}
\fmf{dashes,right}{v2,v7}
\fmf{dashes,left}{v3,v5}
\fmfdot{v1,v2,v3,v5,v6,v7}
\end{fmfgraph*}
}
}

\newcommand{\GraphSigmaThreeI}{
\parbox[c][30mm][c]{40mm}{
\centering
\vspace*{1cm}
\begin{fmfgraph*}(40,25)
\fmfleft{i}
\fmfright{o}
\fmf{plain}{i,v1}
\fmf{plain}{o,v7}
\fmfn{plain}{v}{7}
\fmffreeze
\fmf{dashes,right}{v1,v3}
\fmf{dashes,left}{v2,v7}
\fmf{dashes,left}{v4,v6}
\fmfdot{v1,v2,v3,v4,v6,v7}
\end{fmfgraph*} \\
\vspace*{-0.5cm}
}
}

\newcommand{\GraphSigmaThreeJ}{
\parbox[c][30mm][c]{40mm}{
\centering
\vspace*{1cm}
\begin{fmfgraph*}(40,25)
\fmfleft{i}
\fmfright{o}
\fmf{plain}{i,v1}
\fmf{plain}{o,v7}
\fmfn{plain}{v}{7}
\fmffreeze
\fmf{phantom,left,tag=1}{v3,v7}
\fmf{dashes,right}{v1,v5}
\fmfposition
\fmfipath{p[]}
\fmfiset{p1}{vpath1(__v3,__v7)}
\fmfi{dashes}{subpath (0,1length(p1)/3) of p1}
\fmfi{plain}{point 1length(p1)/3 of p1 .. point 2length(p1)/3 of p1 .. point 1length(p1)/3 of p1}
\fmfi{dashes}{subpath (2length(p1)/3,3length(p1)/3) of p1}
\fmfdot{v1,v3,v5,v7}
\myvert{point 1length(p1)/3 of p1}
\myvert{point 2length(p1)/3 of p1}
\end{fmfgraph*} 
}
}

\newcommand{\GraphSigmaThreeK}{
\parbox[c][30mm][c]{40mm}{
\centering
\vspace*{1cm}
\begin{fmfgraph*}(40,25)
\fmfleft{i}
\fmfright{o}
\fmf{plain}{i,v1}
\fmf{plain}{o,v5}
\fmfn{plain}{v}{5}
\fmffreeze
\fmf{phantom,left,tag=1}{v1,v5}
\fmf{phantom,left,tag=2}{v2,v4}
\fmf{phantom,left,tag=3}{v3,v4}
\fmfposition
\fmfipath{p[]}
\fmfiset{p1}{vpath1(__v1,__v5)}
\fmfiset{p2}{vpath2(__v2,__v4)}
\fmfiset{p3}{vpath3(__v3,__v4)}
\fmfi{dashes}{subpath (0,1.1length(p1)/3) of p1}
\fmfi{plain}{point 1.1length(p1)/3 of p1 .. point 1.9length(p1)/3 of p1 .. point 1.1length(p1)/3 of p1}
\fmfi{dashes}{subpath (1.9length(p1)/3,3length(p1)/3) of p1}
\fmfi{dashes}{point 1length(p2)/2 of p2 .. point 0length(p3) of p3}
\fmfdot{v1,v3,v5}
\myvert{point 1.1length(p1)/3 of p1}
\myvert{point 1.9length(p1)/3 of p1}
\myvert{point 1length(p2)/2 of p2}
\end{fmfgraph*}  \\
\vspace*{-1cm}
}
}

\newcommand{\GraphSigmaThreeL}{
\parbox[c][30mm][c]{40mm}{
\centering
\vspace*{1cm}
\begin{fmfgraph*}(40,25)
\fmfleft{i}
\fmfright{o}
\fmf{plain}{i,v1}
\fmf{plain}{o,v6}
\fmfn{plain}{v}{6}
\fmffreeze
\fmf{dashes,left}{v1,v5}
\fmf{dashes,left}{v2,v4}
\fmf{dashes,right}{v3,v6}
\fmfdotn{v}{6}
\end{fmfgraph*} 
}
}

\newcommand{\GraphSigmaThreeM}{
\parbox[c][30mm][c]{40mm}{
\centering
\vspace*{1cm}
\begin{fmfgraph*}(40,25)
\fmfleft{i}
\fmfright{o}
\fmf{plain}{i,v1}
\fmf{plain}{o,v6}
\fmfn{plain}{v}{6}
\fmffreeze
\fmf{dashes,right}{v1,v3}
\fmf{dashes,left}{v2,v5}
\fmf{dashes,right}{v4,v6}
\fmfdotn{v}{6}
\end{fmfgraph*}
}
}

\newcommand{\GraphSigmaThreeN}{
\parbox[c][30mm][c]{40mm}{
\centering
\vspace*{1cm}
\begin{fmfgraph*}(40,25)
\fmfleft{i}
\fmfright{o}
\fmf{plain}{i,v1}
\fmf{plain}{o,v7}
\fmfn{plain}{v}{7}
\fmffreeze
\fmf{phantom,left,tag=1}{v1,v7}
\fmf{dashes,left}{v3,v5}
\fmfposition
\fmfipath{p[]}
\fmfiset{p1}{vpath1(__v1,__v7)}
\fmfi{dashes}{subpath (0,2length(p1)/5) of p1}
\fmfi{plain}{point 2length(p1)/5 of p1 .. point 3length(p1)/5 of p1 .. point 2length(p1)/5 of p1}
\fmfi{dashes}{subpath (3length(p1)/5,5length(p1)/5) of p1}
\fmfdot{v1,v3,v5,v7}
\myvert{point 2length(p1)/5 of p1}
\myvert{point 3length(p1)/5 of p1}
\end{fmfgraph*} \\
\vspace*{-1cm}
}
}

\newcommand{\GraphSigmaThreeO}{
\parbox[c][30mm][c]{40mm}{
\centering
\vspace*{1cm}
\begin{fmfgraph*}(40,25)
\fmfleft{i}
\fmfright{o}
\fmf{plain}{i,v1}
\fmf{plain}{o,v6}
\fmfn{plain}{v}{6}
\fmffreeze
\fmf{dashes,left}{v1,v4}
\fmf{dashes,right}{v2,v5}
\fmf{dashes,left}{v3,v6}            
\fmfdotn{v}{6}
\end{fmfgraph*}
}
}

\newcommand{\GraphVThreeA}{
\parbox[c][40mm][c]{50mm}{
\centering
\begin{fmfgraph*}(50,25)
\fmfleft{i1,i2}
\fmfright{o1,o2}
\fmf{plain}{i1,v1}
\fmf{plain}{i2,v1}
\fmf{plain}{o1,v6}
\fmf{plain}{o2,v6}
\fmfn{phantom}{v}{6}            
\fmffreeze
\fmf{dashes}{v1,v2}
\fmf{dashes}{v5,v6}
\fmf{plain,left}{v2,v5}
\fmf{plain,right}{v2,v5}
\fmfdotn{v}{6}
\fmf{plain,right}{v3,v4,v3}
\fmf{phantom,left=0.95,tag=1}{w1,w2}
\fmfforce{(0.345w,0.25h)}{w1}
\fmfforce{(0.655w,0.25h)}{w2}
\fmfdotn{w}{2}
\fmfposition
\fmfipath{p[]}
\fmfiset{p1}{vpath1(__w1,__w2)}
\fmfi{dashes}{subpath (0length(p1),0.34length(p1)) of p1}
\fmfi{dashes}{subpath (0.66length(p1),1length(p1)) of p1}
\end{fmfgraph*} \\
\vspace*{-1cm}
}
}

\newcommand{\GraphVThreeB}{
\parbox[c][40mm][c]{50mm}{
\centering
\begin{fmfgraph*}(50,25)
\fmfleft{i1,i2}
\fmfright{o1,o2}
\fmf{plain}{i1,v1}
\fmf{plain}{i2,v1}
\fmf{plain}{o1,v6}
\fmf{plain}{o2,v6}
\fmfn{phantom}{v}{6}
\fmffreeze
\fmf{dashes}{v1,v2}
\fmf{dashes}{v5,v6}
\fmf{plain,left}{v2,v5}
\fmf{plain,right}{v2,v5}
\fmfdot{v1,v2,v5,v6}
\fmf{dashes,right}{w1,w2}
\fmf{dashes,right}{w3,w4}
\fmfforce{(0.6w,0.16h)}{w1}
\fmfforce{(0.4w,0.16h)}{w2}
\fmfforce{(0.655w,0.25h)}{w3}
\fmfforce{(0.345w,0.25h)}{w4}
\fmfdotn{w}{4}
\end{fmfgraph*} \\
\vspace*{-1cm}
}
}

\newcommand{\GraphVThreeC}{
\parbox[c][40mm][c]{50mm}{
\centering
\begin{fmfgraph*}(50,25)
\fmfleft{i1,i2}
\fmfright{o1,o2}
\fmf{plain}{i1,v1}
\fmf{plain}{i2,v1}
\fmf{plain}{o1,v6}
\fmf{plain}{o2,v6}
\fmfn{phantom}{v}{6}
\fmffreeze
\fmf{dashes}{v1,v2}
\fmf{dashes}{v5,v6}
\fmf{plain,left}{v2,v5}
\fmf{plain,right,tag=1}{v2,v5}
\fmfdot{v1,v2,v5,v6}
\fmfposition
\fmfipath{p[]}
\fmfiset{p1}{vpath1(__v2,__v5)}
\myvert{point 1length(p1)/7 of p1}
\myvert{point 3length(p1)/7 of p1}
\myvert{point 4length(p1)/7 of p1}
\myvert{point 6length(p1)/7 of p1}
\fmf{dashes,left}{w1,w2}
\fmf{dashes,left}{w3,w4}
\fmfforce{(0.325w,0.31h)}{w1}
\fmfforce{(0.455w,0.115h)}{w2}
\fmfforce{(0.545w,0.115h)}{w3}
\fmfforce{(0.675w,0.31h)}{w4}
\end{fmfgraph*} \\
\vspace*{-1cm}
}
}

\newcommand{\GraphVThreeD}{
\parbox[c][40mm][c]{50mm}{
\centering
\begin{fmfgraph*}(50,25)
\fmfleft{i1,i2}
\fmfright{o1,o2}
\fmf{plain}{i1,v1}
\fmf{plain}{i2,v1}
\fmf{plain}{o1,v6}
\fmf{plain}{o2,v6}
\fmfn{phantom}{v}{6}
\fmffreeze
\fmf{dashes}{v1,v2}
\fmf{dashes}{v5,v6}
\fmf{plain,left}{v2,v5}
\fmf{plain,right}{v2,v5}
\fmfdot{v1,v2,v5,v6}
\fmf{dashes,right}{w1,w2}
\fmf{dashes,right}{w3,w4}
\fmfforce{(0.5w,0.1h)}{w1}
\fmfforce{(0.675w,0.31h)}{w2}
\fmfforce{(0.6w,0.16h)}{w3}
\fmfforce{(0.4w,0.16h)}{w4}
\fmfdotn{w}{4}
\end{fmfgraph*} \\
\vspace*{-1cm}
}
}

\newcommand{\GraphVThreeE}{
\parbox[c][40mm][c]{50mm}{
\centering
\begin{fmfgraph*}(50,25)
\fmfleft{i1,i2}
\fmfright{o1,o2}
\fmf{plain}{i1,v1}
\fmf{plain}{i2,v1}
\fmf{plain}{o1,v6}
\fmf{plain}{o2,v6}
\fmfn{phantom}{v}{6}
\fmffreeze
\fmf{dashes}{v1,v2}
\fmf{dashes}{v5,v6}
\fmf{plain,left}{v2,v5}
\fmf{plain,right}{v2,v5}
\fmfdot{v1,v2,v5,v6}
\fmf{plain,right}{v3,v4,v3}
\fmf{dashes}{w1,w2}
\fmf{dashes}{w3,w4}
\fmfforce{(0.5w,0.9h)}{w1}
\fmfforce{(0.5w,0.625h)}{w2}
\fmfforce{(0.5w,0.370h)}{w3}
\fmfforce{(0.5w,0.1h)}{w4}
\fmfdotn{w}{4}
\end{fmfgraph*} \\
\vspace*{-1cm}
}
}

\newcommand{\GraphVThreeF}{
\parbox[c][40mm][c]{50mm}{
\centering
\begin{fmfgraph*}(50,25)
\fmfleft{i1,i2}
\fmfright{o1,o2}
\fmf{plain}{i1,v1}
\fmf{plain}{i2,v1}
\fmf{plain}{o1,v6}
\fmf{plain}{o2,v6}
\fmfn{phantom}{v}{6}
\fmffreeze
\fmf{dashes}{v1,v2}
\fmf{dashes}{v5,v6}
\fmf{plain,left,tag=1}{v2,v5}
\fmf{plain,right,tag=2}{v2,v5}
\fmfdot{v1,v2,v5,v6}
\fmfposition
\fmfipath{p[]}
\fmfiset{p1}{vpath1(__v2,__v5)}
\fmfiset{p2}{vpath2(__v2,__v5)}
\fmfi{dashes}{point 1length(p1)/3 of p1 .. point 1length(p2)/3 of p2}
\myvert{point 1length(p1)/3 of p1}
\myvert{point 1length(p2)/3 of p2}
\fmf{dashes,left}{w1,w2}
\fmfforce{(0.545w,0.115h)}{w1}
\fmfforce{(0.675w,0.31h)}{w2}
\myvert{point 4length(p1)/7 of p2}
\myvert{point 6length(p1)/7 of p2}
\end{fmfgraph*} \\
\vspace*{-1cm}
}
}

\newcommand{\GraphVThreeG}{
\parbox[c][40mm][c]{50mm}{
\centering
\begin{fmfgraph*}(50,25)
\fmfleft{i1,i2}
\fmfright{o1,o2}
\fmf{plain}{i1,v1}
\fmf{plain}{i2,v1}
\fmf{plain}{o1,v6}
\fmf{plain}{o2,v6}
\fmfn{phantom}{v}{6}
\fmffreeze
\fmf{dashes}{v1,v2}
\fmf{dashes}{v5,v6}
\fmf{plain,left}{v2,v5}
\fmf{plain,right}{v2,v5}
\fmfdot{v1,v2,v5,v6}
\fmf{dashes,left}{w1,w2}
\fmf{dashes}{w3,w4}
\fmfforce{(0.4w,0.84h)}{w1}
\fmfforce{(0.6w,0.84h)}{w2}
\fmfforce{(0.5w,0.9h)}{w3}
\fmfforce{(0.5w,0.1h)}{w4}
\fmfdotn{w}{4}
\end{fmfgraph*} \\
\vspace*{-1cm}
}
}

\newcommand{\GraphVThreeH}{
\parbox[c][40mm][c]{50mm}{
\centering
\begin{fmfgraph*}(50,25)
\fmfleft{i1,i2}
\fmfright{o1,o2}
\fmf{plain}{i1,v1}
\fmf{plain}{i2,v1}
\fmf{plain}{o1,v6}
\fmf{plain}{o2,v6}
\fmfn{phantom}{v}{6}
\fmffreeze
\fmf{dashes}{v1,v2}
\fmf{dashes}{v5,v6}
\fmf{phantom,left,tag=1}{v2,v5}
\fmf{phantom,right,tag=2}{v2,v5}
\fmfdot{v1,v2,v5,v6}
\fmfposition
\fmfipath{p[]}
\fmfiset{p1}{vpath1(__v2,__v5)}
\fmfiset{p2}{vpath2(__v2,__v5)}
\fmfi{plain,left}{point 1length(p1)/3 of p1 .. point 1length(p2)/3 of p2}
\fmfi{plain,right}{point 2length(p1)/3 of p1 .. point 2length(p2)/3 of p2}
\fmfi{dashes}{subpath (1length(p1)/3,2length(p1)/3) of p1}
\fmfi{dashes}{subpath (1length(p2)/3,2length(p2)/3) of p2}
\fmfi{plain}{subpath (0length(p1)/3,1length(p1)/3) of p1}
\fmfi{plain}{subpath (2length(p1)/3,3length(p1)/3) of p1}
\fmfi{plain}{subpath (0length(p2)/3,1length(p2)/3) of p2}
\fmfi{plain}{subpath (2length(p2)/3,3length(p2)/3) of p2}
\myvert{point 1length(p1)/3 of p1}
\myvert{point 2length(p1)/3 of p1}
\myvert{point 1length(p2)/3 of p2}
\myvert{point 2length(p2)/3 of p2}
\end{fmfgraph*} \\
\vspace*{-1cm}
}
}

\newcommand{\GraphVThreeI}{
\parbox[c][40mm][c]{50mm}{
\centering
\begin{fmfgraph*}(50,25)
\fmfleft{i1,i2}
\fmfright{o1,o2}
\fmf{plain}{i1,v1}
\fmf{plain}{i2,v1}
\fmf{plain}{o1,v6}
\fmf{plain}{o2,v6}
\fmfn{phantom}{v}{6}
\fmffreeze
\fmf{dashes}{v1,v2}
\fmf{dashes}{v5,v6}
\fmf{plain,left,tag=1}{v2,v5}
\fmf{plain,right,tag=2}{v2,v5}
\fmfdot{v1,v2,v5,v6}
\fmfposition
\fmfipath{p[]}
\fmfiset{p1}{vpath1(__v2,__v5)}
\fmfiset{p2}{vpath2(__v2,__v5)}
\fmfi{dashes,left}{point 1length(p1)/3 of p1 .. point 1length(p2)/3 of p2}
\fmfi{dashes,right}{point 2length(p1)/3 of p1 .. point 2length(p2)/3 of p2}
\myvert{point 1length(p1)/3 of p1}
\myvert{point 2length(p1)/3 of p1}
\myvert{point 1length(p2)/3 of p2}
\myvert{point 2length(p2)/3 of p2}
\end{fmfgraph*} \\
\vspace*{-1cm}
}
}

\newcommand{\GraphVThreeJ}{
\parbox[c][40mm][c]{50mm}{
\centering
\begin{fmfgraph*}(50,25)
\fmfleft{i1,i2}
\fmfright{o1,o2}
\fmf{plain}{i1,v1}
\fmf{plain}{i2,v1}
\fmf{plain}{o1,v6}
\fmf{plain}{o2,v6}
\fmfn{phantom}{v}{6}
\fmffreeze
\fmf{dashes}{v1,v2}
\fmf{dashes}{v5,v6}
\fmf{plain,left}{v2,v5}
\fmf{plain,right}{v2,v5}
\fmfdot{v1,v2,v5,v6}
\fmffreeze
\fmf{dashes,right}{w1,w2}
\fmf{dashes,left}{w3,w4}
\fmfforce{(0.4w,0.84h)}{w1}
\fmfforce{(0.6w,0.84h)}{w2}
\fmfforce{(0.4w,0.16h)}{w3}
\fmfforce{(0.6w,0.16h)}{w4}
\fmfdotn{w}{4}
\end{fmfgraph*} \\
\vspace*{-1cm}
}
}

\newcommand{\GraphVThreeK}{
\parbox[c][40mm][c]{50mm}{
\centering
\begin{fmfgraph*}(50,25)
\fmfleft{i1,i2}
\fmfright{o1,o2}
\fmf{plain}{i1,v1}
\fmf{plain}{i2,v1}
\fmf{plain}{o1,v6}
\fmf{plain}{o2,v6}
\fmfn{phantom}{v}{6}
\fmffreeze
\fmf{dashes}{v1,v2}
\fmf{dashes}{v5,v6}
\fmf{plain,left,tag=1}{v2,v5}
\fmf{plain,right,tag=2}{v2,v5}
\fmfdot{v1,v2,v5,v6}
\fmfposition
\fmfipath{p[]}
\fmfiset{p1}{vpath1(__v2,__v5)}
\fmfiset{p2}{vpath2(__v2,__v5)}
\fmfi{dashes}{point 1length(p1)/4 of p1 .. point 3length(p2)/4 of p2}
\fmfi{dashes}{point 3length(p1)/4 of p1 .. point 1length(p2)/4 of p2}
\myvert{point 1length(p1)/4 of p1}
\myvert{point 3length(p1)/4 of p1}
\myvert{point 1length(p2)/4 of p2}
\myvert{point 3length(p2)/4 of p2}
\end{fmfgraph*} \\
\vspace*{-1cm}
}
}

%% file: graphs/graphs_masters.tex
\newcommand{\MasterThreeA}{
    \parbox[c][13mm][c]{15mm}{
        \centering
        \begin{fmfgraph*}(15,15)
            \fmfleft{i}
            \fmfright{o}
            \fmf{plain}{i,v1}
            \fmf{plain}{o,v3}
            \fmfn{phantom}{v}{3}
            \fmffreeze
            \fmfforce{(0.1w,0.5h)}{v1}
            \fmfforce{(0.9w,0.5h)}{v3}
            \fmf{plain,left}{v1,v3}
            \fmf{plain,right}{v1,v3}
            \fmf{plain,left=0.5}{v1,v3}
            \fmf{plain,right=0.5}{v1,v3}
            \fmfdot{v1,v3}
        \end{fmfgraph*}
    }
}

\newcommand{\MasterThreeB}{
    \parbox[c][13mm][c]{15mm}{
        \centering
        \begin{fmfgraph*}(15,15)
            \fmfleft{i}
            \fmfright{o}
            \fmf{plain}{i,v1}
            \fmf{plain}{o,v3}
            \fmfn{phantom}{v}{3}
            \fmffreeze
            \fmfforce{(0.1w,0.5h)}{v1}
            \fmfforce{(0.9w,0.5h)}{v3}
            \fmf{plain,left}{v1,v3}
            \fmf{plain,right}{v1,v3}
            \fmffreeze
            \fmfforce{(0.5w,0.1h)}{v4}
            \fmf{plain,left=0.5}{v1,v4}
            \fmf{plain,left=0.5}{v4,v3}
            \fmfdot{v1,v3,v4}
        \end{fmfgraph*}
    }
}

\newcommand{\MasterThreeC}{
    \parbox[c][13mm][c]{20mm}{
        \centering
        \begin{fmfgraph*}(20,15)
            \fmfleft{i}
            \fmfright{o}
            \fmf{plain}{i,v1}
            \fmf{plain}{o,v4}
            \fmfn{phantom}{v}{4}
            \fmffreeze
            \fmfforce{(0.1w,0.5h)}{v1}
            \fmfforce{(0.9w,0.5h)}{v4}
            \fmfn{plain}{v}{3}
            \fmf{plain,left}{v1,v3}
            \fmf{plain,right}{v1,v3}
            \fmf{plain,left}{v3,v4}
            \fmf{plain,right}{v3,v4}
            \fmfdot{v1,v3,v4}
        \end{fmfgraph*}
    }
}

\newcommand{\MasterThreeD}{
    \parbox[c][13mm][c]{15mm}{
        \centering
        \begin{fmfgraph*}(15,15)
            \fmfleft{i}
            \fmfright{o}
            \fmf{plain}{i,v1}
            \fmf{plain}{o,v3}
            \fmfn{phantom}{v}{3}
            \fmffreeze
            \fmfforce{(0.1w,0.5h)}{v1}
            \fmfforce{(0.9w,0.5h)}{v3}
            \fmf{plain,left}{v1,v3}
            \fmf{plain,right}{v1,v3}
            \fmffreeze
            \fmfforce{(0.5w,0.1h)}{v4}
            \fmfforce{(0.5w,0.9h)}{v5}
            \fmf{plain,left=0.5}{v4,v5}
            \fmf{plain,right=0.5}{v4,v5}
            \fmfdot{v1,v3,v4,v5}
        \end{fmfgraph*}
    }
}

\newcommand{\MasterThreeE}{
    \parbox[c][13mm][c]{25mm}{
        \centering
        \begin{fmfgraph*}(25,15)
            \fmfleft{i}
            \fmfright{o}
            \fmf{plain}{i,v1}
            \fmf{plain}{o,v4}
            \fmfn{phantom}{v}{4}
            \fmffreeze
            \fmfforce{(0.1w,0.5h)}{v1}
            \fmfforce{(0.367w,0.5h)}{v2}
            \fmfforce{(0.633w,0.5h)}{v3}
            \fmfforce{(0.9w,0.5h)}{v4}
            \fmf{plain,left}{v1,v2}
            \fmf{plain,right}{v1,v2}
            \fmf{plain,left}{v2,v3}
            \fmf{plain,right}{v2,v3}
            \fmf{plain,left}{v3,v4}
            \fmf{plain,right}{v3,v4}
            \fmfdotn{v}{4}
        \end{fmfgraph*}
    }
}

\newcommand{\MasterThreeF}{
    \parbox[c][13mm][c]{15mm}{
        \centering
        \begin{fmfgraph*}(15,15)
            \fmfleft{i}
            \fmfright{o}
            \fmf{plain}{i,v1}
            \fmf{plain}{o,v3}
            \fmfn{phantom}{v}{3}
            \fmffreeze
            \fmfforce{(0.1w,0.5h)}{v1}
            \fmfforce{(0.9w,0.5h)}{v3}
            \fmffreeze
            \fmf{plain,left,tag=1}{v1,v3}
            \fmf{plain,right,tag=2}{v1,v3}
            \fmfposition
            \fmfipath{p[]}
            \fmfiset{p1}{vpath1(__v1,__v3)}
            \fmfiset{p2}{vpath2(__v1,__v3)}
            \fmfi{plain}{point 1length(p1)/4 of p1 .. point 3length(p2)/4 of p2}
            \fmfi{plain}{point 3length(p1)/4 of p1 .. point 1length(p2)/4 of p2}
            \myvert{point 1length(p1)/4 of p1}
            \myvert{point 3length(p1)/4 of p1}
            \myvert{point 1length(p2)/4 of p2}
            \myvert{point 3length(p2)/4 of p2}
            \fmfdot{v1,v3}
        \end{fmfgraph*}
    }
}

%% file: graphs/graphs_topologies.tex
\newcommand{\generaloneloop}[2]{
\parbox[c][20mm][c]{20mm}{
\centering
\begin{fmfgraph*}(10,10)
\fmfleft{i}
\fmfright{o}
\fmfleft{ve}
\fmfright{vo}
\fmffreeze
\fmfforce{(-0.3w,0.5h)}{i}
\fmfforce{(1.3w,0.5h)}{o}
\fmfforce{(0w,0.5h)}{ve}
\fmfforce{(1.0w,0.5h)}{vo}
\fmffreeze
\fmf{plain}{i,ve}
\fmf{plain,left,label=$#1$,l.d=0.1h}{ve,vo}
\fmf{plain,left,label=$#2$,label.side=left,l.d=0.1h}{vo,ve}
\fmf{plain}{vo,o}
\fmffreeze
\fmfdot{ve,vo}
\end{fmfgraph*}
}
}

\newcommand{\generaltwoloop}[5]{
\parbox[c][20mm][c]{20mm}{
\centering
\begin{fmfgraph*}(12,10)
\fmfleft{i}
\fmfright{o}
\fmfleft{ve}
\fmfright{vo}
\fmftop{vn}
\fmftop{vs}
\fmffreeze
\fmfforce{(-0.3w,0.5h)}{i}
\fmfforce{(1.3w,0.5h)}{o}
\fmfforce{(0w,0.5h)}{ve}
\fmfforce{(1.0w,0.5h)}{vo}
\fmfforce{(.5w,1.1h)}{vn}
\fmfforce{(.5w,-0.1h)}{vs}
\fmffreeze
\fmf{plain}{i,ve}
\fmf{plain,left}{ve,vo}
\fmf{phantom,left,label=$#1$,l.d=-0.1w}{ve,vn}
\fmf{phantom,right,label=$#2$,l.d=-0.1w}{vo,vn}
\fmf{plain,left}{vo,ve}
\fmf{phantom,left,label=$#3$,l.d=-0.1w}{vo,vs}
\fmf{phantom,right,label=$#4$,l.d=-0.1w}{ve,vs}
\fmf{plain,label=$#5$,l.d=0.05w}{vs,vn}
\fmf{plain}{vo,o}
\fmffreeze
\fmfdot{ve,vn,vo,vs}
\end{fmfgraph*}
}
}

\newcommand{\GraphGeneraThreeLoopLA}{
\parbox[c][20mm][c]{34mm}{
\centering
\begin{fmfgraph*}(30,25)
\fmfleft{i}
\fmfright{o}
\fmf{plain}{i,v1}
\fmf{plain}{v2,o}
\fmf{phantom,right,tension=0.1,tag=1}{v1,v2}
\fmf{phantom,right,tension=0.1,tag=2}{v2,v1}
\fmf{phantom,tension=0.1,tag=3}{v1,v2}
\fmfdot{v1,v2}
\fmfposition
\fmfipath{p[]}
\fmfiset{p1}{vpath1(__v1,__v2)}
\fmfiset{p2}{vpath2(__v2,__v1)}
\fmfi{plain,label=$\alpha_6$,l.d=0.02w}{subpath (0,length(p1)/3) of p1}
\fmfi{plain,label=$\alpha_5$,l.d=0.04w}{subpath (length(p1)/3,2*length(p1)/3) of p1}
\fmfi{plain,label=$\alpha_4$,l.d=0.02w}{subpath (2*length(p1)/3,length(p1)) of p1}
\fmfi{plain,label=$\alpha_3$,l.d=0.02w}{subpath (0,length(p2)/3) of p2}
\fmfi{plain,label=$\alpha_2$,l.d=0.04w}{subpath (length(p2)/3,2*length(p2)/3) of p2}
\fmfi{plain,label=$\alpha_1$,l.d=0.02w}{subpath (2*length(p2)/3,length(p2)) of p2}
\fmfi{plain,label=$\alpha_7$,l.d=0.01w}{point length(p1)/3 of p1 -- point 2*length(p2)/3 of p2}
\fmfi{plain,label=$\alpha_8$,l.d=0.01w,label.side=left}{point 2*length(p1)/3 of p1 -- point length(p2)/3 of p2}
\myvert{point length(p1)/3 of p1}
\myvert{point 2*length(p1)/3 of p1}
\myvert{point length(p2)/3 of p2}
\myvert{point 2*length(p2)/3 of p2}
\end{fmfgraph*}
}
}

\newcommand{\GraphGeneraThreeLoopBE}{
\parbox[c][20mm][c]{34mm}{
\centering
\begin{fmfgraph*}(30,25)
\fmfleft{i}
\fmfright{o}
\fmf{plain}{i,v1}
\fmf{plain}{v2,o}
\fmf{phantom,right,tension=0.1,tag=1}{v1,v2}
\fmf{phantom,right,tension=0.1,tag=2}{v2,v1}
\fmf{phantom,tension=0.1,tag=3}{v1,v2}
\fmfdot{v1,v2}
\fmfposition
\fmfipath{p[]}
\fmfiset{p2}{vpath1(__v1,__v2)}
\fmfiset{p1}{vpath2(__v2,__v1)}
\fmfiset{p3}{vpath3(__v2,__v1)}
\fmfi{plain,label=$\alpha_3$,l.d=0.02w}{subpath (0,length(p1)/4) of p1}
\fmfi{plain,label=$\alpha_2$,l.d=0.04w}{subpath (length(p1)/4,3*length(p1)/4) of p1}
\fmfi{plain,label=$\alpha_1$,l.d=0.02w}{subpath (3*length(p1)/4,length(p1)) of p1}
\fmfi{plain,label=$\alpha_5$,l.d=0.02w}{subpath (0,length(p2)/2) of p2}
\fmfi{plain,label=$\alpha_4$,l.d=0.02w}{subpath (length(p2)/2,length(p2)) of p2}
\fmfi{plain,label=$\alpha_7$,label.side=left,l.d=0.001w}{point length(p1)/4 of p1 -- point length(p3)/2 of p3}
\fmfi{plain,label=$\alpha_6$,label.side=right,l.d=0.001w}{point 3*length(p1)/4 of p1 -- point length(p3)/2 of p3}
\fmfi{plain,label=$\alpha_8$,l.d=0.02w}{point length(p2)/2 of p2 -- point length(p3)/2 of p3}
\myvert{point length(p1)/4 of p1}
\myvert{point 3*length(p1)/4 of p1}
\myvert{point length(p3)/2 of p3}
\myvert{point length(p2)/2 of p2}
\end{fmfgraph*}
}
}

\newcommand{\GraphGeneraThreeLoopNP}{
\parbox[c][20mm][c]{34mm}{
\centering
\begin{fmfgraph*}(30,25)
\fmfleft{i}
\fmfright{o}
\fmf{plain}{i,v1}
\fmf{plain}{v2,o}
\fmf{phantom,right,tension=0.1,tag=1}{v1,v2}
\fmf{phantom,right,tension=0.1,tag=2}{v2,v1}
\fmf{phantom,tension=0.1,tag=3}{v1,v2}
\fmfdot{v1,v2}
\fmfposition
\fmfipath{p[]}
\fmfiset{p2}{vpath1(__v1,__v2)}
\fmfiset{p1}{vpath2(__v2,__v1)}
\fmfi{plain,label=$\alpha_3$,l.d=0.02w}{subpath (0,length(p1)/4) of p1}
\fmfi{plain,label=$\alpha_2$,l.d=0.04w}{subpath (length(p1)/4,3*length(p1)/4) of p1}
\fmfi{plain,label=$\alpha_1$,l.d=0.02w}{subpath (3*length(p1)/4,length(p1)) of p1}
\fmfi{plain,label=$\alpha_6$,l.d=0.02w}{subpath (0,length(p2)/4) of p2}
\fmfi{plain,label=$\alpha_5$,l.d=0.04w}{subpath (length(p2)/4,3*length(p2)/4) of p2}
\fmfi{plain,label=$\alpha_4$,l.d=0.02w}{subpath (3*length(p2)/4,length(p2)) of p2}
\fmfi{plain,label=~~$\alpha_7$,label.side=left,l.d=0.05w}{point length(p1)/4 of p1 -- point length(p2)/4 of p2}
\fmfi{plain,label=~~$\alpha_8$,label.side=left,l.d=+0.05w}{point 3*length(p1)/4 of p1 -- point 3*length(p2)/4 of p2}
\myvert{point length(p1)/4 of p1}
\myvert{point 3*length(p1)/4 of p1}
\myvert{point length(p2)/4 of p2}
\myvert{point 3*length(p2)/4 of p2}
\end{fmfgraph*}
}
}

\newcommand{\GraphTolopogyLfour}{
\parbox[c][20mm][c]{20mm}{
\centering
\begin{fmfgraph*}(20,20)
\fmfleft{i}
\fmfright{o}
\fmf{plain}{i,v1}
\fmf{plain}{o,v2}
\fmfdot{v1,v2}
\fmffreeze
\fmfforce{(0.1w,0.5h)}{v1}
\fmfforce{(0.9w,0.5h)}{v2}
\fmf{plain,left=1,tag=1}{v1,v2}
\fmf{plain,right=1,tag=2}{v1,v2}
\fmfposition
\fmfipath{p[]}
\fmfiset{p1}{vpath1(__v1,__v2)}
\fmfiset{p2}{vpath2(__v2,__v1)}
\fmfi{plain}{point 1*length(p1)/3 of p1 -- point 1*length(p2)/3 of p2}
\fmfi{plain}{point 1*length(p1)/2 of p1 -- point 1*length(p2)/2 of p2}
\fmfi{plain}{point 2*length(p1)/3 of p1 -- point 2*length(p2)/3 of p2}
\myvert{point 1*length(p1)/3 of p1}
\myvert{point 1*length(p1)/2 of p1}
\myvert{point 2*length(p1)/3 of p1}
\myvert{point 1*length(p2)/3 of p2}
\myvert{point 1*length(p2)/2 of p2}
\myvert{point 2*length(p2)/3 of p2}
\end{fmfgraph*}
}
}

\newcommand{\GraphTolopogyHfour}{
\parbox[c][20mm][c]{20mm}{
\centering
\begin{fmfgraph*}(20,20)
\fmfleft{i}
\fmfright{o}
\fmf{plain}{i,v1}
\fmf{plain}{o,v2}
\fmfdot{v1,v2}
\fmffreeze
\fmfforce{(0.1w,0.5h)}{v1}
\fmfforce{(0.9w,0.5h)}{v2}
\fmf{plain,left=1,tag=1}{v1,v2}
\fmf{plain,right=1,tag=2}{v1,v2}
\fmf{phantom,tag=3}{v1,v2}
\fmfposition
\fmfipath{p[]}
\fmfiset{p1}{vpath1(__v1,__v2)}
\fmfiset{p2}{vpath2(__v2,__v1)}
\fmfiset{p3}{vpath3(__v2,__v1)}
\fmfi{plain}{point 1*length(p3)/4 of p3 .. point 3*length(p3)/4 of p3}
\fmfi{plain}{point 1*length(p1)/3 of p1 .. point 1*length(p2)/3 of p2}
\fmfi{plain}{point 2*length(p1)/3 of p1 .. point 2*length(p2)/3 of p2}
\myvert{point 1*length(p3)/4 of p3}
\myvert{point 3*length(p3)/4 of p3}
\myvert{point 1*length(p1)/3 of p1}
\myvert{point 1*length(p2)/3 of p2}
\myvert{point 2*length(p1)/3 of p1}
\myvert{point 2*length(p2)/3 of p2}
\end{fmfgraph*}
}
}

\newcommand{\GraphTolopogyFfour}{
\parbox[c][20mm][c]{20mm}{
\centering
\begin{fmfgraph*}(20,20)
\fmfleft{i}
\fmfright{o}
\fmf{plain}{i,v1}
\fmf{plain}{o,v2}
\fmfdot{v1,v2}
\fmffreeze
\fmfforce{(0.1w,0.5h)}{v1}
\fmfforce{(0.9w,0.5h)}{v2}
\fmf{plain,left=1,tag=1}{v1,v2}
\fmf{plain,right=1,tag=2}{v1,v2}
\fmf{phantom,tag=3}{v1,v2}
\fmf{phantom,tag=4}{w1,w2}
\fmfforce{(0.1w,0.5h)}{w1}
\fmfforce{(0.5w,0.3h)}{w2}
\fmfposition
\fmfipath{p[]}
\fmfiset{p1}{vpath1(__v1,__v2)}
\fmfiset{p2}{vpath2(__v2,__v1)}
\fmfiset{p3}{vpath3(__v1,__v2)}
\fmfiset{p4}{vpath4(__w1,__w2)}
\fmfi{plain}{point 1*length(p1)/4 of p1 -- point 1*length(p3)/2 of p3}
\fmfi{plain}{point 3*length(p1)/4 of p1 -- point 1*length(p3)/2 of p3}
\fmfi{plain}{point 1*length(p3)/2 of p3 -- point 1*length(p2)/2 of p2}
\fmfi{plain}{point 1*length(p1)/8 of p1 -- point 1*length(p4)/1 of p4}
\myvert{point 1*length(p1)/4 of p1}
\myvert{point 1*length(p3)/2 of p3}
\myvert{point 3*length(p1)/4 of p1}
\myvert{point 1*length(p3)/2 of p3}
\myvert{point 1*length(p3)/2 of p3}
\myvert{point 1*length(p2)/2 of p2}
\myvert{point 1*length(p1)/8 of p1}
\myvert{point 1*length(p4)/1 of p4}
\end{fmfgraph*}
}
}

\newcommand{\GraphTolopogyDfour}{
\parbox[c][20mm][c]{20mm}{
\centering
\begin{fmfgraph*}(20,20)
\fmfleft{i}
\fmfright{o}
\fmf{plain}{i,v1}
\fmf{plain}{o,v2}
\fmfdot{v1,v2}
\fmffreeze
\fmfforce{(0.1w,0.5h)}{v1}
\fmfforce{(0.9w,0.5h)}{v2}
\fmf{plain,left=1,tag=1}{v1,v2}
\fmf{plain,right=1,tag=2}{v1,v2}
\fmf{plain,left=1,tag=3}{w1,w2}
\fmf{plain,right=1,tag=4}{w1,w2}
\fmfforce{(0.345w,0.5h)}{w1}
\fmfforce{(0.655w,0.5h)}{w2}
\fmfposition
\fmfipath{p[]}
\fmfiset{p1}{vpath1(__v1,__v2)}
\fmfiset{p2}{vpath2(__v2,__v1)}
\fmfiset{p3}{vpath3(__w2,__w1)}
\fmfiset{p4}{vpath4(__w2,__w1)}
\fmfi{plain}{point 1*length(p1)/4 of p1 .. point 1*length(p3)/4 of p3}
\fmfi{plain}{point 3*length(p1)/4 of p1 .. point 3*length(p3)/4 of p3}
\fmfi{plain}{point 1*length(p4)/2 of p4 .. point 1*length(p2)/2 of p2}
\myvert{point 1*length(p1)/4 of p1}
\myvert{point 1*length(p3)/4 of p3}
\myvert{point 3*length(p1)/4 of p1}
\myvert{point 3*length(p3)/4 of p3}
\myvert{point 1*length(p4)/2 of p4}
\myvert{point 1*length(p2)/2 of p2}
\end{fmfgraph*}
}
}

\newcommand{\GraphTolopogyCfour}{
\parbox[c][20mm][c]{20mm}{
\centering
\begin{fmfgraph*}(20,20)
\fmfleft{i}
\fmfright{o}
\fmf{plain}{i,v1}
\fmf{plain}{o,v2}
\fmfdot{v1,v2}
\fmffreeze
\fmfforce{(0.1w,0.5h)}{v1}
\fmfforce{(0.9w,0.5h)}{v2}
\fmf{plain,left=1,tag=1}{v1,v2}
\fmf{plain,right=1,tag=2}{v1,v2}
\fmf{phantom,tag=3}{v1,v2}
\fmfposition
\fmfipath{p[]}
\fmfiset{p1}{vpath1(__v1,__v2)}
\fmfiset{p2}{vpath2(__v2,__v1)}
\fmfiset{p3}{vpath3(__v1,__v2)}
\fmfi{plain}{point 1*length(p1)/8 of p1 -- point 5*length(p2)/8 of p2}
\fmfi{plain}{point 1*length(p3)/2 of p3 -- point 1*length(p2)/2 of p2}
\fmfi{plain}{point 1*length(p1)/4 of p1 -- point 1*length(p3)/2 of p3}
\fmfi{plain}{point 3*length(p1)/4 of p1 -- point 1*length(p3)/2 of p3}
\myvert{point 1*length(p1)/8 of p1}
\myvert{point 5*length(p2)/8 of p2}
\myvert{point 1*length(p3)/2 of p3}
\myvert{point 1*length(p2)/2 of p2}
\myvert{point 1*length(p1)/4 of p1}
\myvert{point 1*length(p3)/2 of p3}
\myvert{point 3*length(p1)/4 of p1}
\myvert{point 1*length(p3)/2 of p3}
\end{fmfgraph*}
}
}

\newcommand{\GraphTolopogyJfour}{
\parbox[c][20mm][c]{20mm}{
\centering
\begin{fmfgraph*}(20,20)
\fmfleft{i}
\fmfright{o}
\fmf{plain}{i,v1}
\fmf{plain}{o,v2}
\fmfdot{v1,v2}
\fmffreeze
\fmfforce{(0.1w,0.5h)}{v1}
\fmfforce{(0.9w,0.5h)}{v2}
\fmf{plain,left=1,tag=1}{v1,v2}
\fmf{plain,right=1,tag=2}{v1,v2}
\fmfposition
\fmfipath{p[]}
\fmfiset{p1}{vpath1(__v1,__v2)}
\fmfiset{p2}{vpath2(__v2,__v1)}
\fmfi{plain}{point 1*length(p1)/4 of p1 -- point 2*length(p2)/4 of p2}
\fmfi{plain}{point 2*length(p1)/4 of p1 -- point 1*length(p2)/4 of p2}
\fmfi{plain}{point 2*length(p1)/3 of p1 -- point 2*length(p2)/3 of p2}
\myvert{point 1*length(p1)/4 of p1}
\myvert{point 2*length(p2)/4 of p2}
\myvert{point 2*length(p1)/4 of p1}
\myvert{point 1*length(p2)/4 of p2}
\myvert{point 2*length(p1)/3 of p1}
\myvert{point 2*length(p2)/3 of p2}
\end{fmfgraph*}
}
}

\newcommand{\GraphTolopogyBfour}{
\parbox[c][20mm][c]{20mm}{
\centering
\begin{fmfgraph*}(20,20)
\fmfleft{i}
\fmfright{o}
\fmf{plain}{i,v1}
\fmf{plain}{o,v2}
\fmfdot{v1,v2}
\fmffreeze
\fmfforce{(0.1w,0.5h)}{v1}
\fmfforce{(0.9w,0.5h)}{v2}
\fmf{plain,left=1,tag=1}{v1,v2}
\fmf{plain,right=1,tag=2}{v1,v2}
\fmf{phantom,tag=3}{v1,v2}
\fmfposition
\fmfipath{p[]}
\fmfiset{p1}{vpath1(__v1,__v2)}
\fmfiset{p2}{vpath2(__v2,__v1)}
\fmfiset{p3}{vpath3(__v2,__v1)}
\fmfi{plain}{point 2*length(p1)/7 of p1 -- point 1.02*length(p3)/3 of p3}
\fmfi{plain}{point 2*length(p1)/4 of p1 -- point 1.02*length(p3)/3 of p3}
\fmfi{plain}{point 1.02*length(p3)/3 of p3 -- point 2*length(p2)/5 of p2}
\fmfi{plain}{point 2*length(p1)/3 of p1 -- point 2*length(p2)/3 of p2}
\myvert{point 2*length(p1)/7 of p1}
\myvert{point 1.02*length(p3)/3 of p3}
\myvert{point 2*length(p1)/4 of p1}
\myvert{point 1.02*length(p3)/3 of p3}
\myvert{point 1.02*length(p3)/3 of p3}
\myvert{point 2*length(p2)/5 of p2}
\myvert{point 2*length(p1)/3 of p1}
\myvert{point 2*length(p2)/3 of p2}
\end{fmfgraph*}
}
}

\newcommand{\GraphTolopogyGfour}{
\parbox[c][20mm][c]{20mm}{
\centering
\begin{fmfgraph*}(20,20)
\fmfleft{i}
\fmfright{o}
\fmf{plain}{i,v1}
\fmf{plain}{o,v2}
\fmfdot{v1,v2}
\fmffreeze
\fmfforce{(0.1w,0.5h)}{v1}
\fmfforce{(0.9w,0.5h)}{v2}
\fmf{plain,left=1,tag=1}{v1,v2}
\fmf{plain,right=1,tag=2}{v1,v2}
\fmf{plain}{w1,w2}
\fmfforce{(0.33w,0.72h)}{w1}
\fmfforce{(0.7w,0.3h)}{w2}
\fmfdot{w1,w2}
\fmfposition
\fmfipath{p[]}
\fmfiset{p1}{vpath1(__v1,__v2)}
\fmfiset{p2}{vpath2(__v2,__v1)}
\fmfi{plain}{point 1*length(p1)/6 of p1 -- point 5*length(p1)/6 of p1}
\fmfi{plain}{point 2*length(p1)/3 of p1 -- point 2*length(p2)/3 of p2}
\myvert{point 1*length(p1)/6 of p1}
\myvert{point 5*length(p1)/6 of p1}
\myvert{point 2*length(p1)/3 of p1}
\myvert{point 2*length(p2)/3 of p2}
\end{fmfgraph*}
}
}

\newcommand{\GraphTolopogyNfour}{
\parbox[c][20mm][c]{20mm}{
\centering
\begin{fmfgraph*}(20,20)
\fmfleft{i}
\fmfright{o}
\fmf{plain}{i,v1}
\fmf{plain}{o,v2}
\fmfdot{v1,v2}
\fmffreeze
\fmfforce{(0.1w,0.5h)}{v1}
\fmfforce{(0.9w,0.5h)}{v2}
\fmf{plain,left=1,tag=1}{v1,v2}
\fmf{plain,right=1,tag=2}{v1,v2}
\fmfposition
\fmfipath{p[]}
\fmfiset{p1}{vpath1(__v1,__v2)}
\fmfiset{p2}{vpath2(__v2,__v1)}
\fmfi{plain}{point 1*length(p1)/4 of p1 -- point 2*length(p2)/4 of p2}
\fmfi{plain}{point 2*length(p1)/4 of p1 -- point 3*length(p2)/4 of p2}
\fmfi{plain}{point 3*length(p1)/4 of p1 -- point 1*length(p2)/4 of p2}
\myvert{point 1*length(p1)/4 of p1}
\myvert{point 2*length(p1)/4 of p1}
\myvert{point 3*length(p1)/4 of p1}
\myvert{point 1*length(p2)/4 of p2}
\myvert{point 2*length(p2)/4 of p2}
\myvert{point 3*length(p2)/4 of p2}
\end{fmfgraph*}
}
}

\newcommand{\GraphTolopogyMfour}{
\parbox[c][20mm][c]{20mm}{
\centering
\begin{fmfgraph*}(20,20)
\fmfleft{i}
\fmfright{o}
\fmf{plain}{i,v1}
\fmf{plain}{o,v2}
\fmfdot{v1,v2}
\fmffreeze
\fmfforce{(0.1w,0.5h)}{v1}
\fmfforce{(0.9w,0.5h)}{v2}
\fmf{plain,left=1,tag=1}{v1,v2}
\fmf{plain,right=1,tag=2}{v1,v2}
\fmfposition
\fmfipath{p[]}
\fmfiset{p1}{vpath1(__v1,__v2)}
\fmfiset{p2}{vpath2(__v2,__v1)}
\fmfi{plain}{point 2*length(p1)/4 of p1 -- point 2*length(p2)/4 of p2}
\fmfi{plain}{point 1*length(p1)/4 of p1 -- point 3*length(p2)/4 of p2}
\fmfi{plain}{point 3*length(p1)/4 of p1 -- point 1*length(p2)/4 of p2}
\myvert{point 1*length(p1)/4 of p1}
\myvert{point 2*length(p1)/4 of p1}
\myvert{point 3*length(p1)/4 of p1}
\myvert{point 1*length(p2)/4 of p2}
\myvert{point 2*length(p2)/4 of p2}
\myvert{point 3*length(p2)/4 of p2}
\end{fmfgraph*}
}
}

\newcommand{\GraphTolopogyEfour}{
\parbox[c][20mm][c]{20mm}{
\centering
\begin{fmfgraph*}(20,20)
\fmfleft{i}
\fmfright{o}
\fmf{plain}{i,v1}
\fmf{plain}{o,v2}
\fmfdot{v1,v2}
\fmf{plain}{w1,w2}
\fmffreeze
\fmfforce{(0.1w,0.5h)}{v1}
\fmfforce{(0.9w,0.5h)}{v2}
\fmfforce{(0.5w,0.7h)}{w1}
\fmfforce{(0.5w,0.3h)}{w2}
\fmf{plain,left=1,tag=1}{v1,v2}
\fmf{plain,right=1,tag=2}{v1,v2}
\fmfposition
\fmfipath{p[]}
\fmfiset{p1}{vpath1(__v1,__v2)}
\fmfiset{p2}{vpath2(__v2,__v1)}
\fmfi{plain}{point 1*length(p1)/6 of p1 .. point 5*length(p1)/6 of p1}
\fmfi{plain}{point 1*length(p2)/6 of p2 .. point 5*length(p2)/6 of p2}
\myvert{point 1*length(p1)/6 of p1}
\myvert{point 5*length(p1)/6 of p1}
\myvert{point 1*length(p2)/6 of p2}
\myvert{point 5*length(p2)/6 of p2}
\end{fmfgraph*}
}
}

%% file: texcode/graphs_settings.tex
\newcommand{\marrow}[5]{%
\fmfcmd{style_def marrow#1
expr p = drawarrow subpath(1/4, 3/4)of p shifted 6 #2 withpen pencircle scaled 0.4;
LaTeX_text(point 0.5 of p shifted 10 #2, #3, "#4");
enddef;}
\fmf{marrow#1,tension=0}{#5}
}

\newcommand{\Marrow}[6]{%
\fmfcmd{style_def marrow#1
expr p = drawarrow subpath (1/4, 3/4) of p shifted #6 #2 withpen pencircle scaled 0.4;
label.#3(btex #4 etex, point 0.5 of p shifted #6 #2);
enddef;}
\fmf{marrow#1,tension=0}{#5}}

\def\myvert#1{\fmfiv{decor.shape=circle,decor.filled=full,decor.size=2thick}{#1}}

\fmfset{dash_len}{2mm} 
\fmfset{arrow_len}{2.5mm} 
\fmfset{arrow_ang}{19} 

\newcommand*\circled[1]{\tikz[baseline=(char.base)]{\node[thick, shape=circle,draw,fill=white,inner sep=2pt] (char) {\tiny #1};}}

\newcommand\custompoint[3]{
\node[] at (#1,#2) {|};
\fill[draw=none,fill=black,opacity=0.3] ($(0,#2)+(0,-0.1)$) rectangle ($(#1,#2)+(0,0.1)$);
\node[above=0.1cm,anchor=south west] at ($(#1,#2)+(-0.2,0)$) {#3};
\node[below=0.2cm] at (#1,#2) {#1};
}

\newcommand\customtick[2]{
\node[] at (#1,#2) {|};
\node[below=0.3cm] at (#1,#2) {#1};
}

\newcommand\fmfsetbaseline{
\fmf{plain}{base1,base2}
\fmfforce{(0w,0h)}{base1}
\fmfforce{(0w,1h)}{base2}
}

%% file: texcode/shortcuts.tex
\newcommand{\com}[1]{\textcolor{magenta}{{#1}}} 
\newcommand{\coms}[1]{\textcolor{red}{{#1}}} 

\newcommand{\eg}{{\it e.g.}}
\newcommand{\ie}{{\it i.e.}}

\newcommand{\I}{\ensuremath{{\mathrm{i}}}}
\newcommand{\D}{\ensuremath{\mathrm{d}}}
\newcommand{\Ord}{{\rm O}}
\newcommand{\eps}{\ensuremath{\varepsilon}}
\newcommand{\al}{\alpha}
\newcommand{\bra}{\langle }
\newcommand{\ket}{\rangle }
\newcommand{\abs}[1]{\left|#1\right|}
\def\ra{{\rightarrow}}
\newcommand{\fsl}[1]{{\centernot{#1}}} 

\newcommand{\EP}{\mathcal{E}} 
\newcommand{\xib}{\bar{\xi}} 
\newcommand{\Lpt}{\tilde{L}} 
\newcommand{\KK}[1]{\mathcal{K}\!\left[\!\!#1\!\!\right]} 
\newcommand{\pE}{p_{{}_{\!E}}} 
\newcommand{\kE}{k_{{}_{\!E}}} 

\newcommand{\cE}{\ensuremath{\mathcal{E}}}
\newcommand{\cL}{\ensuremath{\mathcal{L}}}
\newcommand{\cC}{\ensuremath{\mathcal{C}}}
\newcommand{\cN}{\ensuremath{\mathcal{N}}}
\newcommand{\C}{\mathbb{C}}
\newcommand{\N}{\mathbb{N}}
\newcommand{\Q}{\mathbb{Q}}
\newcommand{\R}{\mathbb{R}}
\newcommand{\Z}{\mathbb{Z}}

\def\be{\begin{equation}}
\def\ee{\end{equation}}
\def\ba#1\ea{\begin{align}#1\end{align}}
\def\bs{\begin{subequations}}
\def\es{\end{subequations}}
\def\nonum{\nonumber}

\makeatletter
\newcommand{\vast}{\bBigg@{4}}
\newcommand{\Vast}{\bBigg@{5}}
\makeatother

\newcommand*\widefbox[1]{\fbox{\hspace{0.25cm}#1\hspace{0.25cm}}} 
\newcommand\titlemath[1]{\texorpdfstring{\(#1\)}{xx}} 
\newcommand\phantomfrac{\phantom{$\cfrac{1}{1}$}\!\!\!} 
\newcommand\myclearpage{\checkoddpage\ifoddpage\else\newpage\mbox{}\fi}
\newcommand\mypart[1]{\myclearpage\part{\textsc{\textbf{#1}}}}

%% file: plots/plots.tex
\newcommand{\Plotetafit}{
\begin{tikzpicture}[scale=1]
\begin{axis}[
xmin = 0, xmax = 10,
ymin = 0.8, ymax = 1,
grid = both,
minor tick num = 1,
major grid style = {lightgray},
minor grid style = {lightgray!25},
width = 0.65\textwidth,
height = 0.4\textwidth,
xlabel = {Loop order ($L$)},
ylabel = {Anomalous stiffness $\eta(\text{P}_4)$},
legend cell align = {left}
]
\addplot[
domain = 0:10,
samples = 200,
smooth,
thick,
black,
] {0.834677 + 0.195333 * exp(-0.445534*x)};

\addplot[
domain = 0:10,
samples = 20,
smooth,
thick,
black,
dashed,
] {0.835};

\addplot[
smooth,
thick,
black,
only marks,
mark=square
] file[skip first] {plots/plot.dat};

\legend{$\eta(\text{P}_4)=\eta_{\text{all-order}}(\text{P}_4)+0.195 e^{-0.446 L} $, $\eta_{\text{all-order}}(\text{P}_4)=0.8347$}
\end{axis}
\end{tikzpicture}
}